\documentclass[review]{elsarticle}
\usepackage{booktabs} 
\usepackage{comment}
\usepackage{url}
\usepackage{amsfonts}
\usepackage{pifont}
\usepackage{caption}
\usepackage{subcaption}
\usepackage{balance}
\usepackage{setspace}
\usepackage[linesnumbered,ruled,vlined,lined,boxed]{algorithm2e}
\usepackage[acronym,toc,shortcuts]{glossaries}
\usepackage{color}
\usepackage{listings} 
\usepackage{enumitem}
\usepackage{refcount}
\usepackage{textcomp} 
\usepackage{cleveref} 
\usepackage{amsmath,amsthm,amssymb}
\usepackage{pdflscape} 
\usepackage{multirow}
\usepackage{float} 
\usepackage[export]{adjustbox} 
\usepackage{xcolor,pifont}
\newcommand*\colourcheck[1]{%
  \expandafter\newcommand\csname #1check\endcsname{\textcolor{#1}{\ding{55}}}%
}
\usepackage[title]{appendix} 
\PassOptionsToPackage{normalem}{ulem}
\usepackage{ulem}
\providecolor{added}{rgb}{0,0,1}
\providecolor{deleted}{rgb}{1,0,0}

\usepackage[disable]{todonotes} 
\newcommand{\mi}[1]{\mathit{#1}}

\newcommand{\ml}[1]{\todo[fancyline,color=yellow,size=\tiny]{\textbf{ML: }#1 \color{black}}}
\newcommand{\mli}[1]{\todo[inline,color=yellow]{\textbf{}#1 \color{black}}}

\SetKwProg{Function}{function}{}{end function}
\SetKwInOut{KwVar}{Variables}
\SetKwInOut{KwDescribe}{Description}
\SetKw{KwTo}{to}
\SetKw{KwAnd}{and}
\SetKwIF{If}{ElseIf}{Else}{if}{then}{else if}{else}{end if}
\SetKwFor{For}{for}{do}{end for}
\SetKwFor{ForEach}{for each}{do}{end for}
\SetKwFor{ForAll}{for all}{do}{end for}
\SetKwFor{ForAllP}{for all}{do in parallel}{end for}

\newcommand{\code}[1]{{\fontfamily{lmtt}\small\selectfont#1}}
\newcommand{\system}{\textsc{Tcep}\xspace}
\newcommand{\eg}{e.g.,\xspace}
\newcommand{\ie}{i.e.,\xspace}

\lstdefinelanguage{Scala}{
	morekeywords={},
	morekeywords=[2]{
		WHERE,JOIN,ON,WINDOW,SLIDING,NOT,
		DEMAND,LATENCY,DEMAND,PROXIMITY,WHEN,
		abstract,case,catch,class,def,
		do,else,extends,false,final,finally,
		for,if,implicit,import,lazy,match,mixin,
		new,null,object,override,package,
		private,protected,requires,return,sealed,
		super,this,throw,trait,true,try,
		type,val,var,while,with,yield,
		window, join, on, where, within, demand, latency, proximity, when},
	sensitive=true,
	morecomment=[l]{//},
	morecomment=[n]{/*}{*/},
	morecomment=[s][identifierstyle]{`}{`},
	morestring=[b]",
	morestring=[b]',
	morestring=[b]"""
}

\lstdefinelanguage{JavaScript}{
	morekeywords={
		break,const,continue,delete,do,while,export,for,in,function,
		if,else,import,in,instanceOf,label,let,new,return,switch,this,
		throw,try,catch,typeof,var,void,with,yield
	},
	sensitive=true,
	morecomment=[l]{//},
	morecomment=[s]{/*}{*/},
	morestring=[b]",
	morestring=[d]'
}
\theoremstyle{definition}
\newtheorem{definition}{Definition}[section]
\newtheorem{theorem}{Theorem}[section]
\lstnewenvironment{codenv}{\lstset{language=Scala}}{}
\lstnewenvironment{framedcodenv}{\lstset{language=Scala,frame=btlr}}{}
\makeglossaries
\newacronym{CEP}{CEP}{Complex Event Processing}
\newacronym{TCEP}{\textsc{Tcep}}{Transitive-CEP}
\newacronym{OP}{OP}{Operator Placement}
\newacronym{QoS}{QoS}{Quality of Service}
\newacronym{DCEP}{DCEP}{Distributed Complex Event Processing}
\newacronym{GA}{GA}{Genetic Algorithms}
\newacronym{IoT}{IoT}{Internet of Things}
\newacronym{MFGS}{MFGS}{Moving Fine-Grained State}
\newacronym{SMS}{SMS}{Seamless Minimal State}

\journal{Journal of Computer and System Sciences}

\title{
TCEP: Transitions in Operator Placement to Adapt to Dynamic Network Environments}

\author[tuda]{Manisha Luthra}
\ead{manisha.luthra@kom.tu-darmstadt.de}

\author[tuda,groningen]{Boris Koldehofe}
\ead{b.koldehofe@rug.nl}

\author[tuda]{Niels Danger}

\author[tuda]{Pascal Weisenberger}
\author[tuda,stgallen]{Guido Salvaneschi}

\author[athens]{Ioannis Stavrakakis}
\ead{ioannis@di.uoa.gr}

\address[tuda]{Technical University of Darmstadt, Germany}
\address[groningen]{University of Groningen, Netherlands}
\address[stgallen]{University of St Gallen, Switzerland}
\address[athens]{National and Kapodistrian University of Athens, Greece}

\renewcommand{\i}{\textit{(i)}~}
\newcommand{\ii}{\textit{(ii)}~}
\newcommand{\iii}{\textit{(iii)}~}
\newcommand{\iv}{\textit{(iv)}~}

\newboolean{komversion}
\setboolean{komversion}{true} 
\ifthenelse{\boolean{komversion}}{
    \usepackage{eso-pic}
    \AddToShipoutPictureBG{
        \AtPageUpperLeft{%
            \raisebox{-3\baselineskip}{
\makebox[\paperwidth]{\colorbox{yellow!40}{\begin{minipage}{0.85\paperwidth}\centering\small
                            {Manisha Luthra, Boris Koldehofe, Niels Danger, Pascal Weisenburger, Guido Salvaneschi, Ioannis Starvrakakis. \emph{TCEP: Transitions in Operator Placement to Adapt to Dynamic Network Environments.} This is a post-peer-review manuscript accepted for publication in the Journal of Computer and System Sciences, Special Issue on Algorithmic Theory of Dynamic Networks and its Applications.}
                        \end{minipage}}}}%
                    }
                    \AtPageLowerLeft{%
                        \raisebox{3\baselineskip}{
\makebox[\paperwidth]{\fbox{\begin{minipage}{0.85\paperwidth}\scriptsize
                                        {The documents distributed by this server have been provided by the contributing authors as a means to ensure timely dissemination of scholarly and technical work on a non-commercial basis. Copyright and all rights therein are maintained by the authors or by other copyright holders, not withstanding that they have offered their works here electronically. It is understood that all persons copying this information will adhere to the terms and constraints invoked by each author's copyright. These works may not be reposted without the explicit permission of the copyright holder. © 2021. This manuscript version is made available under the CC-BY-NC-ND 4.0 license http://creativecommons.org/licenses/by-nc-nd/4.0/}
                                    \end{minipage}}}}%
                    }
    }
}{}    

\begin{document}

\begin{abstract}

\ac{DCEP} is a commonly used paradigm to detect and act on situational changes of many applications, including the \ac{IoT}.
\ac{DCEP} achieves this using a simple specification of analytical tasks on data streams called operators and their distributed execution on a set of infrastructure.
The adaptivity of \ac{DCEP} to the dynamics of \ac{IoT} applications is essential and very challenging in the face of changing demands concerning Quality of Service.
In our previous work, we addressed this issue by enabling transitions, which allow for the adaptive use of multiple operator placement mechanisms.
In this article, we extend the transition methodology by optimizing the costs of transition and analyzing the behaviour using multiple operator placement mechanisms.
Furthermore, we provide an extensive evaluation on the costs of transition imposed by operator migrations and learning, as it can inflict overhead on the performance if operated uncoordinatedly.

\end{abstract}

\maketitle

\section{Introduction} \label{sec:Introduction}

The unprecedented growth in IoT devices has enabled multiple applications in connected vehicles, financial trading, and industrial manufacturing. 
Cisco predicts that there will be 29.3 billion IoT devices by 2023, and among those, connected vehicles will be the fastest-growing application type~\cite{Cisco}.
IoT applications, especially involving highly mobile components such as connected vehicles, incorporate inherent dynamics in the environment and the required \ac{QoS} demands.
Such applications need to continuously adapt their system's components to meet specific QoS demands related to environmental conditions.
An essential aspect in the adaptation cycle of IoT applications is detecting situational changes that trigger actions at the distributed application components-- 
for instance, detecting and reacting to the change in the density of the vehicles depending on the time of the day, such as rush hours vs regular hours.

\ac{DCEP} is a prevalent and frequently applied paradigm to detect and act on such situational changes.
DCEP analyzes data streams from many distinct data sources and detects event patterns, named \emph{complex events}, corresponding to the situational changes to which IoT applications need to adapt.
The logic to detect such situational changes is modelled using a data flow graph, commonly referred to as an \emph{operator graph}. 
An operator graph represents the computational units that help detect complex events, named \emph{operators}, which are interconnected by data streams. 
DCEP needs to ensure that events of interest or complex events are delivered while meeting the specified \ac{QoS} demands of the IoT application.
For instance, a connected car application that shares contextual information between multiple vehicles for time-critical and safety-critical decisions has a latency demand of delivering information in less than 30ms~\cite{IEEEV2X}.
A central mechanism of a DCEP system towards fulfilling such \ac{QoS} demands is an \ac{OP} mechanism that dictates the assignment of the operators on the resources of the IoT infrastructure.
The placement of operators on resources, such as on the things (IoT devices) at the edge, or resources at the fog~\cite{C2:Luthra2019a}, or inside data centers, helps accomplish the specified \ac{QoS} demands, such as low latency, bandwidth efficiency, or reliable delivery.

Typically, \ac{DCEP} systems rely on a single \ac{OP} mechanism optimized for one \ac{QoS} demand or combining multiple demands.
For instance, \ac{OP} mechanisms have been widely researched to minimize latency~\cite{Ahmad2004}, to reduce load~\cite{Zhou2006OP, Starks2015,Eidenbenz2016OP}, to minimize network usage (bandwidth-delay product)~\cite{Pietzuch2006,Rizou2010,Schilling2011}, and to preserve trust and privacy~\cite{Dwarakanath2017OP}.
Some authors even combine multiple \ac{QoS} demands in a multi-objective optimization formulation to find Pareto-optimal solutions for operator placement~\cite{op/TPDS/Nardelli2019,placement/sigmetrics/cardellini2017}.
However, under changing QoS demands, which are unknown beforehand, current DCEP systems fail to find a suitable \ac{OP} mechanism because they are restricted to a single \ac{OP} mechanism. 
Furthermore, the \ac{OP} mechanisms are specialized for given environmental conditions, such as %
the mobility of producers or consumers -- stationary or highly mobile. 

Current \ac{OP} mechanisms are known to have trade-offs regarding supported \ac{QoS} demands dependent on the given environmental conditions.
The reasons are twofold, 
\i the conflicting nature of \ac{QoS} demands, such as minimizing latency but limiting the overhead in assigning operators, and 
\ii because \ac{OP} mechanisms favor specific environmental conditions, such as high mobility vs low mobility of connected vehicles.
Instead of aiming for a single universal mechanism supporting all kinds of QoS demands and environmental conditions, we pursue in this article the idea of dynamically changing mechanisms at runtime by introducing and analyzing an adaptation technique named \emph{transition}~\cite{Alt2019}.
The \emph{transition} facilitates dynamic change of mechanisms to benefit ideally from the best suitable mechanisms required under specific environmental conditions.

Introducing transitions in a seamless and non-disruptive manner, \ie without any interruption in the output into \ac{DCEP}, is a highly challenging task and requires careful choice of system mechanisms.
In this article, we aim to solve this challenge in the context of \ac{OP} mechanisms.
A critical issue that we address is to efficiently migrate operator graphs while maintaining the correctness of the results and imposing minimum costs into the DCEP system.
Naively approaching the problem will lead to high overhead in terms of state transfer for stateful operators and communication overhead, which eventually leads to a failure in terms of fulfilment of \ac{QoS} demands. 
Therefore, a systematic selection of an operator placement mechanism is required to fulfil the \ac{QoS} demands.

In this article, we extend our previous findings on \system~\cite{Luthra2018}\footnote{TCEP and its programming model are made publicly available for use. \url{https://luthramanisha.github.io/TCEP/} [Accessed on 21.04.2021]} by
\i proposing a programming model that enables analysis of distinct \ac{OP} mechanisms and their adaptation for various \ac{QoS} demands, 
\ii determining optimal discrete-time points when to perform operator migrations such that the cost is minimal as part of the cost-efficient algorithm, and 
\iii adaptively selecting \ac{OP} mechanisms while maintaining a low overhead using genetic learning methods.

In more detail, this article provides the following \ml{R2: provides following -> provides the following done} contributions:
\begin{enumerate}[leftmargin=*]
\item We formalize the problem of \emph{transitions} for operator placement problem in the DCEP system, considering distinct \ac{QoS} demands of applications, and present the definition of the \emph{cost} that needs to be considered in performing \emph{transitions} in \ac{OP} mechanisms.
\item We present a \emph{programming model} that enables the development of \ac{OP} mechanisms with specific QoS demands, which is used to support \emph{seamless transitions}. 
\item We present and analyze the \emph{genetic learning}-based method for adaptively planning \emph{transitions} between \ac{OP} mechanisms to meet dynamically changing QoS demands and changes in the network environment.
\item We present and analyze two transition algorithms to facilitate the dynamic change of \ac{OP} mechanisms in a \emph{non-disruptive} and \emph{seamless} manner while maintaining the correctness of the results.
\item We present an extensive evaluation to analyze the behaviour of state-of-the-art \ac{OP} mechanisms using distinct queries and analyze their performance on the distributed set of fog-cloud infrastructure, including GENI~\cite{GeniArticle2014}, CloudLab~\cite{CloudLab}, and MAKI~\cite{MAKI2015} resources. 
Furthermore, we analyze the performance of 
\i \emph{mechanism transitions} under dynamics of environmental conditions, 
\ii proposed \emph{transition} algorithm in terms of costs imposed, and 
\iii costs incurred by genetic learning-based selection algorithm.
\end{enumerate}

Our extensive evaluations of \system show in the context of presented traffic congestion detection queries that the \emph{transitions} can be performed in the range between $0.85-2$ seconds while maintaining 100\% throughput in detecting the complex event due to the minimal costs in terms of time and overhead.

The remainder of this article is structured as follows.
We provide a brief introduction to DCEP using an example of the traffic control scenario and motivate \emph{mechanism transitions} by a preliminary evaluation in Section~\ref{sec:motivation}.
We introduce the \system system model in Section~\ref{sec:model} and present the problem statement in \Cref{sec:problem}.
We present the design of \system in Section~\ref{sec:design} and evaluate the \system system in
Section~\ref{sec:evaluation}. 
Finally, in Sections~\ref{sec:relatedwork} and \ref{sec:conclusion}, we
present the related work and conclude our paper, respectively.

\section{The need for transition of \ac{OP} mechanisms}\label{sec:motivation}

To demonstrate and motivate the need for exchanging \ac{OP} mechanisms through transitions, we  first introduce a typical use-case of \ac{CEP} in the context of a traffic control application that is consistently used in this article until later on as part of the evaluation.
Furthermore, we show significant shortcomings of current \ac{CEP} systems for the scenario by performing an initial evaluation study on state-of-the-art placement mechanisms.

\subsection{Complex Event Processing} \label{subsec:CEP}
CEP is a powerful paradigm that detects patterns in the incoming data streams to derive higher-level events such as traffic congestion. 
Consider a traffic control application in an IoT scenario that processes information from different producers such as \emph{smart vehicles} and \emph{radar sensors}. These producers generate continuous data streams comprising of event tuples of the following form--
\emph{vehicle sensor}: $<ts, section\_id, vehicle\_id, vehicle\_speed>$ and \\ \emph{radar sensor}: $<ts, section\_id, no\_vehicles, avg\_speed>$. CEP allows specification of the higher-level events such as traffic congestion in the form of a \emph{query}. A query comprises computational units called \emph{operators} such as \emph{filter}, \emph{join}, and \emph{sequence} that can specify transformations on the data streams. 
CEP operators can be classified as \emph{stateless} such as filter operator, and \emph{stateful} such as window-join, window-aggregate and sequence operators. 
\emph{Stateless} operators perform computation only on the current input tuples, while the \emph{stateful} operators perform computation on the current and past input tuples depending on the semantics of the operator. 
The number of past tuples considered for computation in a stateful operator is typically formulated using a \emph{window} based on time or tuple size. 
While there exist multiple window types, we consider a sliding window in our running example in this article\footnote{Yet, the proposed system is not restricted to sliding windows and can be applied to other types of windows such as tumbling windows.}. Here, slide size refers to the number of event tuples shifted in a given window such that new event tuples from the data stream are included in the next window cycle. 
Moreover, the window size refers to the number of events tuples to be considered for the computation in the current window cycle.  

\begin{figure*}
\begin{subfigure}{\linewidth}
\centering
		\includegraphics[width=\linewidth]{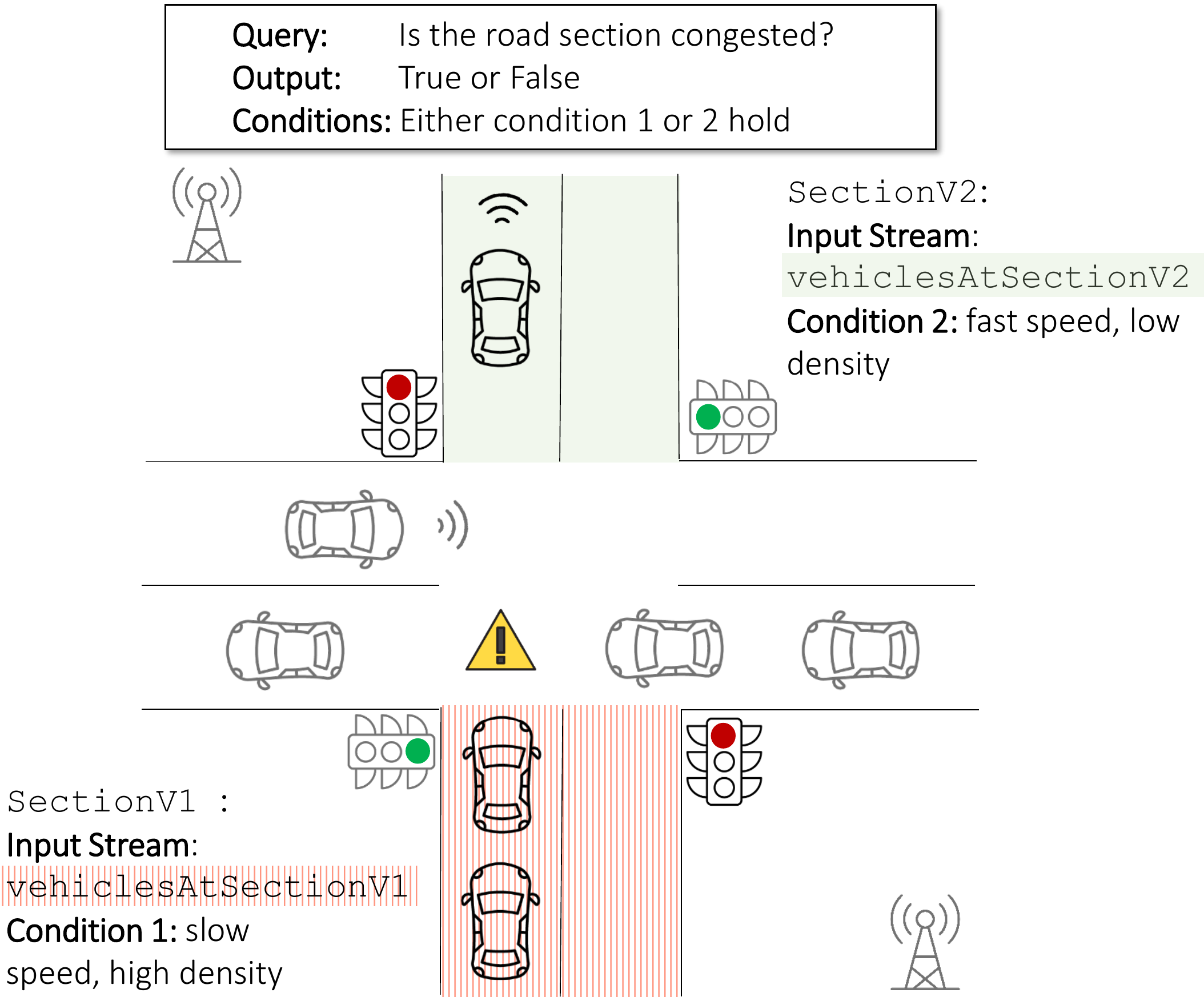}
		\setlength{\abovecaptionskip}{0pt}
		\caption{Congestion detection under dynamic environmental conditions performed by query in (b).}
		\label{fig:querysemantics}
	\end{subfigure}
	\hspace*{\fill}%
	\begin{subfigure}{\linewidth}
	\begin{codenv}
case class VehiclesAtSection(sectionId: Int, avgVehiclesDensity: Long, avgVehiclesSpeed Long, time: Long)(*@\label{lin:roadsection}@*)
val vehiclesAtSectionV1: Stream[VehiclesAtSection] = ...(*@\label{lin:sectionX}@*)
val vehiclesAtSectionV2: Stream[VehiclesAtSection] = ...(*@\label{lin:sectionY}@*)
val congestedAdjacentRoadSections = Query[RoadSections]
  ((vehiclesAtSectionV1 where { v1 =>(*@\label{lin:filterX-start}@*)
     v1.avgVehiclesSpeed < NormalSpeedThreshold &&
     v1.avgVehiclesDensity > HighTrafficThreshold(*@\label{lin:filterX-end}@*)
   })
   -> (*@\label{lin:seuqence-operator}@*)
   (vehiclesAtSectionV2 where { v2 =>(*@\label{lin:filterY-start}@*)
     v2.avgVehiclesSpeed > NormalSpeedThreshold &&
     v2.avgVehiclesDensity < HighTrafficThreshold(*@\label{lin:filterY-end}@*)
   })
   within 1.min (*@\label{lin:time-window}@*)
   where { case (v1, v2) => v2.time > v1.time } (*@\label{lin:filter-last}@*)
   demand QOS_DEMAND) (*@\label{lin:qos-demand}@*)
   \end{codenv}
	\setlength{\abovecaptionskip}{0pt}
   \caption{Query to detect congestion at a road section.}\label{code:samplequery}
\end{subfigure}
\setlength{\abovecaptionskip}{4pt}
\setlength{\belowcaptionskip}{-10pt}
\caption{Traffic congestion control scenario highlighting the change in environmental conditions at the two road sections \texttt{SectionV1} and \texttt{SectionV2} necessitates different \ac{OP} mechanisms for dynamic user environments.}
\label{fig:queryIoTscenario}
\end{figure*}

In the running example, a stateful operation like joining of data streams observed at \texttt{SectionV1} (shaded in lines with red) and \texttt{SectionV2} (shaded in green), which are two road sections of a crossing, result in \emph{composite data streams} seen in \Cref{fig:queryIoTscenario}b: \texttt{vehiclesAtSectionV1} and \texttt{vehiclesAtSectionV2}.
Another example of a stateful operator used in detecting a traffic congestion event is when the sequence of Condition 1 followed by Condition 2 as described in \Cref{fig:queryIoTscenario}a takes place, which is detailed in the next section. 
We note that traffic detection in real applications is more complex than the provided example. 
However, for simplicity and better understandability, we refer to the above example.

CEP can be realized in two ways: \emph{centralized} or \emph{distributed}. While the processing at a single node (centralized) is beneficial for some scenarios, \ac{DCEP} is particularly useful for large scale scenarios as in this work. In this work, we focus on DCEP that comprises multiple nodes, which collaboratively process the query. 
We further detail the problem using the traffic control application in the next section. 

\subsection{Case Study: IoT Traffic Control Application} \label{subsec:case_study}

In this section using the traffic control application introduced in the above \Cref{subsec:CEP}, we show that under the dynamics of environmental conditions, 
 state-of-the-art placement mechanisms~\cite{Pietzuch2006,Starks2015} fail to fulfil \ac{QoS} demands while detecting a traffic congestion event under dynamically changing environmental conditions. 

Let us consider a continuous query\footnote{\label{fn:adaptiveCEP}in the \textsc{AdaptiveCEP} query language written in Scala~\cite{Weisenburger2017}.} to detect that a road section on a crossing is congested, as seen in Figure~\ref{fig:queryIoTscenario}b.
Any consumer can pose a query for a specific road section on the crossing, say \texttt{SectionV1}. 
Examples for a consumer could be emergency services, traffic lights, and all vehicles near \texttt{SectionV1}, which are interested in getting traffic updates. 
The query specifies conditions, such as high traffic density and low vehicle speed on \texttt{SectionV1} and its crossed road section, \texttt{SectionV2}.
The query specifies a sequence (Line~\ref{lin:seuqence-operator}) of such conditions
for \texttt{SectionV1} (Lines \ref{lin:filterX-start}--\ref{lin:filterX-end}) and \texttt{SectionV2} (Lines \ref{lin:filterY-start}--\ref{lin:filterY-end}).
The composite data streams \texttt{vehiclesAtSectionV1} and
\texttt{vehiclesAtSectionV2} are assumed to contain information on the average speed and density. This is done by employing transformation of data streams from heterogeneous sources such as sensor nodes in the \ac{IoT} infrastructure, \eg speed sensors, radar sensors, and road side units, as seen in the previous section.
The complex event: ``congestion of road \texttt{SectionV1}" is successfully detected when the sequence of conditions on \texttt{SectionV1} and \texttt{SectionV2} in a temporal timespan of one minute (Line~\ref{lin:time-window}) indicates \i~dense traffic and slow vehicles for \texttt{SectionV1} and \ii~sparse traffic and fast vehicles for \texttt{SectionV2}, respectively.
  
The execution of the query is performed in a distributed manner on the available resources in the \ac{IoT} infrastructure, such as vehicles, that can directly communicate using techniques like V2X~\cite{v2x/v2xand5g/chen2017} and device-to-device communication~\cite{Dwarakanath2016}. 
The mapping of the operators to these resources is done through an \ac{OP} mechanism, which must account for the \texttt{QoS\_DEMAND} specified within the query. As part of the query specification\footnotemark[\getrefnumber{fn:adaptiveCEP}], these demands such as low latency can be specified according to the users' requirements.

A premise underlying our work is that the same \ac{OP} mechanism cannot accommodate conflicting \ac{QoS} demands.
Therefore, we analyze the ability to fulfil specific QoS demands for the query in Figure~\ref{code:samplequery}
for two popular state-of-the-art \ac{OP} placement mechanisms: \emph{Relaxation}~\cite{Pietzuch2006} and \emph{Mobile DCEP}~\cite{Starks2015}.
The key idea of the \emph{Relaxation} mechanism is to place operators based on a model referred to as a latency space. 
The latency space allows determining communication delays between resources in the \ac{IoT} environment, and the mechanism uses the relation to find a near-optimal embedding of an operator graph with respect to end-to-end latency. 
In contrast, \emph{Mobile DCEP} avoids the overhead in maintaining any topological information, which needs to be updated frequently in a highly dynamic environment. 
Instead, the placement decisions are based on devices within the communication range capable of forming a device-to-device network closer to the data sources. In this way, the authors achieve a sub-optimal embedding of the operator graph at low control message overhead.

We analyzed the above two mechanisms in an \ac{IoT} environment with mobile \ac{IoT} resources (i.e., vehicles in this scenario) under the two crucial QoS demands
\i~end-to-end latency defined as the total time required to detect events, and \ii~control message overhead needed to establish stable communication between the placed operators.  

\begin{figure}[t]
\hspace*{\fill}%
  \subcaptionbox{Relaxation achieves lower end-to-end latency than Mobile DCEP.\label{fig:latency}}{\includegraphics[width=0.48\linewidth]{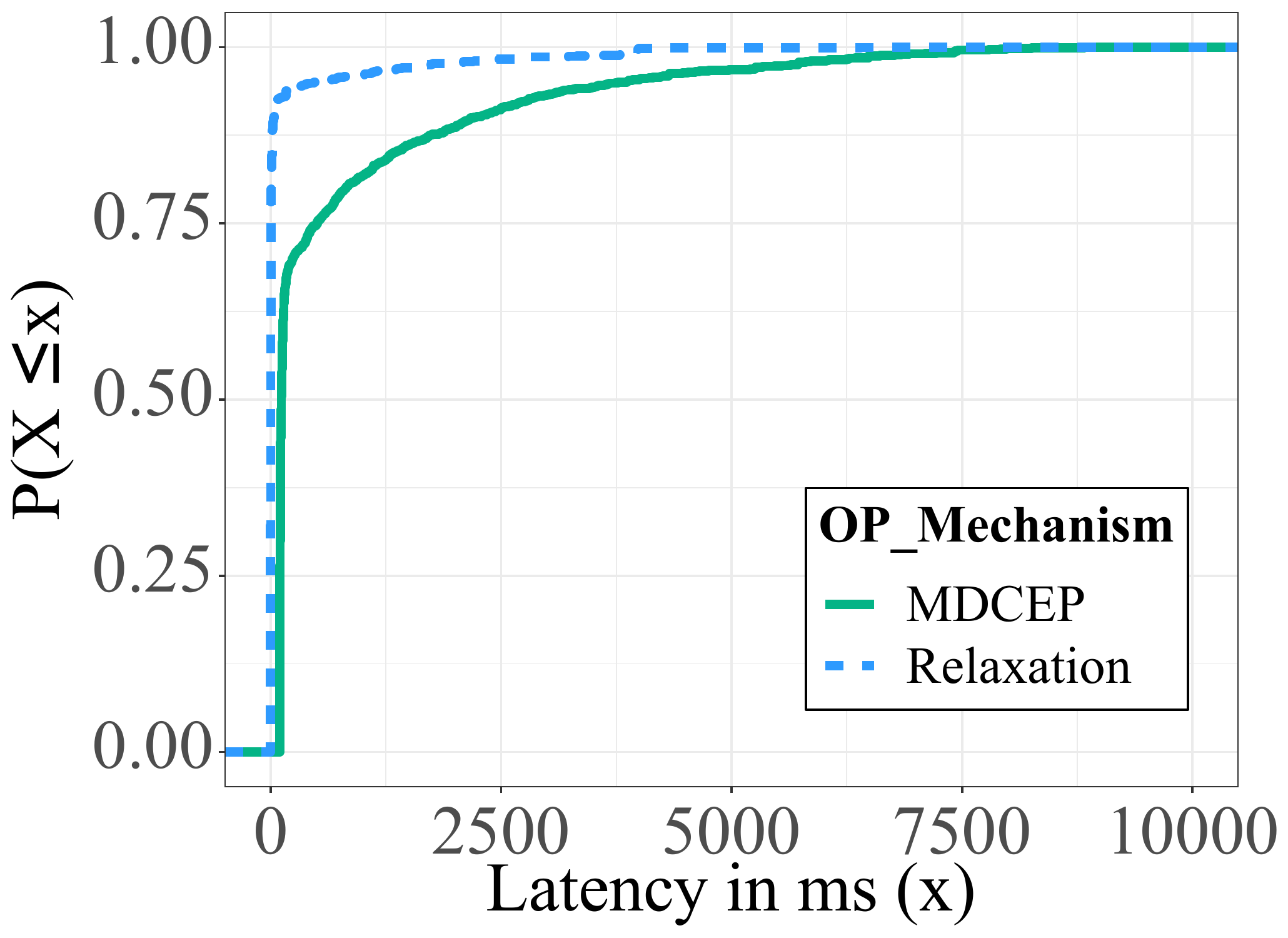}}\hfill%
\subcaptionbox{Mobile DCEP achieves lower message overhead than Relaxation.\label{fig:overhead}}{\includegraphics[width=0.48\linewidth]{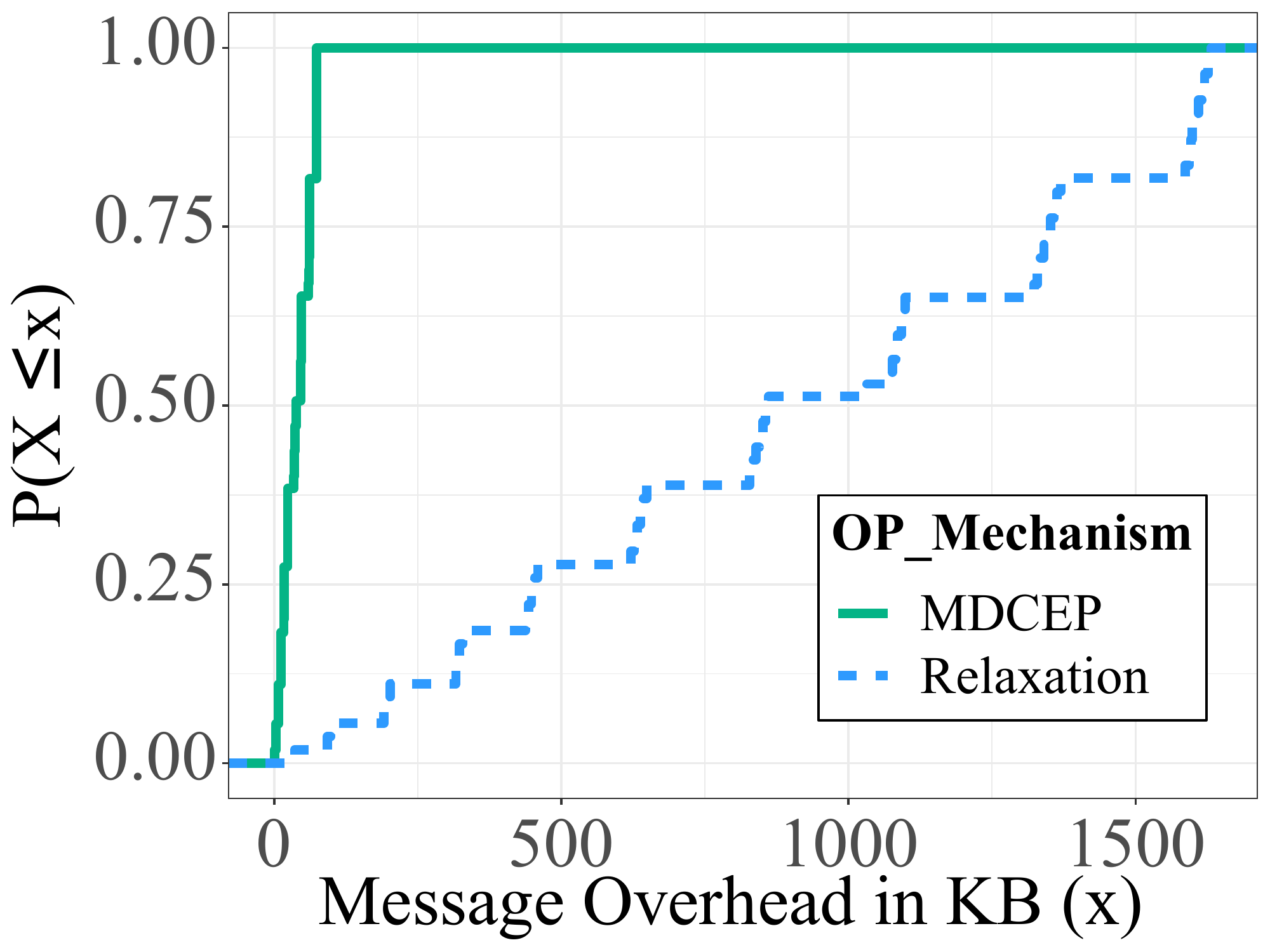}}
\caption{Performance comparison of 
Relaxation~\cite{Pietzuch2006} and Mobile DCEP~\cite{Starks2015} \ac{OP} mechanisms for 50 incrementally deployed queries.}
\label{fig:preliminaryevaluation}
\end{figure} 

Figure~\ref{fig:preliminaryevaluation} shows the measurements on end-to-end latency and control message overhead achieved by the \ac{OP} mechanisms in a dynamic mobile environment for 50 incrementally deployed queries given in \Cref{code:samplequery}. The details on the evaluation configuration can be found later in \Cref{sec:evaluation}.
The cumulative distribution function (CDF) of latency under an increasing number of deployed queries confirms that \emph{Relaxation} achieves consistently very low latency less than 100\,ms for most of the queries, i.e., 80\,\% of the query workload, as seen in \Cref{fig:preliminaryevaluation}a. This is consistent with the findings of
Pietzuch et al.~\cite{Pietzuch2006}. 
However, the control message overhead to coordinate the placement, in this case, to build the latency space, is increasing quickly with the number of deployed queries up to 1500\,KB on average, as seen in \Cref{fig:preliminaryevaluation}b. 
In contrast, \emph{Mobile DCEP} achieves little message overhead for all queries in the order of few bytes allowing for a very stable \ac{OP}, but many queries suffer a long end-to-end latency of $\sim$7.5\,s on average. 

The above preliminary evaluation shows that different \ac{QoS} demands require building on different \ac{OP} mechanisms.
Most importantly, depending on the changing environmental conditions -- high or low mobility and high or low query workload -- different mechanisms must fulfil the specific \ac{QoS} demands.
In a less dynamic environment concerning node mobility, such as with slow-moving vehicles, we measured a significantly lower control overhead for \emph{Relaxation}, and hence it can be used to achieve low latency in condition 1. 
However, when changing from condition 1 (with lower dynamics) to condition 2 (with higher dynamics), a transition from \emph{Relaxation} to \emph{Mobile DCEP} is essential. 
Controlling the overhead improves the stability of the \ac{OP} under the increased dynamics. 
In the presence of a dynamic environment and conflicting QoS demands, it becomes imperative to adapt OP mechanisms, which is the focus of this work.

\section{System Model} \label{sec:model}

In this section, we introduce the system model we use in describing the concepts of \system.
In particular, we introduce the operator graph that models event processing to detect complex events, the IoT resource model that describes the placement infrastructure, the node model that describes the entities participating in the processing of events, \ac{OP} mechanisms and transition model for adapting \ac{DCEP}, and \ac{QoS} demand model which IoT applications use to express their requirements.

\subsection{\system Model} \label{sec:TCEPmodel}

\begin{table}[]
\small
\centering
\begin{tabular}{lp{7cm}}
\hline
Notation & Meaning \\
\hline
  $P$       &  Set of event producers ($p \in P$)   \\
  $C$       &  Set of event consumers ($c \in C$)  \\
  $B$       &  Set of brokers ($b \in B$)  \\
  $D$       &  Continuous data stream   \\
  $E$       & Set of event tuples ($e \in E$) \\
  $\Omega$  &   Set of CEP operators ($\omega \in \Omega$)  \\
  $G$       &  Operator graph           \\
  $f_{\omega}$  &   Processing logic of an operator \\
  $B_I$  &   Input buffer of an operator    \\
  $B_O$  &   Output buffer of an operator   \\
  $M_1 \ldots M_N$ &  \ac{OP} mechanisms \\
  $\alpha$  & Mapping function of the operator $\omega$ \\
  $T$    &   Transition function defining a dynamic change of mechanisms \\
  $en(t)$    &   Environmental conditions dependent on time $t$ \\
\hline
\end{tabular}
\caption{Notations and their meaning.}
\label{tab:notations}
\end{table}

\system consists of \i a set of event producers ($P$), which produce continuous data streams ($D$), \ii a set of event consumers ($C$), which express a complex event on the incoming data streams, and \iii a set of event brokers ({$B$}), which host a set of operators ($\Omega$) to process and forward events. Event consumers specify complex events that represent an event pattern by means of a continuous \ac{DCEP} query.
The query induces a directed acyclic \textit{operator graph} $G = (\Omega \cup P \cup C, \text{ }D)$, comprising of operators, producers, consumers and
data streams, s.t., $D \subseteq (P \cup \Omega) \times (C \cup \Omega)$.
\mli{R3: * Sec 3.1 At the end of page, where the operator graph G is defined, vertexes seem to also include producers and consumers, but it is stated that "each vertex corresponds to an operator". Make sure all definitions are consistent. done}

\begin{figure}[H]
	\centering
	\includegraphics[width=0.8\linewidth]{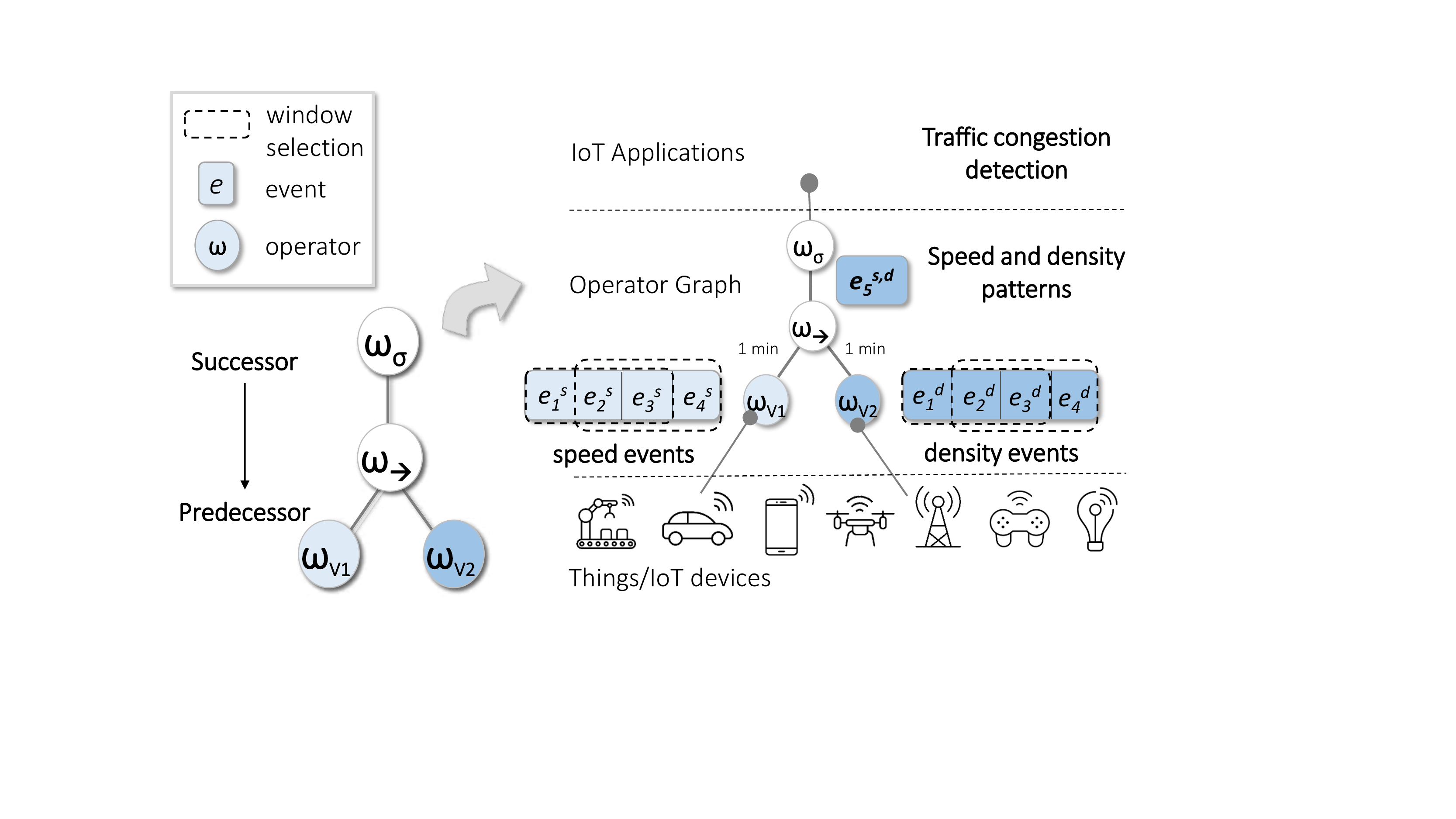}
	\setlength{\abovecaptionskip}{5pt}
	\setlength{\belowcaptionskip}{0pt}
	\caption{Operator graph for the query in Figure~\ref{code:samplequery}.\mli{R1: Figure 3 could be placed closer to page 11 done}}
	\label{fig:operatorgraph}
\end{figure}

The operator graph dictates the execution plan specific to the query given by the event consumer.
Figure~\ref{fig:operatorgraph} illustrates an operator graph for detecting traffic congestion at road sections corresponding to the query in Figure~\ref{code:samplequery}.
The data flow of the events in the operator graph is given from bottom to top of the graph. Here, the operators down the hierarchy are the predecessors (producers are at the last level), while the operators up the hierarchy are the successors (consumers are at the top level).
Operators $\omega_{V1}$ and $\omega_{V2}$ correspond to the window-aggregate operators of the two input streams from the road sections $V1$ and $V2$. Operators $\omega_{\rightarrow}$ and $\omega_\sigma$ denote sequence and selection operators, respectively.
Each operator $\omega$ dictates a processing logic $f_{\omega}$. The data stream encapsulates a set of event tuples $E$, where each tuple is of the form $e = \{(k_1, v_1), \ldots, (k_{last^e}, v_{last^e})\}$. Here, $k$ refers to the name of the tuple and $v$ refers to the tuple value. In \system, we assume that the data streams arrive in the order indicated by the timestamp in the event tuple~\cite{Gulisano2010} and the system nodes are equipped with clocks that can be synchronized using a clock synchronization protocol such as Network Time Protocol~\cite{Mills1991}.

\begin{definition}{\emph{Operator Buffers and State.}}
The function $f_{\omega}$ processes ordered input data streams
from the operator's input buffer $B_I$ and produce output events stored in the operator's output buffer $B_O$. An operator either works based on the fixed computational parameters that are immutable (e.g., filter and stream operators) or it works on a dynamically changing computational state that is mutable (e.g., window and sequence operators), depending on the internal logic of the operator~\cite{Ottenwaelder2013}. A mutable operator can dynamically change the selection of events determined by an operator-specific \emph{selection policy} and \emph{consumption policy} of window and sequence operators~\cite{Chakravarthy1994Snoop}.
\end{definition}

For instance, in \Cref{fig:operatorgraph}, the operator $\omega_{V1}$ specifies a selection policy for a sliding window size of three subsequent speed events $\{e^{s}_{1}, e^{s}_2, e^{s}_3\}$ on the incoming speed data stream.
In a subsequent transformation step, operator $\omega_{V1}$ applies the processing function on the updated selection of events $\{e^{s}_2, e^{s}_3, e^{s}_4\}$ after sliding one event.
Each transformation step produces zero or more events as output.
Events are evicted from the incoming data streams after each transformation step by means of a \emph{consumption policy}.
In this example, the slide size defines the consumption policy, e.g., $e^{s}_1$ is evicted when the subsequent transformation step with $\{e^{s}_2, e^{s}_3, e^{s}_4\}$ is performed.

\begin{figure}[t]
	\centering
	\includegraphics[width=0.8\linewidth]{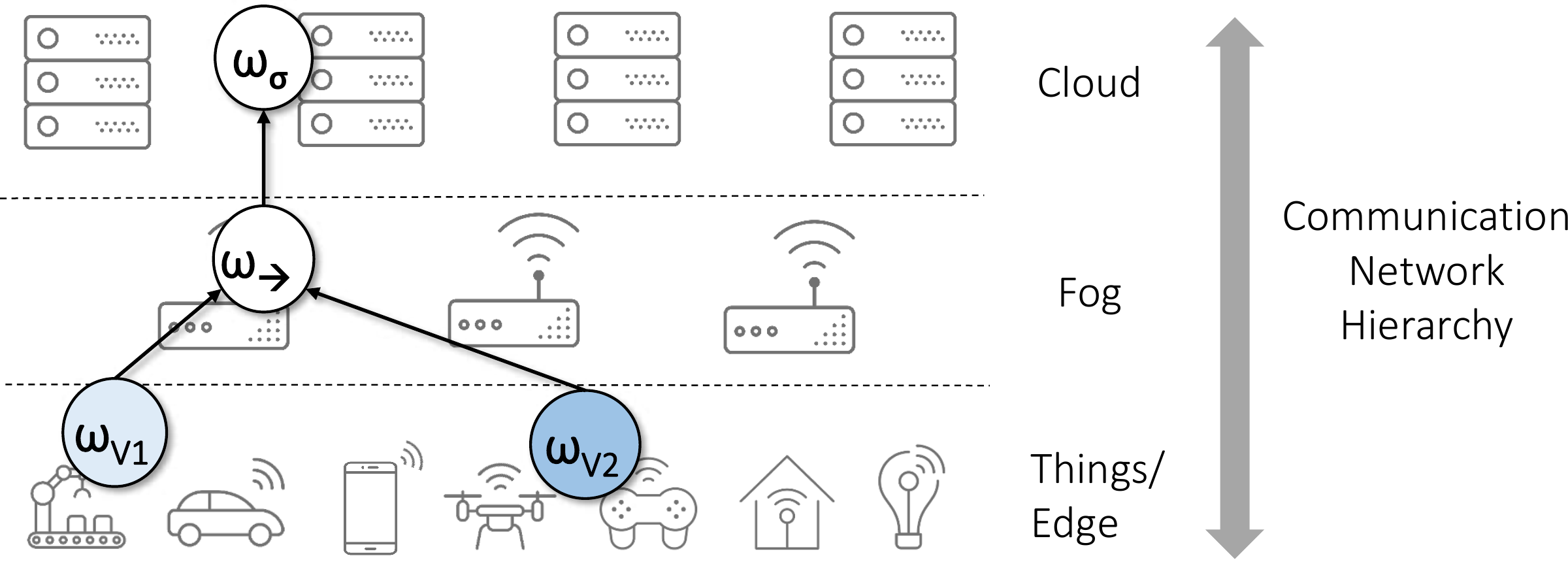}
	\setlength{\abovecaptionskip}{5pt}
	\setlength{\belowcaptionskip}{0pt}
	\caption{Example operator graph deployment and \system execution environment on the IoT network resources.}
	\label{fig:deployment}
\end{figure}

\subsection{IoT Resource Model} \label{subsec:networkmodel}
Although \system is not limited to a specific network topology and resource model, we will focus on the resources commonly considered in the context of IoT. Consider a hierarchical network infrastructure illustrated in \Cref{fig:deployment}. The figure presents three layers: \i (mobile) Things referring to IoT devices interconnected over wireless communication, \ii a layer of resources at the fog that offer a low-latency link to the Things in physical proximity, \iii and a fixed network layer comprising distributed resources in data centers or cloud. It is important to note that cloud and fog resources are assumed to communicate via a fixed IP infrastructure or novel ICN architectures~\cite{C2:Luthra2019b}. In contrast, IoT devices and edge resources can form different wireless network topologies, including device-to-device communication~\cite{Dwarakanath2016} between IoT devices or V2X~\cite{v2x/v2xand5g/chen2017} between vehicles.

Things represent producers and consumers in the \ac{IoT} scenario, while operators can be placed on any three layers. The end-to-end latency for this resource model is influenced by the physical proximity of resources and the computational power of resources. In general, we assume higher resource availability and processing power in the cloud. In contrast, IoT devices have resource constraints because they are battery-powered. Fog nodes are computationally more powerful than mobile nodes. For things, the availability of spatially nearby fog resources is restricted. For instance, IoT devices like Raspberry Pis and smartphones are resource-constrained and less powerful than computational resources at the fog locations such as micro data centers. Moreover, the availability of a fog location nearby an IoT device is not ascertained.
Each operator $\omega$ is encapsulated in a container on the computational resources of the \ac{IoT} infrastructure, as defined in \Cref{def:container}.
\subsection{Node Model} \label{subsec:nodemodel}
\begin{figure}[H]
	\centering
	\includegraphics[width=0.7\linewidth]{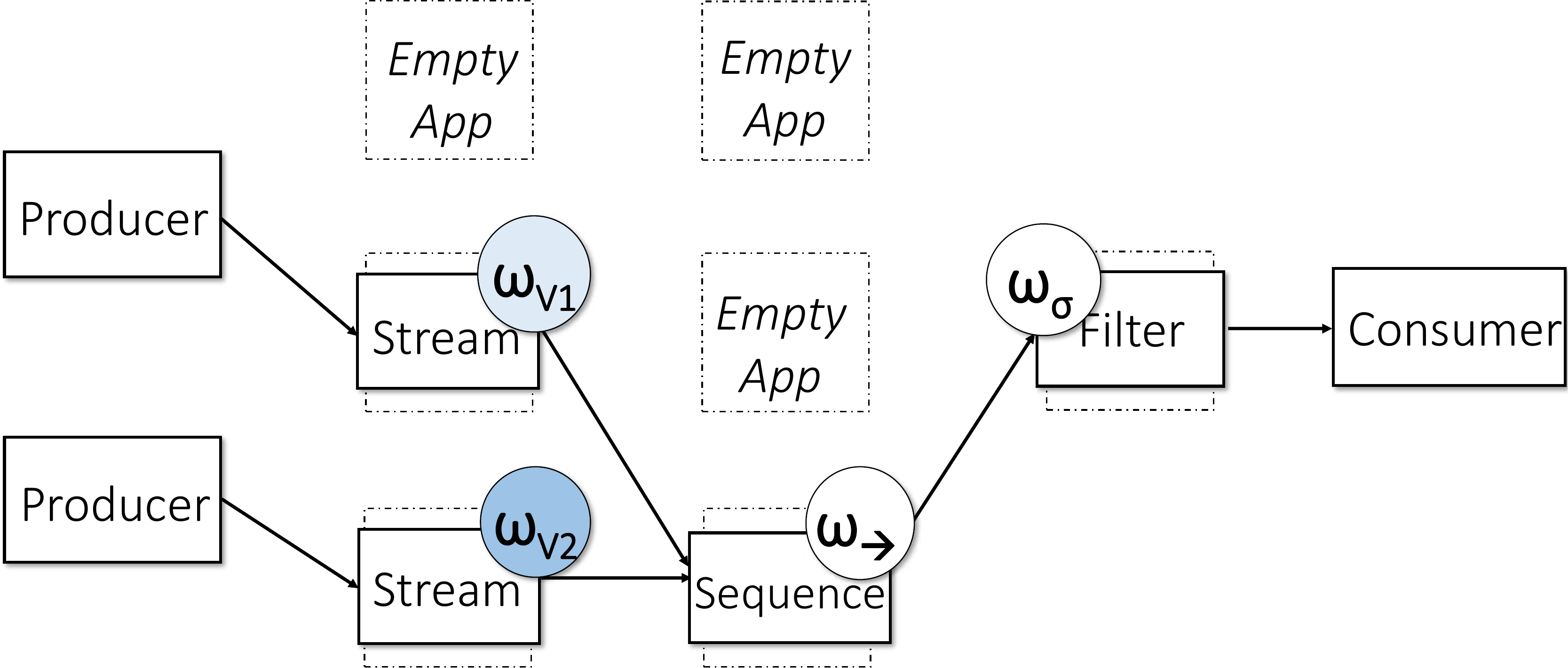}
	\caption{\system node model. The solid contour indicates pinned operators, while the dotted indicates unpinned operators.}
	\label{fig:system_model}
\end{figure}

A node acts as a host to the system entities producers, consumers, or brokers. Nodes refer to resources of the IoT resource model over which a producer, consumer, or broker can be executed. Note that the mapping of brokers on the node can change dynamically due to the dynamics in the environment. The nodes form an overlay network imposed by the interconnection of the operator graph on top of the \ac{IoT} resource model.
\Cref{fig:system_model} illustrates such an overlay network for the query and operator graph introduced in \Cref{subsec:case_study}. Here, the rectangular boxes denote the nodes, and the operator graph is executed on them.

\begin{definition} {\emph{Containers.}} A \system container enables the flexible movement of nodes in the IoT resource model.
\label{def:container}
\end{definition}
\system differentiates between pinned entities, \ie producers and consumers, from unpinned entities, \ie \ac{DCEP} operators. This is accomplished using \emph{static} and \emph{dynamic} containers. As the name suggests, the static containers are pinned to one node, while the dynamic containers are unpinned, meaning these support migration of operators on different nodes at runtime. An example of a static container is a producer and consumer, while a broker can be pinned or unpinned to a dynamic container named \emph{Empty App} (cf.~\Cref{fig:system_model}). Although \emph{EmptyApp} can hold more than one operator, they are free to move between other \emph{EmptyApp} without harming the other operators being executed on the same node. In this way, we enable flexible operator deployment and operator migrations on the fog-cloud infrastructure.

\subsection{OP Mechanism and Transition Model}
The \system follows a modular design as a composition of multiple \ac{OP} mechanisms $M_1, M_2,\ldots, M_N$.

\begin{definition}{\emph{\ac{OP} mechanism.}}
An \ac{OP} mechanism determines \emph{where} and \emph{how} to map an operator graph $G$ on a set of given brokers $B=\{ b_1, b_2, \ldots, b_{last^B}\}$ in the IoT resource model.
The mapped network of brokers is well known as an operator network. We define the mapping of the operator network as follows:
\end{definition}
\begin{equation}
    \begin{aligned}
    &&& \alpha: \Omega \times B \rightarrow \{0,1\}, s.t. \\
    &&& \alpha_{i,j} = \begin{cases}

      1, & \mbox{if $\omega_i$ is placed on $b_j$} \\
      0, & \mbox{if $\omega_i$ is not placed on $b_j$} \text{ .} \\
    \end{cases}
    \end{aligned}
\end{equation}
\begin{definition}{\emph{Transition.}}
In this work, we define the concept of a transition for \ac{OP} mechanisms, denoted as $T: M_A \rightarrow M_B$.
A transition $T$ performs a switch from a mechanism $M_A$ to $M_B$, e.g., \ac{OP} mechanisms at run time.
\end{definition}

The goal is to perform a transition in a \emph{seamless} or non-disruptive manner and to avoid oscillations during a transition. By \emph{seamless} execution of transition, we mean no disruption in delivering complex events during the lifecycle of a continuous query (cf. \Cref{sec:problem}). By oscillations, we mean that given the dynamics in the environmental conditions, the system may decide in a short interval to transit to a different mechanism multiple times. \system prevents oscillations by maintaining a balance in exploring multiple \ac{OP} mechanism vs exploiting best \ac{OP} mechanisms (cf. \Cref{sec:placement_performance}).

\subsection{QoS Demand Model} \label{sec:qosmodel}
An essential principle of an \ac{OP} mechanism is to find a mapping of an operator graph to brokers that optimally satisfies an objective function of \ac{QoS} demands,
such as end-to-end latency, bandwidth, and control message overhead. \system allows specification of one or more \ac{QoS} demands ($QoS$) and changing them at run time. The dynamics in the environmental conditions ($en_1,en_2,\ldots,en_{last^{en}}$), such as varying workload and mobility, influence the fulfilment of such \ac{QoS} demands.

In this work, we consider two crucial performance metrics influencing the decision of operator placement in a dynamic environment: \emph{end-to-end latency} and \emph{control message overhead}.

\begin{definition}\emph{End-to-end latency.} \label{def:e2elatency}
It is the time taken to \i receive the required primary events for the query at the placed nodes, \ii process the query, \iii emit a complex event, and \iv transmit the complex event through the network path between the given event producers $P$ to the given consumers $C$.
\end{definition}

It is important to note that end-to-end latency can be time-varying due to the dynamic nature of the network and the placement update of the operators.
In case multiple producers or consumers are involved, then latency is measured from the producer with the maximum network delay to the consumer, as explained in the example below.

To better understand, let us revisit the example scenario introduced in \Cref{sec:motivation}. We assume that two producers \texttt{vehiclesAtSectionV1} and \texttt{vehiclesAtSectionV2}, and a single consumer is interested in detecting congestion. Now, consider the path from $p_1:$ \texttt{vehiclesAtSectionV1} and $p_2:$ \texttt{vehiclesAtSectionV2} via some broker vehicles $b_1, b_2, \ldots, b_{last^B}$ to the consumer $c$. We assume the position of the consumer is at Section V1 when the query is triggered, and the \ac{OP} was determined at the aforementioned producer and broker network path. In this case, the end-to-end latency is the sum of the network delay observed on the path $p_2, p_1, b_1, b_2, \ldots, b_{last^B}, c$ and the execution time of the query on these nodes in the path.
In case multiple consumers, say $c_1$ and $c_2$, are interested in the same query, the end-to-end latency is given by the interval between the first primary event production at $p_1$ and the complex event reception at the consumer $c_1$ or $c_2$. Note, even when the query has been placed at the same set of brokers, the end-to-end latency for the different consumers will depend on the consumer's location, and hence it could be different for each consumer.

\begin{definition}{\emph{Control message overhead}} \label{def:cmOverhead}
The number of control messages sent to assign all the operators $\omega \in \Omega$ of a query to the brokers $b \in B$. In essence, it is given by the overhead caused in exchanging messages to place the query on the IoT resource model.
\end{definition}

Using the above definition of control message overhead and the assumptions on the traffic control scenario in \Cref{def:e2elatency}, let us demonstrate the meaning of control message overhead. To fulfil an objective, such as end-to-end latency, an \ac{OP} mechanism such as Relaxation~\cite{Pietzuch2006} maintains a latency cost space to find out network paths with minimum end-to-end latency. However, to build such a cost space, many messages have to be exchanged between the considered nodes for placement and the \ac{OP} coordinator. Furthermore, to place an operator graph, acknowledgements on the assignment of operators on nodes are sent. We refer to the number of such control messages for \ac{OP} as control message overhead.
Some \ac{OP} mechanisms like MDCEP~\cite{Starks2015} aim to minimize this metric on the cost of sub-optimal \ac{OP} concerning metrics, like end-to-end latency, to prevent overhead on resource-constrained \ac{IoT} nodes.

\section{Problem Statement} \label{sec:problem}

Consider the availability of $N$-different \ac{OP} mechanisms that can be selected to execute and place a query on the \ac{IoT} network resources. Dependent on the environmental conditions $en(t)$ at time $t$, the
\ac{QoS} demands of consumers, say $QoS_{|en(t)}$ are changing.
 Furthermore, the ability and cost of an \ac{OP} in terms of resource requirements to fulfil the \ac{QoS} demands are changing over time.

The \system system aims to ensure that the QoS demands of queries are fulfilled despite changing environmental conditions using the IoT resource model.
Therefore, we determine for changing environmental conditions
$en(t)$ and corresponding $QoS_{|en(t)}$ demands a sequence of points in time, say $t_1, \ldots, t_{n}$
and a sequence of \ac{OP} mechanisms $M(t_1), \ldots, M(t_n)$ on which a transition $T_i:M(t_i) \rightarrow M(t_{i+1})$ is initiated at time $t_i$.
It is important to note that while performing a transition, several operator migrations must take place. The operator migrations impose a high cost because of state migrations in terms of time and overhead.
Moreover, the transition needs to be performed in a non-disruptive manner, i.e., even during the transition, the \ac{QoS} demands of a query need to be satisfied. Consequently, state migrations have to take place in a cost-efficient manner.

We define the objective function
of the \emph{transition problem} considering two key cost factors, namely, the costs imposed in terms of
transition time $C_{Time}(T_i)$
and transition overhead $C_{Overhead}(T_i)$. The transition time is defined as the time it takes to select a new target placement mechanism $M(t_{i+1})$~($Time_{select}$), to find a placement $\alpha$ dependent on $M(t_{i+1})$~($Time_{\alpha}$), and to migrate $j$ operators $\omega_j \in \Omega \text{, } \forall j \in [1,num^j]$ to the target brokers~($Time_{mig.(\omega_j)}$) dependent on $\alpha$. Thus, we define the cost in terms of transition time as:

\begin{equation}
    C_{Time}(T_i) = Time_{select} + Time_{\alpha} + \sum_{j=1}^{num^j} Time_{mig.(\omega_j)} \text{ .}
     \label{eq:cost_Ttime}
\end{equation}

Similarly, the transition overhead is given by the overall number of messages exchanged
in order to perform a transition, including the
\i~selection of a placement mechanism ($Overhead_{select}$),
\ii~the placement ($Overhead_{\alpha}$),
\iii~and migration of the operators including their state ($Overhead_{mig.(\omega_j)}$). Formally, it is defined as follows:

\begin{equation}
     C_{Overhead}(T_i) = Overhead_{select} + Overhead_{\alpha} + \sum_{j=1}^{num^j} Overhead_{mig.(\omega_j)} \text{ .}
     \label{eq:cost_Toverhead}
\end{equation}

The transition problem in this paper, therefore, is to minimize a weighted sum of normalized values\footnote{using mean normalization method.}
transition time ($\hat{C}_{Time}(T_i)$) and transition overhead ($\hat{C}_{Overhead}(T_i)$) in order to meet the \ac{QoS} demands under the execution of transitions as stated below:

\begin{equation}
    \begin{aligned}
    &&& \min \left[w_t * \hat{C}_{Time}(T_i) + w_o * \hat{C}_{Overhead}(T_i)\right] \\
    &&& s.t. \text{ }
      \alpha(t) \text{ satisfies } QoS_{|en(t)} \text{ under the execution of } T_i  \\
      &&&  C_{Time}(T_i), C_{Overhead}(T_i), QoS_{|en(t)} \in \mathbb{R}^{+} \text{ .}
  \end{aligned}
  \label{eq:tr_cost_objective}
\end{equation}

Here, $w_t$, $w_o \ge 0$, $w_t + w_o = 1$, denote weights for transition time and overhead, respectively.

\section{The TCEP System Design} \label{sec:design}
\begin{figure}
\centering
\includegraphics[width=\linewidth]{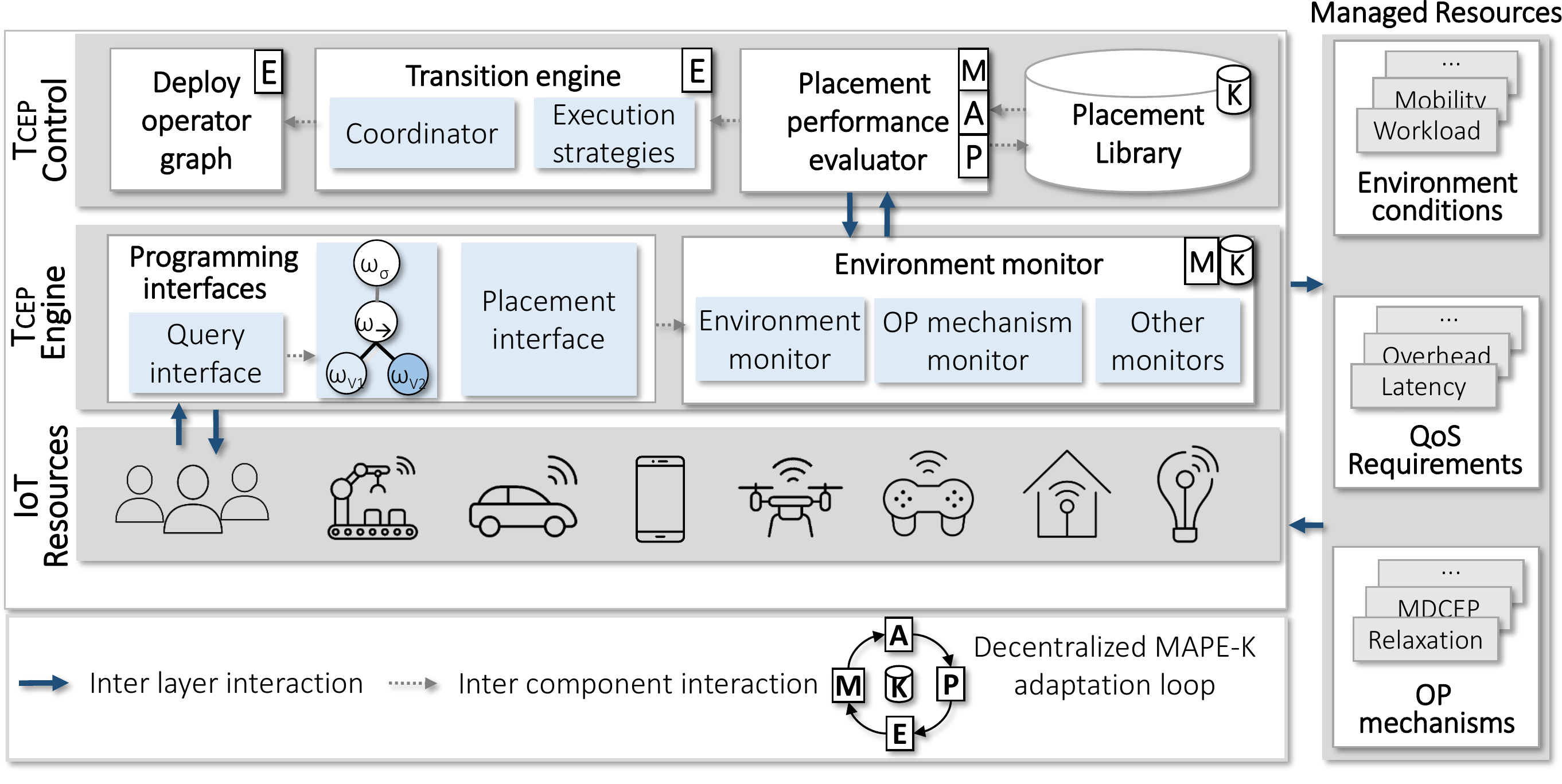}
\caption{The TCEP system design.
}
\label{fig:TCEP}
\end{figure}

\paragraph{\textbf{Conceptual Overview}} The four key components of the \system system are represented in Figure~\ref{fig:TCEP}.
The \emph{IoT resources} layer includes event consumers, which can pose queries with specific \ac{QoS} demands; event producers, which generates continuous data streams that are to be processed; and event brokers, which process the data streams to derive results.
The \emph{TCEP engine} layer provides a programming environment to create and execute queries and \ac{OP} mechanisms on the infrastructure of the IoT resources layer.
Moreover, the \system engine provides mechanisms for monitoring the performance of the \ac{OP} mechanism and the environmental conditions.
The \emph{TCEP control} layer utilizes and manages a library of state-of-the-art \ac{OP} mechanisms in a so-called placement library.
Here, the Transition engine decides when and how to perform a transition.
It is also responsible for coordinating the transition, \ie performing operator migrations building on the proposed transition execution algorithms.

Furthermore, the placement performance evaluator decides which placement mechanism to select for a transition.
The deploy operator graph component performs the deployment of the operator graph on the infrastructure resources.
Finally, \emph{Managed Resources} represent the resources monitored and controlled by the \system system, such as environmental conditions, performance metrics, and \ac{OP} mechanisms.

A brief overview on the execution of a continuous query and its adaptation using transitions in \system is presented as follows.
An event consumer poses a query using the query interface of the \emph{\system engine} programming interfaces.
The query is then transformed into an operator graph. The placement interface could be used to develop an \ac{OP} mechanism.
After transforming the query into an operator graph and selecting an \ac{OP} mechanism, the placement performance evaluator deploys the query on the \ac{IoT} resources based on the previously collected statistics on the query.
If the environment monitor finds a change in the environmental conditions or the \ac{QoS} requirements, a transition is triggered. The transition engine manages the adaptation, and the operator graph is redeployed by the deploy operator graph component.

\paragraph{\textbf{Decentralized MAPE-K adaptation loop}} \system follows the well-known \emph{MAPE-K}~\cite{Kephart2003} loop for adaptation. The four processes of the loop, \emph{Monitor} (M), \emph{Analyze} (A), \emph{Plan} (P), \emph{Execute} (E), and \emph{Knowledge} (K) are realized in a decentralized manner (cf. Figure~\ref{fig:TCEP}) in the control layer and within the \system engine to manage the resources depicted in the lowest layer. In the following, we provide the definitions of these components for the \system system.

\emph{Monitor (M)}: This function provides mechanisms to collect, aggregate, filter, and report details on the managed resources. Examples of monitoring information are environmental conditions such as mobility of cars and workload, performance metrics related to a query such as \ac{QoS} metrics latency and bandwidth observed on the links, and performance metrics related to the transition of an \ac{OP} mechanism: transition time and overhead. Hence, the decentralized monitoring components lie within the environment monitor and placement performance evaluator, responsible for collecting and aggregating the above monitoring information.

\emph{Analyze (A)}: This provides mechanisms that correlate and model complex situations. These mechanisms allow the transition engine to learn about the managed resources and predict future situations. For instance, the placement performance evaluator implements a fitness score mechanism that measures the performance of the \ac{OP} mechanism, which is used to predict the next suitable operator placement for the respective environmental conditions.

\emph{Plan (P)}: It provides mechanisms that construct the actions needed to achieve the goals and objectives. For instance, the placement performance evaluator determines if a change to a new \ac{OP} mechanism would help fulfil the \ac{QoS} demands.

\emph{Execute (E)}: It provides mechanisms to manage the necessary changes required for the adaptation. It is responsible for carrying out the transition itself. For instance, the transition coordinator generates a plan on the operator graph transition, and the transition engine performs the transition.

\emph{Knowledge (K)}: The data shared across the above four functions are stored as shared knowledge. This includes \ac{OP} mechanisms in the placement library, monitoring information on the performance, among others.

In the following sections, we will focus on four research questions, namely:
\begin{enumerate}[leftmargin=12mm]
\item [RQ 1] How to specify an operator placement and its performance characteristics?
\item [RQ 2] How to adaptively select an \ac{OP} mechanism for a transition?
\item [RQ 3] How to realize transitions in a seamless manner?
\item [RQ 4] How to decide when to perform a transition?
\end{enumerate}

\paragraph{\textbf{Structure}} The following sections detail on the functionality of the aforementioned MAPE-K processes handled by the different components of \system in a decentralized manner. \Cref{subsec:prog-model} presents (RQ 1) a programming model for specifying \ac{QoS} demands in a query and OP mechanisms.
\Cref{sec:placement_performance} presents (RQ 2) the \emph{genetic learning} algorithm for an adaptive selection of \ac{OP} mechanism such that the \ac{QoS} demands are satisfied.
Section~\ref{sec:transitionEngine} addresses (RQ 3) and (RQ 4) by presenting seamless and concurrent execution of a transition while considering a minimal state for a cost-efficient transition.

 In \Cref{fig:TCEP}, \Cref{subsec:prog-model} is illustrated as the Programming interfaces component, Section~\ref{sec:placement_performance} as Placement performance evaluator, and Section~\ref{sec:transitionEngine} as Transition engine. \mli{R1: the last paragraph presents sections out of order. fixed the order}

\subsection{Programming Model} \label{subsec:prog-model}
The programming model provides a means for developers to implement novel \ac{OP} mechanisms while utilizing IoT resources.
Existing works~\cite{placement/sigmetrics/cardellini2017, Starks2017, Lakshmanan2008} focus on proposing \ac{OP} mechanisms for a diversity of \ac{QoS} demands.
However, none of them provides a common API for the development of novel OP mechanisms\footnote{For a detailed discussion on related work, we refer the readers to \Cref{sec:relatedwork}.}.
This section introduces the major components of the \system programming model:
\i~\ac{QoS} monitors that is an integral part of the programming model as each OP mechanism observes some \ac{QoS} metrics (cf. \Cref{subsubsec:qos_monitors}), and
\ii~\ac{OP} interface that provides methods to develop unique \ac{OP} mechanism.

\subsubsection{QoS Monitors} \label{subsubsec:qos_monitors}

As prominently discussed in the literature~\cite{Lakshmanan2008, Starks2017, Liu2020/RMSurvey}, our programming model characterizes the existing \ac{OP} mechanisms based on the placement decision into two main categories: \i centralized and \ii decentralized. A centralized OP mechanism assumes global knowledge on the network and the nodes (specific \ac{QoS} demands) to host an operator on a physical node.  
In contrast, a decentralized mechanism assumes only partial knowledge of the network, and hence the placement decision is decentralized. For instance, a cluster head assigns an operator on each node of the cluster.
It is known that finding an optimal placement from the number of possible resources is an NP-complete problem~\cite{npcomplete/sigmod/srivastava2005}.
Furthermore, the assignment varies with the \ac{QoS} demands in consideration for the cost objective function. Hence, there exist many solutions and heuristics towards the \ac{OP} problem.

Both kinds of placement heuristics assume monitoring knowledge on the network and host information. The \system programming model provides explicit extensible monitors for commonly used network and host information metrics such as latency, bandwidth and CPU load. These metrics are measured from end-to-end, meaning the cumulative latency or bandwidth observed while data streams traverse the path from producer to consumer.
The measurements are accumulated step by step, and hence individual measurements can also be fetched easily. The monitoring information is collected by every node separately and aggregated on the decision node based on the placement characteristics.
In centralized OP mechanisms, the QoS monitors transfer the observed metric to a centralized node responsible for the placement decision.
While for decentralized mechanisms, we provide decentralized monitoring solutions such as Vivaldi~\cite{Dabek2004}, which is prominently used in several \ac{OP} mechanisms~\cite{Pietzuch2006, Rizou2010, Cardellini2016, placement/sigmetrics/cardellini2017} that handles the dissemination of monitoring information for placement decision.

\begin{table}[t]
\centering
\footnotesize
\begin{tabular}{lp{6cm}}
\toprule
Method                 & Description                                                                       \\
\midrule
\texttt{getPlacementMetrics()} & Determines the \ac{QoS} demands that must be optimized           \\
\texttt{configurePlacement()} & Resets placement parameters. It is called initially and on reconfiguration            \\
\texttt{findPlacementNode()}           & Finds placement node determined based on the \ac{QoS} metrics \\
\texttt{findPossibleNodes()} & Retrieves all nodes that can host operators \\
\texttt{initialVirtualOperatorPlacement()} &
 Centralized mechanisms treat all operators at once during the initial placement instead of one by one by using a heuristic to find optimal locations in the virtual space \\
\bottomrule
\end{tabular}
\setlength{\abovecaptionskip}{5pt}
\setlength{\belowcaptionskip}{0pt}
\caption{\system placement API for developing OP mechanisms.}
\label{tab:placement_api}
\end{table}

\subsubsection{\ac{OP} Interface} \label{subsubsec:op_interface}

\Cref{tab:placement_api} lists the foremost API of the \system programming model used to implement \ac{OP} mechanisms in TCEP.
\code{PlacementStrategy} API defines these methods for \ac{OP} mechanisms in order \i to formulate a single objective and multi-objective optimization function for centralized \ac{OP} mechanisms, \ii to define heuristics for decentralized \ac{OP} mechanisms, and finally, \iii to make \ac{OP} mechanisms exchangeable at runtime to enable transitions.

An \ac{OP} mechanism needs to represent a cost objective function dictating the \ac{QoS} demands. An example of a cost objective function is to minimize the end-to-end latency from the producers to the consumers.
Each mechanism, centralized and decentralized, must define a cost objective function for the \ac{QoS} demands that need be optimized. The cost objective function can comprise a single or multiple \ac{QoS} demands, \eg latency, CPU load and bandwidth utilization. The objective function depends on the runtime measurements from the \ac{QoS} monitors defined above, which are used to determine placement decisions on physical hosts of IoT resources.
\code{getPlacementMetrics()} method is used (cf. \Cref{tab:placement_api}) to fetch monitoring information related to the objective function. Consequently, this helps in formulating the cost function. The specific way of solving the placement problem (optimally or sub-optimally) using heuristics is defined in the specific implementations of the \ac{OP} mechanisms.

In \Cref{tab:OPoverview}, we define the currently available implementations of \ac{OP} in \system. We define the heuristic approaches used by the respective OP mechanism--for example, the Relaxation mechanism \cite{Pietzuch2006} uses a spring relaxation technique, while the MOPA mechanism \cite{Rizou2010} uses an approximation for the Weber problem, though both aim for the same \ac{QoS} metric: bandwidth-delay product. Also, in optimal solutions of \ac{OP}, the optimization problem can be solved using different methods.
The heuristic used also varies based on the nature of the objective function (convex or concave) and the scenario at hand. Hence, the in \system programming model, we segregate the implementation of a specific optimization approach of the \ac{OP} mechanism from the common interfaces.

\begin{table}[]
\footnotesize
\begin{tabular}{p{1.5cm}p{1.5cm}p{2cm}p{4cm}l}

\toprule
\ac{OP} Mechanism & Placement Decision & Optimization Goal             & Approach                                              & §~\ref{subsubsec:opmechanisms} \\
\midrule
Relaxation~\cite{Pietzuch2006}          & Centralized    & bandwidth-delay$^2$ product (BDP) & Spring relaxation technique                           & (1) \\
MOPA~\cite{Rizou2010}                & Centralized       & bandwidth-delay product       & Approximation for Weber Problem                       &  (2) \\
Global Optimal      & Centralized      & bandwidth-delay product & Optimally finds node with minimum BDP                 &  (3) \\
MDCEP~\cite{Starks2015}               & Decentralized       & control message overhead, latency      & Place on nearest neighbours unless producer or consumer &  (4) \\
Producer-Consumer   & Decentralized        & hops                          & Always host on the producer or consumer                   &  (5) \\
Random              & Decentralized        & -                             & Random allocation                                     & (6) \\
\bottomrule
\end{tabular}
\caption{Design space of \ac{OP} mechanisms.}
\label{tab:OPoverview}
\end{table}

The placement parameters are initialized using the \code{configurePlacement()} method, which is invoked in the beginning and each reconfiguration, \eg during periodic updates of the same \ac{OP} mechanism. \code{findPossibleNodes()} and \code{findPlacementNode()} methods determine the possible nodes where the operator can be deployed depending on the cost function and the optimal or sub-optimal (depending on the placement mechanism) node for the deployment, respectively.
Some centralized mechanisms behave differently when performing \ac{OP} initially and on reconfiguration, such as the Relaxation~\cite{Pietzuch2006} mechanism. This mechanism places all operators of the query at once based on the virtual coordinate space using \code{initialVirtualOperatorPlacement()}, and the physical placement is performed using \code{findHost()} since no operator is physically deployed only using virtual placement.
However, on reconfiguration, only the physical placement is changed.
In contrast, decentralized mechanisms only implement the \code{findHost()} since their behaviour is the same during initial placement and transitions. 
Having understood the functionality of the programming model and the monitors of the \emph{\system engine} layer, we detail the placement performance evaluator component in the following subsection.


\subsection{Placement Performance Evaluator}
\label{sec:placement_performance}
This component measures the performance of the \ac{OP} mechanisms continuously and analyze their behavior. 
A \emph{lightweight online learning} algorithm is employed to statistically determine which mechanism best meets the \ac{QoS} demands, building on a selection strategy of genetic algorithms~\cite{Whitley1989}.
\emph{Lightweight} refers to the fact that learning does not rely on any training set but only uses statistics collected online during the execution. 
This component uses the online learned model 
to select an appropriate \ac{OP} mechanism with the best performance based on the ranking provided by the learning algorithm. 
The environment monitor component keeps track of the performance behaviour (\ac{QoS} demands and environmental conditions via \ac{QoS} monitor and other monitors, respectively) and reports any changes to this component -- e.g., if the \ac{QoS} demand specified in the query is violated. 
When no empirical statistics are available during initialization, the target placement mechanism is determined by comparing the respective \ac{QoS} demand with the specified optimization objective(s) of the placement mechanism. If more than one placement mechanism exists for the respective \ac{QoS} demand, then the selection is performed in a round-robin fashion.

In the remaining section, we first define a heuristic fitness function to evaluate the performance of an \ac{OP} mechanism during its execution. Then, we define an adaptive selection of an \ac{OP} mechanism based on the observed statistics and the fitness function. 

\subsubsection*{\textbf{Heuristic Fitness Score for \ac{OP} Mechanism.}}\label{subsec:score}
We measure the performance of the current \ac{OP} mechanism in execution for each continuous query at regular intervals. 
The collected performance statistics are then used for comparison between different \ac{OP} mechanisms.
To quantify the performance, we measure the fitness of each \ac{OP} mechanism that is in execution per query. 
We define the heuristic fitness function with the objective to maximize the number of times an \ac{OP} mechanism fulfils the current \ac{QoS} demands. 
This means that if an \ac{OP} mechanism fulfils \ac{QoS} demands 
$x >max$ times between the time interval $t_s$ (when the query was first submitted) and $t_t$ (when the transition is triggered), 
then this mechanism is selected for the next execution. 
For each \ac{QoS} demand, we update the fitness score at regular intervals until the next transition.
The score provides information on how well the \ac{OP} mechanism had performed over time, compared to the mechanisms that were in an execution before (when the query was first submitted). 
The goal is to find the best mechanism for the respective \ac{QoS} demands by utilizing the collected statistical information. This goal is accomplished by maintaining the scores of the respective \ac{OP} mechanisms $M_{i,qos_j} (t_t)$ for each \ac{QoS} demand $qos_j$, and updating the score at the occurrence of a transition at time $t_t$.
Since an \ac{OP} mechanism can incorporate multiple \ac{QoS} demands, for instance, in a multi-objective optimization function, the score is determined separately for each \ac{QoS} demand. 
For each \ac{OP} mechanism $M_i$, we maintain a score function $Score_{(M_{i,qos_j})}(t_{t})$  obtained based on the evaluation of each \ac{QoS} demand $qos_j$. The score $M_{i,qos_j}(t_{t})$ is normalized for each \ac{OP} mechanism $M_i$, based on the \emph{mean normalization method} to make the scores comparable. 
We compute the fitness score based on the statistics collected from executing \ac{OP} mechanism $i$ (with subscript $i$), which is then compared to other mechanisms executed from time $t_s$ (when the query was first submitted) until time $t_t$ (when the transition is triggered), given as $t_{s,t}$: 

\begin{equation}\label{eq:score}
\begin{aligned}
M_{i,qos_j}(t_t) = \frac{\mu_{i,qos_j}(t_{s,t}) - \mu_{qos_j}(t_{s,t})}{max_{qos_j}(t_{s,t}) - min_{qos_j}(t_{s,t})} \cdot (1-decay) + \\ M_{i,qos_j}(t_t-1) \cdot decay \text{ .}
\end{aligned}
\end{equation}

In Equation~\ref{eq:score}, $\mu_{qos_j}(t_{s,t})$, $max_{qos_j}(t_{s,t})$, and $min_{qos_j}(t_{s,t})$ denote the mean, maximum and minimum score values for \emph{all} the \ac{OP} mechanisms, respectively, that have been used until time $t_t$
 considering the \ac{QoS} demand $j$. $\mu_{i,qos_j}(t_{s,t})$ represents the mean score value of \ac{OP} mechanism $M_i$ until time $t_t$ considering the \ac{QoS} demand $qos_j$. 
$M_{i,qos_j}(t_t-1)$ is the last score of \ac{OP} mechanism $M_i$, and a $decay$ factor is used to exponentially reduce the effect of old statistics to prioritize the data that is recently collected. The decay factor ranges of $[0, 0.5]$, such that more preference is given to current statistics. The initial value of decay is set to 0 , and it is updated once a transition is performed by a factor dependent on the number of \ac{OP} mechanisms to be explored. For instance, if there are $10$ \ac{OP} mechanisms, then the decay is incremented by $0.05$.   
The overall score of an \ac{OP} mechanism is computed based on all the statistics collected on the \ac{QoS} demands fulfilled by the \ac{OP} mechanism. The score is the sum of the normalized scores for each \ac{QoS} demand $qos_j \in [qos_1, qos_2, \ldots, qos_k] $, where $k$ is the maximum possible \ac{QoS} demands considered by \ac{OP} mechanism $M_i$: 
\[
Score_{(Mi)}(t_t) = \sum_{j=1}^{k} M_{i,qos_j}(t_{s,t}) \text{ .} \] 

\subsubsection*{\textbf{Adaptive Selection of \ac{OP} Mechanism.}} \label{subsec:adaptive}
 The adaptive selection of an \ac{OP} mechanism is performed once each \ac{OP} mechanism has been defined with a fitness score.   
We adopt the \emph{Linear Ranking Selection Strategy}~\cite{Whitley1989}, 
a selection method from \ac{GA}. The ranking based method is suitable for our \ac{OP} mechanism selection problem since it 
allows us 
\i to perform a relative analysis suitable for the heuristic fitness function that indicates which \ac{OP} mechanism is better, and 
\ii by an appropriate selection pressure it favours exploitation over exploration avoiding selecting worse \ac{OP} mechanisms. 
More specifically, by only using the fitness values of the \ac{OP} mechanisms, the linear ranking method selects the best \ac{OP} mechanism for the given \ac{QoS} demands, which is a perfect choice since our goal is to compare \ac{OP} mechanism relatively. The selection pressure defines the intensity of search focused towards the best \ac{OP} mechanism. 
By reducing the selection pressure, the diversity of the \ac{OP} mechanism increases, while increasing the selection pressure focuses on the reduced search space of selected best \ac{OP} mechanisms. 
This explains the idea of exploration vs exploitation using the ranking method. 
Theoretically, using the linear ranking method, we can compute the appropriate selection pressure $\mathcal{S}$ using the average fitness distribution $\mathcal{\overline{M}}$ before selection and expected average fitness distribution $\mathcal{\overline{M^*}}$ for given fitness values $f_1, \ldots, f_{F}, ( {F} \leq \mathcal{N})$ as follows: 

\[ \mathcal{\overline{M}} = \frac{1}{\mathcal{N}} \sum_{k=f_1}^{f_{F} }\overline{s}(f), \]
\[ \mathcal{\overline{M^*}} = \frac{1}{\mathcal{N}} \sum_{k=f_1}^{f_{F}} \overline{s^*}(f),  \]



\begin{equation} \label{eq:selectionpressure}
 \mathcal{S} = \frac{\mathcal{\overline{M^*}} - \mathcal{\overline{M}}}{\overline{\sigma}} \text{ .}  
\end{equation}


Here,  $\overline{s}(f)$ and $\overline{s^*}(f)$ are the fitness distribution and expected fitness distribution of the \ac{OP} mechanisms, respectively. The notation $\mathcal{N}$ denotes the size of fitness distribution and $\overline{\sigma}$ denotes the standard deviation of the fitness distribution $\overline{s}(f)$. All functions assumed to be continuous are denoted with an overline, and the fitness values for the \ac{OP} mechanisms are assumed to be sorted ($f_1 < f \leq f_F$)~\cite{Blickle1996}.

After \ac{OP} mechanisms are sorted according to their fitness values, the ranks are assigned to them. Rank $R$ is assigned to the best \ac{OP} mechanism, while rank 1 is assigned to the worst. 
The selection probability $P_i$ is linearly assigned according to the rank as follows:
 
 \begin{equation}\label{eq:selectionprobability}
P_i = \frac{1}{R} \left( \eta^- + (\eta^+ - \eta^-) \frac{i-1}{R-1} \right); i \in [1,R] \text{ .}
 \end{equation}
 
In Equation~\ref{eq:selectionprobability}, $ \frac{\eta^-}{R} $ is the probability that the worst \ac{OP} mechanism is selected and $ \frac{\eta^+}{R} $ the probability that the best \ac{OP} mechanism is selected. Since \ac{OP} mechanisms in the placement library are constant during runtime, the conditions $ \eta^+ = 2 - \eta^-$ and $\eta^- \geq 0$ must be fulfilled. Also, note that all the \ac{OP} mechanisms are ranked differently, i.e., they have distinct selection probability -- although they can have the same fitness score~\cite{Blickle1996}.
The probability of the \ac{OP} mechanism to be selected is proportional to its fitness function score. The worst probability and the best probability are calculated as the minimum and maximum of the probability distribution function $\eta$:

\begin{equation}
 \eta_i = \frac{Score_{(M_i)}}{\sum_{i} Score_{(M_i)}} \text{ .}
\end{equation}

The selection of the mechanism means the inclusion of it in the reduced search space, which gives well-performing \ac{OP} mechanism a higher probability than the lower ones, i.e.,
we prefer \ac{OP} mechanisms that were classified to perform better (exploitation of the learning algorithm). However, sometimes we also select worse \ac{OP} mechanisms to update their score (exploration).
Assuming that the fitness distribution follows a Gaussian distribution, and using ~\Cref{eq:selectionpressure}, it can be proved (cf. Proof in Appendix~\ref{proof:selcmethod}) that the selection pressure for the ranking method can be computed as follows: 

\begin{equation} \label{eq:selectionpressureRmethod}
 \mathcal{S_R (\eta^-)} =  (1 -  \eta^-) \frac{1}{\sqrt{\pi}}  \text{ .}
\end{equation}

Once all the \ac{OP} mechanisms get assigned a rank based on their performance for a query, the \system system can decide whether the currently running mechanism $M_i$ should be used again or changing to another \ac{OP} mechanism yields better performance. We use a simple Radix sort to rank the \ac{OP} mechanisms in linear time so that the comparison is cheap. 
The complexity of the sorting dominates the complexity of the selection algorithm, i.e., $\mathcal{O}(\mathcal{N})$, where $\mathcal{N}$ is the size of the fitness distribution function of \ac{OP} mechanisms.
Furthermore, the following challenges are considered while selecting the next \ac{OP} mechanism: \i In the beginning, we allow some degree of exploration so that all the \ac{OP} mechanisms get a chance to prove themselves. Therefore, a round-robin selection is used for the adaptive selection of an \ac{OP} mechanism initially. Furthermore, we allow exploration of alternate \ac{OP} mechanisms at random intervals during the execution to give a chance to perhaps better-performing \ac{OP} mechanism.
\ii Adapting too often might cause oscillations (back and forth) while also skewing the results of the used \ac{OP} mechanism. Therefore, we empirically set the delay threshold
between consecutive transitions to give the new \ac{OP} mechanism enough time so that the performance evaluator can correctly assess its behaviour.


\subsection{Transition Engine} \label{sec:transitionEngine}
The \system transition engine coordinates how a transition is performed over the life cycle of a transition~\cite{Frommgen2015a}, i.e., from its invocation to its completion. The two transition algorithms define the life cycle of a transition. 
This component, therefore, is a core of the \system system. 

We first provide a high-level view of the requirements for the transition phase. 
A transition from one \ac{OP} mechanism to another involves several distributed entities of \system. 
The transition execution must be coordinated such that it is consistently performed across these entities. 
Thus, the \emph{transition coordinator} maintains and orchestrates the transition life cycle. \system currently supports two transition algorithms (detailed below).
The difference in the life cycle of the proposed transition algorithms lies in the \emph{seamlessness}, i.e., how smooth the transition is performed and how much is the cost in terms of time and overhead ($C_{Time}(T)$ and $C_{Overhead}(T)$) as defined below.

During the execution of a transition, the target \ac{OP} mechanism determines a set of target brokers for the new placement. As a result, all the operators have to migrate to the target brokers to comply with the new placement logic. While the coordinator performs operator migrations, it must continue satisfying the \ac{QoS} demands by the event consumers, which is the primary goal.
Operator migrations in this realm have been widely studied in the literature, such as stop and restart strategies~\cite{Zhu2004, Zaharia2010} as well as partial pause and resume strategies~\cite{Gulisano2010, Mai2018/pvldb/Chi}. Here, the former completely stops the execution to migrate the operator to start executing at a target broker, while the latter partially pauses the execution of the concerned operator only. However, none of the approaches addresses seamless and cost-efficient operator migrations while using multiple \ac{OP} mechanisms. 
To do this, we specifically look into costs associated with performing a transition in terms of time and overhead. The transition execution algorithm dictates how {\it cost-efficient} operator migrations are performed while fulfilling the \ac{QoS} demands.
Considering these requirements, we present two transition execution algorithms that \i coordinate the transition, \ii perform operator migrations while ensuring the correctness and completeness of the delivered \emph{complex} events to the consumers, and \iii perform the {\it live} and {\it seamless} transition. \par

\subsubsection*{\textbf{\ac{MFGS} Sequential Transition.}}  \label{subsubsec:MFGS}
In this algorithm, the transition coordinator initiates operator migrations in a specific order, i.e., in a bottom-up fashion (cf. Algorithm~\ref{algo:seqTransition}: Lines~\ref{algline:funstart}-\ref{algline:else}). This means an operator is only migrated after all its predecessors were successfully migrated. Here, the dependency of operators follows a bottom-up fashion, where leaf operators are predecessors of their successors or dependent operators as we go level up in the operator graph. The operator migrations are performed in a sequential and breadth-first manner one at a time to the target brokers (Lines~\ref{algline:bottomup}-\ref{algline:ophead}).

In the next step, the coordinator determines the target broker with the help of the newly selected \ac{OP} mechanism (Line~\ref{algline:findbroker}). It is important to note that the target \ac{OP} mechanism is predetermined by the placement performance evaluator component (cf. Section~\ref{sec:placement_performance}). Consequently, an operator $\omega$ may need to be migrated to a new target broker (Line~\ref{algline:if}-\ref{algline:copyexenv}). For operator migrations, a minimum state is extracted, which corresponds to the intermediate state discussed in detail in the next paragraph (Line~\ref{algline:intstate}). Afterwards, this state is sent to the target broker to start executing the operator with the minimum migrated state. 

The target broker subscribes to its producers or predecessors to receive data streams starting from the time of reception of the intermediate state (Line~\ref{algline:transferstate}). 
When the migration is complete, the target broker will send an acknowledgement, including the sequence number of the first output event to the source broker and the coordinator (Line~\ref{algline:ifstart}).
After the source broker has been acknowledged, it will stop its execution, and the target \ac{OP} mechanism will continue at the target broker (Line~\ref{algline:stopex}). %
We start the transition at time $t_i$, to sequentially perform $m$ operator migrations  until the transition is completed at time $t_e$. 
The recursive function performs the operator migration by traversing bottom-up the operator graph (Line~\ref{algline:reccall}). 
If the operator migration is not successful for some reason-- the \ac{IoT} resource becomes unavailable-- and the acknowledgement is not received, the process is repeated until a new target broker is found and the operator is migrated.  (Line~\ref{algline:else}). 
In Line~\ref{algline:else}, we assume a consumer specified parameter $m$ that determines the maximum number of repetitions\footnote{This is very unlikely to happen that the target node is not found again and again.} of this loop and guarantees termination after $m$ tries.

\begin{algorithm}[H]
\footnotesize
\KwVar{%
\begin{tabular}{lll}
$\mi{OList}$ & $\leftarrow$ & bottom-up list of set of operators \\ 
$\mi{\omega}$ &  $\leftarrow$ &  current operator to be migrated \\
$\mi{producers}$ &  $\leftarrow$  &  list of producers connected to $\omega$ \\ 
$\mi{targetMechanism}$ &   $\leftarrow$ &  target \ac{OP} mechanism \\ 
$\mi{targetBroker}$ &   $\leftarrow$ &  target broker host of $\omega$ \\ 
$\phi_{Int}$ &  $\leftarrow$ &  intermediate state of $\omega$
\end{tabular}
}
\BlankLine
\Function{$\textsc{Init-MFGS-SequentialTransition}()$}
{ \label{algline:funstart}
	$ \mi{OList} \leftarrow \textsc{bottomUpAsList}(\Omega)$\;\label{algline:bottomup}
	$\textsc{MFGS-Sequentialalgorithm}(\mi{OList}.\textsc{head} , \mi{targetMechanism})$\label{algline:ophead}
}
\Function{$\textsc{MFGS-SequentialTransition}(\mi{\omega} , \mi{targetMechanism})$} {
		$ \mi{targetBroker} \leftarrow \mi{targetMechanism}.\textsc{findtargetBroker($\omega$)}$\;\label{algline:findbroker}
		\If{ $ \mi{targetBroker} \neq \mi{\omega}.\textsc{sourceBroker}$} { \label{algline:if}
			$ \mi{\omega}.\textsc{copyExecutionEnvironment}\mi{(targetBroker)}$\;\label{algline:copyexenv}
			$ \phi_{Int.} \leftarrow \mi{\omega}.\textsc{computeIntermediateState}()$\;\label{algline:intstate}
			$ \mi{targetBroker}.\textsc{StartExecutionWithData} (\mi{producers}, \linebreak
		 \phi_{Int.})$\;\label{algline:transferstate}
		 	\If{$ \mi{\omega}.\textsc{next}().\textsc{receivedACK}(\mi{timeout}, \mi{retries})$} {\label{algline:ifstart}
			$ \textsc{StopExecution}(\mi{\omega}.\textsc{sourceBroker})$\;\label{algline:stopex}	
			$ \textsc{MFGS-SequentialTransition}(\mi{\omega}.\textsc{next}(), \linebreak \mi{targetMechanism})$\;\label{algline:reccall}			
			}
			\Else {
			$ \textsc{MFGS-SequentialTransition}(\mi{\omega}, \linebreak \mi{targetMechanism})$\;\label{algline:else}
			}
		}	
	
}
\caption{Moving Fine-Grained State Sequential Transition. }
\label{algo:seqTransition}
\end{algorithm}

\begin{figure}[H]
\centering
\includegraphics[width=0.8\linewidth]{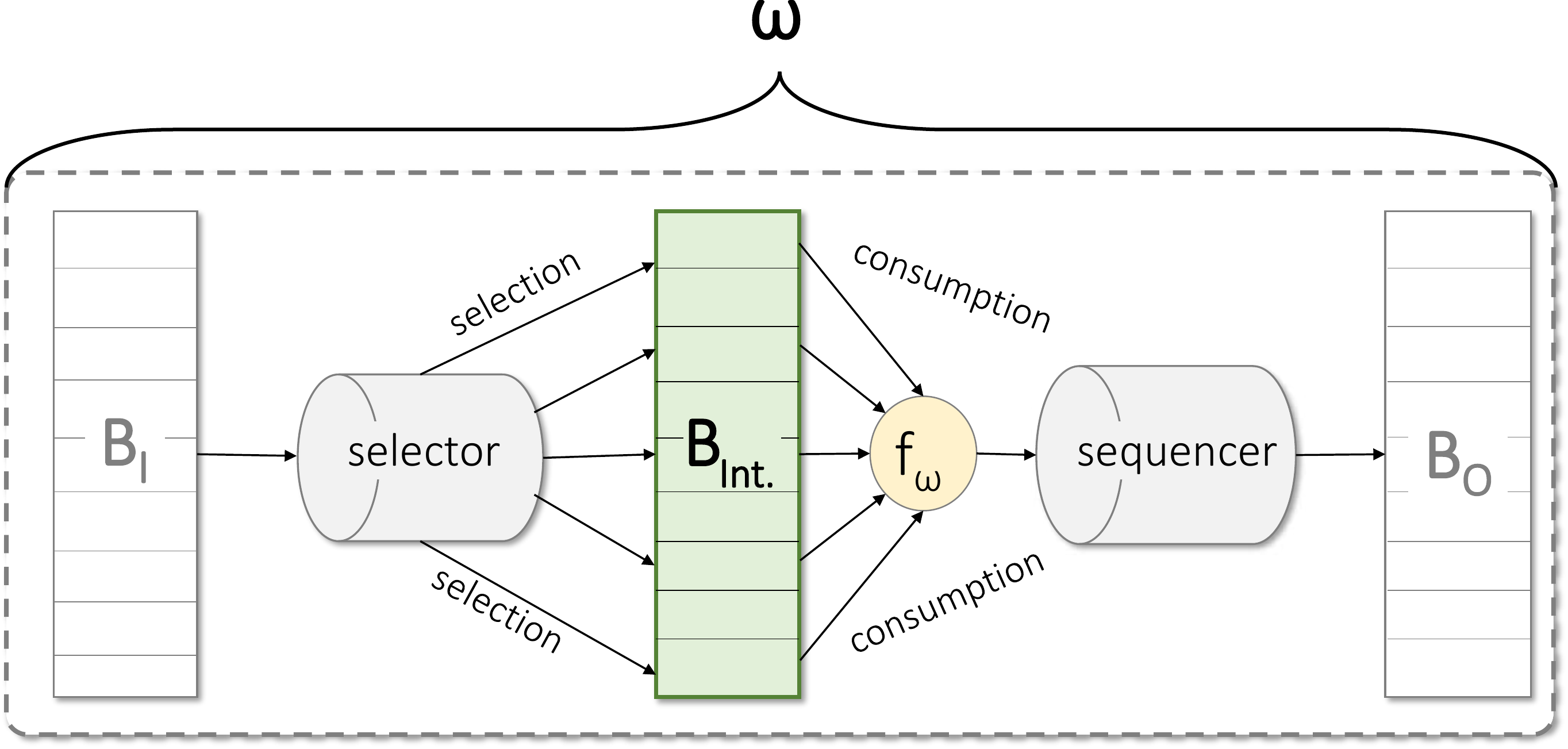}
\caption{Intermediate buffer represented in the operator state model~\cite{PhDWermund2018}.}\label{fig:opexecutionmodel}
\end{figure}

\paragraph{Cost-efficient Operator Migrations}
The \system transition engine computes the fine-grained computational state of an operator  for {\it cost-efficient} operator migrations. 
We improve on the operator state model introduced in ~\cite{Dwarakanath2016,PhDWermund2018} by proposing cost-efficient and seamless operator migrations such that minimal state is transferred at discrete time steps, which are optimal for costs as we explain in the following subsection in seamless and minimal state concurrent transition. Furthermore, our migration model considers dependencies between operators during migration, hence providing a means to migrate an entire operator graph consistently.
In the operator state model (cf. Figure~\ref{fig:opexecutionmodel}), the input events are cached in the input buffer ($B_I$) selected by the \emph{selector} to map the output events determined by the correlation function of the operator ($f_{\omega}$). 
Next, the \emph{selector} handles the removal of events from the input buffer $B_i$ when the same are either consumed or discarded by the correlation function $f_{\omega}$. 
The resulting output or complex events are stamped with a sequence number ($SN$) by the \emph{sequencer} and appended into the output buffer $B_O$ which is then forwarded to the $\omega$'s successor. 
The events which the successor operators have already acknowledged are removed from the output buffer $B_O$.
Although the state model is applicable to many modern CEP systems, such as Apache Flink\footnote{Apache Flink network stack. \url{https://flink.apache.org/2019/06/05/flink-network-stack.html} [Accessed on 18.4.2021]}, which assumes the presence of buffers, with a few adaptations in the internal structures, it can be applied to other CEP system, \eg those that do not assume buffers~\cite{Geethakumari2017/fpga}.

Conventionally, a {CEP} system transfers the internal state $\phi_{\omega}$ that comprise the input buffer $B_I$, the \emph{selector}, the correlation function $f_{\omega}$, the \emph{sequencer}, and the output buffer $B_O$. \system transfers the content of the intermediate buffer $B_{Int}$ instead of the entire state $\phi_{\omega}$.  
The content of $B_{Int}$ contains those events on which the correlation function $f_{\omega}$ is applied to obtain the complex events. 
For example, for a {window-aggregate} operator, the content of $B_{Int}$ will be the events contained in the window, and those are selected to be aggregated by the correlation function (sum, min, or max). This set of events are updated each time the output events are generated, e.g., once the window slides (for a sliding window operator) or the related event is either consumed (inserted into the output buffer $B_O$) or discarded by the correlation function. 

The target broker must subscribe timely to the required incoming data streams to optimize the migration cost and completion time. Consider the intermediate state $\phi_\omega (t_i)$ of operator $\omega$ migrated at time $t_i$ comprises 
$B_{Int}$, the correlation function $f_{\omega}$, and the state of the sequencer (Line~\ref{algline:transferstate}). Here, $B_{Int}$ replays the events that were selected for correlation before the source broker went down (Line~\ref{algline:transferstate}). 
At time $t_i - \delta_M$, the target broker subscribes to the input events from the producers or the predecessor operator. 
Here, $\delta_M$ is a small value to ensure that the target broker receives input events before the processing starts. All input events to the target broker until the source broker is executing are discarded (Line~\ref{algline:ifstart}). 
It is important to note that a careful selection of $\delta_M$ value is essential so that the target broker does not miss any input event. In case the value is very big, there will be an overlap in the execution of the source and the target broker. The duplicates are discarded; however, it results in an unnecessary overhead that should be avoided. 

Contrarily, if the $\delta_M$ value is very small, there is a slight chance that the target broker might miss some of the input events. However, this is very unlikely to happen. Nevertheless, we address this problem by proposing a seamless transition algorithm where the state overhead is further minimized and the correctness of the events is guaranteed, as discussed in the  subsection of seamless minimal state concurrent transition.

\paragraph{Properties}  
We analyze the transition time and present an asymptotic upper bound on the cost ($C_{Time}(T)$). The transition time is bounded by the time required by the algorithm to iterate over all operators sequentially and to transfer the intermediate state of each operator (Lines~\ref{algline:transferstate} to~\ref{algline:reccall}). Therefore, the overall transfer time can be bounded by the transfer time of the entire intermediate operator state $\phi_\Omega$ and the time to iterate over all operators, which yields $\mathcal{O}\left(|\Omega|+ |\phi_\Omega|\right)$.
Here, $\phi_\Omega$ denotes the intermediate state of the set of operators $\Omega$ within the operator graph\footnote{$\Omega$ here stands for the set of operators as previously defined in the notations, not to be confused with the generic notation on asymptotic lower bound of an algorithm.}. 

In this algorithm, we reduce the time required to perform an operator graph transition by transferring a minimum amount of state. However, the processing of an operator at the target broker does not occur unless the source broker is in execution. This means that while the selected state is being transferred (i.e., it is on the wire), some events sent to the target broker remains unprocessed. No output events are produced unless the intermediate state is transferred. Although with this transition algorithm, a minimum amount of state is achieved yet, state transfer involves costs in terms of time  and resources.
Another problem is the sequential transfer of operators. While sequential transfer does not consume much network resources, it is very time-consuming. To solve these issues, we propose a second transition algorithm. 

\par

\subsubsection*{\textbf{\ac{SMS} Concurrent Transition.}} \label{subsubsec:SMS}

In contrast to the above algorithm, this algorithm allows for more than one operator migrations simultaneously (cf. Algorithm~\ref{algo:expoTransition}: Lines~\ref{alg2line:funstart}-\ref{alg2line:reccall}). At each level $l=0$ to $m$ of the operator graph $G$, the coordinator triggers at most $2^l$ operator migrations (for binary operator graph) performed in a bottom-up fashion (Line~\ref{alg2line:forlevel}). 
The benefit of concurrent operator migrations is perceived in the cost computation that is later analyzed in the properties of the algorithm.  
The operator migrations begin when the coordinator transfers the execution environment
(Line~\ref{alg2line:copy}). 
The coordinator determines an optimal time $t_i$ for each operator $\omega$
 when the operator state is minimal
so that the transition consumes minimum resources (Line~\ref{alg2line:minimalState}). For this, we assume the events follow a time order of arrival~\cite{Zhu2004}. 
 The selection of time $t_i$ is such that for each operator $\omega$, \ac{SMS} algorithm waits until the operator $\omega$ is purged from its old state (Line~\ref{alg2line:wait}), \ie until $B_{Int}$ and  $f_{\omega}$ are purged from their old state. 

\IncMargin{1em}
\begin{algorithm}[H]
\footnotesize
\KwVar{%
\begin{tabular}{llp{5.5cm}}
$\mi{producers}$ & $\leftarrow$ & \text{list of the event producers} \\ 
$\mi{OGlevel}$ & $\leftarrow$ & \text{operator graph level for migration} \\ $\mi{targetMechanism}$ & $\leftarrow$ & \text{target \ac{OP} mechanism} \\ 
$\mi{targetBroker}$ & $\leftarrow$ & \text{target broker host of current operator} \\ $\phi_{\mi{sequencer}}$ & $\leftarrow$ & \text{state of sequencer} \\ 
$\mi{waitTime}$ & $\leftarrow$ & time taken until current operator is purged from its state
\end{tabular}
} 
\BlankLine
\Function{$\textsc{SMS-ConcurrentTransition}(\mi{OGlevel} , \mi{targetMechanism})$}{\label{alg2line:funstart}
	\ForAllP {$\mi{\omega} \in \mi{OGlevel}$} { \label{alg2line:forlevel}
		$ \mi{targetBroker} \leftarrow \mi{targetMechanism}.\textsc{findtargetBroker($\omega$)}$\;\label{alg2line:targetBroker}
		\If{ $ \mi{targetBroker} \neq \mi{\omega}.\textsc{sourceBroker}$} {\label{alg2line:ifstart}
			$ \mi{\omega}.\textsc{copyExecutionEnvironment}\mi{(targetBroker)}$\;\label{alg2line:copy}
			$ \textsc{NTPClockSynchronization,}(\mi{targetBroker}, \linebreak  \mi{\omega}.\textsc{sourceBroker})$\;\label{alg2line:clocksync}
			$ \mi{minimalStateTime} \leftarrow \mi{\omega}.\textsc{determineMinimalStateTime}()$\;\label{alg2line:minimalState}
			$ \mi{waitTime} \leftarrow \textsc{waitUntil}(\mi{minimalStateTime}))$\;\label{alg2line:wait}
			$ \phi_{\mi{sequencer}} \leftarrow \mi{\omega}.\textsc{lastSN}$\;\label{alg2line:stateseq}
			$ \mi{targetBroker}.\textsc{StartExecutionWithData} (\mi{producers}, 
		\linebreak \phi_{\mi{sequencer}})$\;
			$ \mi{targetBroker}.\textsc{determineReferencePoint} \linebreak(\mi{minimalStateTime})$\;\label{alg2line:refpoint}
			\If{$ \mi{\omega}.\textsc{parent}().\textsc{receivedACK}(\mi{timeout}, \mi{retries})$} {\label{alg2line:ack}
				$ \textsc{StopExecution}(\mi{\omega}.\textsc{sourceBroker})$\;
				$ \textsc{SMS-ConcurrentTransition}(\mi{OGlevel.\textsc{next}()}, \linebreak \mi{targetMechanism})$\;
			}
			\Else {
				$ \textsc{SMS-ConcurrentTransition}(\mi{OGlevel}, \linebreak \mi{targetMechanism})$\;\label{alg2line:reccall}
			}
		}
	}		
}
\caption{Seamless Minimal State Concurrent Transition.}
\label{algo:expoTransition}
\end{algorithm}
\DecMargin{1em}
 
For example, in a {window-aggregate} operator, the target broker waits until the last event of the window is processed,
$w + \delta_{S}$, where $w$ is the window size, and $\delta_{S}$ is a small value to ensure that $t_i$ is greater than any time instant of input events to the source broker.
Time $t_i$ is chosen as the \emph{transition start time}. We call this time the \emph{minimal state} time of an operator ($t_{imin}(\omega)$).
The target broker starts its execution with the minimal state (the last $SN$) simultaneously at the \emph{transition start time}, while the successor operators at the higher level are still under execution by the former \ac{OP} mechanism. Thus, in this algorithm, the transition coordinator allows the execution of two \ac{OP} mechanisms concurrently. This allows us to deal with the output disruption discussed as follows.

\begin{figure}[H]
\centering
\includegraphics[width=\linewidth]{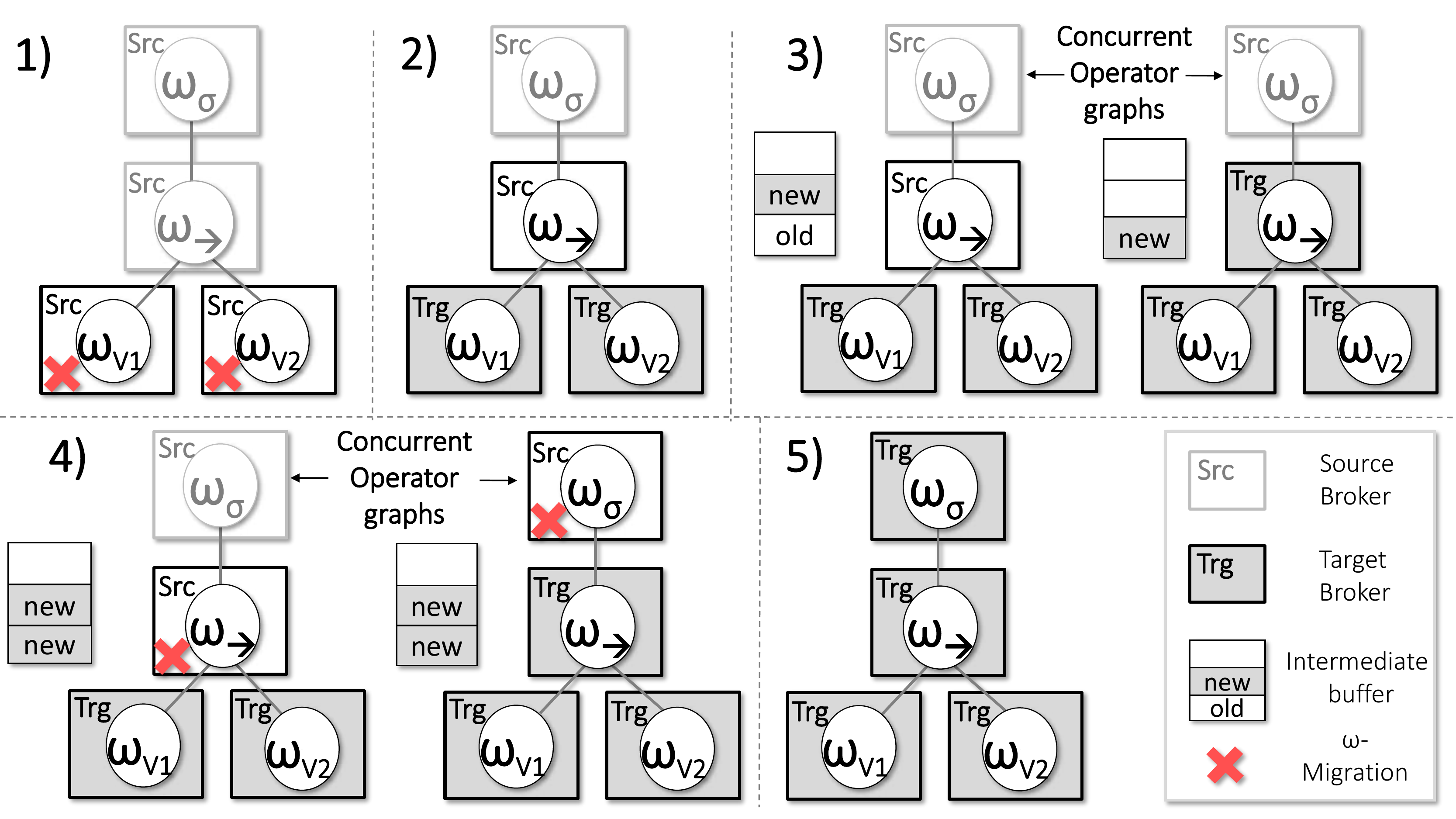} 
\caption{Sequence of operator migrations in the operator graph for SMS transition algorithm.}
\label{fig:opSequenceMovement}
\end{figure}

\paragraph{Seamless and Concurrent Operator Migrations}
To explain the concurrent operator migrations, we refer to the operator graph from our example scenario in Figure~\ref{fig:opSequenceMovement}. \emph{Src} box refers to the placement of an operator at the source broker, and the \emph{Trg} box refers to the placement at the target broker. Steps 3 and 4 show the $B_{Int}$ buffer of the sequence operator with the event tuples being processed. The first step shows the initial placement, while the last one shows the final placement after migration. 
The concurrent execution of two \ac{OP} mechanisms (cf. step 2 to 3 in  Figure~\ref{fig:opSequenceMovement}) enables seamless execution in this algorithm. However, migrations do not interfere with each other, while the operator network gradually transforms the placement (cf. step 4). The transition coordination is accomplished atomically in the \system transition engine. 

To better understand the cost of concurrent operator migrations, we analyze the reception of input events at both source and target brokers after transition start time $t_i$.
For an operator $\omega$, the state $\varphi_\omega(t_i)$ at transition time will only comprise the state of the sequencer (containing the $SN$ of the first event to be produced at the target broker) (Line~\ref{alg2line:stateseq}). 
The basic idea of this transition algorithm is that at the transition start time $t_{imin(\omega)}$, the input buffer $B_I$ and the output buffer $B_O$ are shared among the source and the target brokers until it is safe to discard the source broker. Both source and target broker for a stateful operator $\omega$ run concurrently while all the old tuples in the intermediate buffer $B_{Int}$ of the source broker are gradually purged (cf. step 3 and 4). For instance, in a sliding window operator, with a window size of $S$ events and slide size of 1 event, the old tuples still have to be retrieved until the target broker has received a full window size of $S$ events. During this time, the output is continually produced by both the brokers, while duplicates are discarded using the reference point method later explained. When the intermediate buffer is purged completely, then the source broker is discarded. This is because the target broker now has all the new tuples that exist in the source broker.
The source brokers of stateless operators are gradually replaced by their targets, as illustrated in the figure with a red cross (\textcolor{red}{\ding{53}}). 

To deal with the clock drift between the two clocks of the source and the target brokers, we perform distributed clock synchronization using standard Network Time Protocol (NTP)~\cite{Mills1991} at both ends (Line~\ref{alg2line:clocksync}). To avoid duplicates in the output events due to concurrent processing, we use the reference point method~\cite{Bercken1996} (Line~\ref{alg2line:refpoint}). We treat the start timestamp of the results of the target broker as a reference point. Such timestamp is then compared to the \emph{transition start time} $t_i$. If the reference point is larger than $t_i$, then the complex event is sent to the output buffer $B_O$. 

\paragraph{Correctness of the results} We assess correctness on two aspects as widely done in the literature~\cite{Zhu2004, Monte2020/sigmod/statemgmt}: the output is complete, and there are no duplicates in the output. \Cref{fig:opSequenceMovement} shows the transfer of the operator graph in 1) through 5) steps using the SMS algorithm. The stateless operators are transferred straightaway, while stateful operators run in parallel using the SMS algorithm until all the old tuples are purged. Furthermore, while the predecessor operators are migrated, successors still use the former \ac{OP} mechanism for resolving the query.
We must also ensure that there are no duplicate output tuples, as we can see in step 3): the sequence operator leads to duplicate output tuples from the source and target operator, respectively. 
A naive approach is to discard all the input as old tuples that results from the source broker. 
However, this would lead to incorrect results, as seen in step 3): the old tuple might be a true sequence that will remain undetected if dropped. 
To solve this issue, though we have the source and target brokers in execution concurrently, we drop events from target brokers unless all the events in the $B_{Int}$ buffer are new and the source broker could be stopped. 
For instance, in Step 3, we retrieve the output result from the source broker holding the sequence operator, while in step 4, we can safely discard the source broker since all the tuples in the state are new. 

\paragraph{Properties}
In this algorithm, we partition the transition at discrete time steps such that for each operator migration $M_i$, we determine the \emph{minimal state time} as described before. 
This approach ensures a live and seamless transition without service disruption, thanks to minimal consumption of resources. Due to the concurrent transfer, the number of nodes in the new operator network increases exponentially over time with the increase of the size of the operator graph $G$. Therefore, the total transition time of this algorithm is within $\mathcal{O}\left(log(|\Omega|)+C\right)$, here $C = |\varphi_\Omega|$ that is constant (state of the sequencer) for a given set of operators $\Omega$. 


\section{Evaluation}\label{sec:evaluation}
\mli{R1: In general, the questions the authors want to address are valid. Their answers, though, are way narrower than they should}
In the evaluation of \system, we aim to answer the following questions:

\begin{enumerate}[leftmargin=*]
\item Is the programming model able to simply express existing operator placement mechanisms?
\item Does the mechanism transition concept satisfy changing \ac{QoS} demands for dynamic environmental conditions?
\item Can a transition for the \ac{OP} mechanism be performed in a live and seamless manner?
\item What is the cost involved in the execution of a transition, and is the cost acceptable?
\end{enumerate}

To answer the above questions, we evaluate \system in four ways:
\i~In \Cref{subsec:performanceplacement}, we evaluate the \system programming model in terms of the development of \ac{OP} mechanisms and validating their performance.
\ii~In Section~\ref{subsec:TCEPimpact}, we evaluate the ability of \system to meet \ac{QoS} demands with respect to latency and message overhead.
\iii~In Section~\ref{subsec:TCEPperformance}, we evaluate the stability of the system subject to transitions and the cost imposed by the distinct transition algorithms proposed in Section~\ref{sec:design}.
\iv~In \Cref{subsec:learningcosts}, we evaluate the costs of the genetic learning algorithm in terms of selection and performing a transition.

In the following sections, we first describe our evaluation execution environment, including details on the implementation of \system, the evaluation setup, and then present our evaluation findings.

\begin{table}
{\setstretch{0.9}
\centering
\footnotesize
\vspace{-10pt}
\begin{tabular}{lp{7cm}}
\toprule
Number of producers          & $1-9$                            \\
Number of brokers            & $1-9$                           \\
Number of consumers          & $1-2$                            \\
Number of queries            & $1-50$                           \\
Type of queries              & Stream (Q1), Filter (Q2), Conjunction (Q3), Join (Q4), \underline{Congestion detection} (Q5) (\Cref{fig:og-queries})                                                     \\
QOS\_DEMANDS                 & \underline{latency}, message overhead, network usage, hops\\
\midrule
\ac{OP} mechanisms & \underline{Relaxation~\cite{Pietzuch2006}}, MOPA algorithm~\cite{Rizou2010}, \underline{MDCEP~\cite{Starks2017}}, \\
&  Global Optimal,  Producer-Consumer, Random \\
Transition execution algorithms & \underline{{MFGS}-Sequential}, {MFGS}-Concurrent, \\
 &  {SMS}-Sequential, {SMS}-Concurrent \\
 Placement selection algorithm & \underline{Genetic learning-based}, Requirement-based \\
 \bottomrule
\end{tabular}
\caption{Configuration parameters for the evaluation. \emph{The default/mostly used parameters are underlined}.}´
\label{tab:evaluation}
\vspace{2pt}
}
\end{table}

\subsection{Evaluation Environment and Setup} \label{subsec:ev_setup}

\paragraph{\textbf{Implementation}} The implementation of \system builds on an adaptive complex event processing system proposed in~\cite{Weisenburger2017}.
In particular, \system builds on the AdaptiveCEP programming model for specifying \ac{QoS} demands at run time (cf. the query in Figure~\ref{code:samplequery}). We provide the runtime environment based on the Akka actor system~\cite{Akka} and Akka Cluster to build a distributed network of containers for easy deployment in the edge-IoT scenario.
The Docker container helps encapsulate a runtime environment to enable the deployment of operators on the IoT resources.
Furthermore, we realized extensions in the form of a placement module that integrates state-of-the-art \ac{OP} mechanisms~\cite{Pietzuch2006,Starks2015} and measure the resulting \ac{OP} performance.

\begin{figure}[H]
    \centering
    \includegraphics[width=0.7\linewidth]{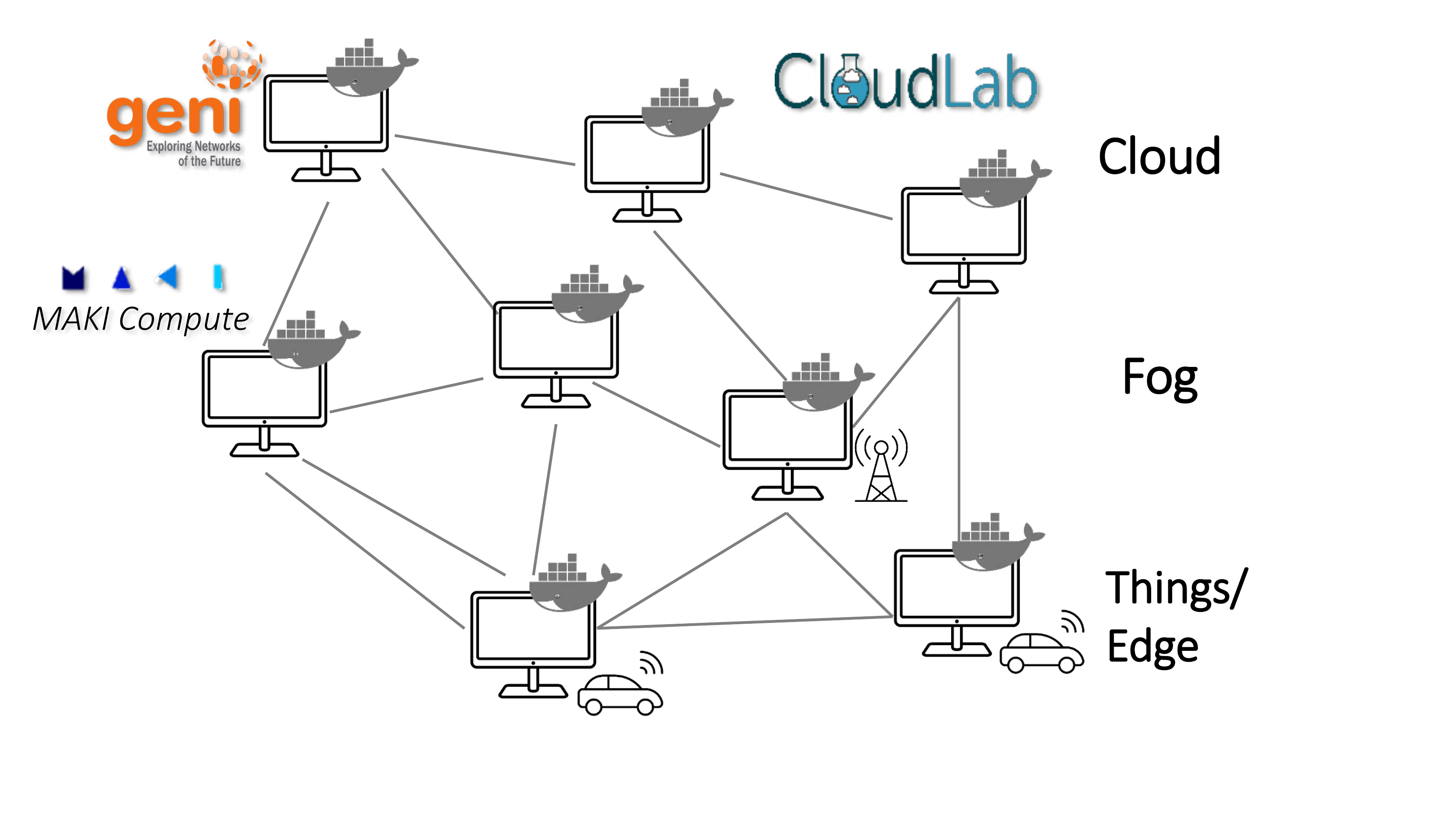}
    \caption{Our setup comprises 8 physical machines of publicly available network infrastructures running our virtualized DCEP system in docker containers.}
    \label{fig:setup}
\end{figure}

\begin{figure}[H]
    \centering
    \includegraphics[width=0.6\linewidth]{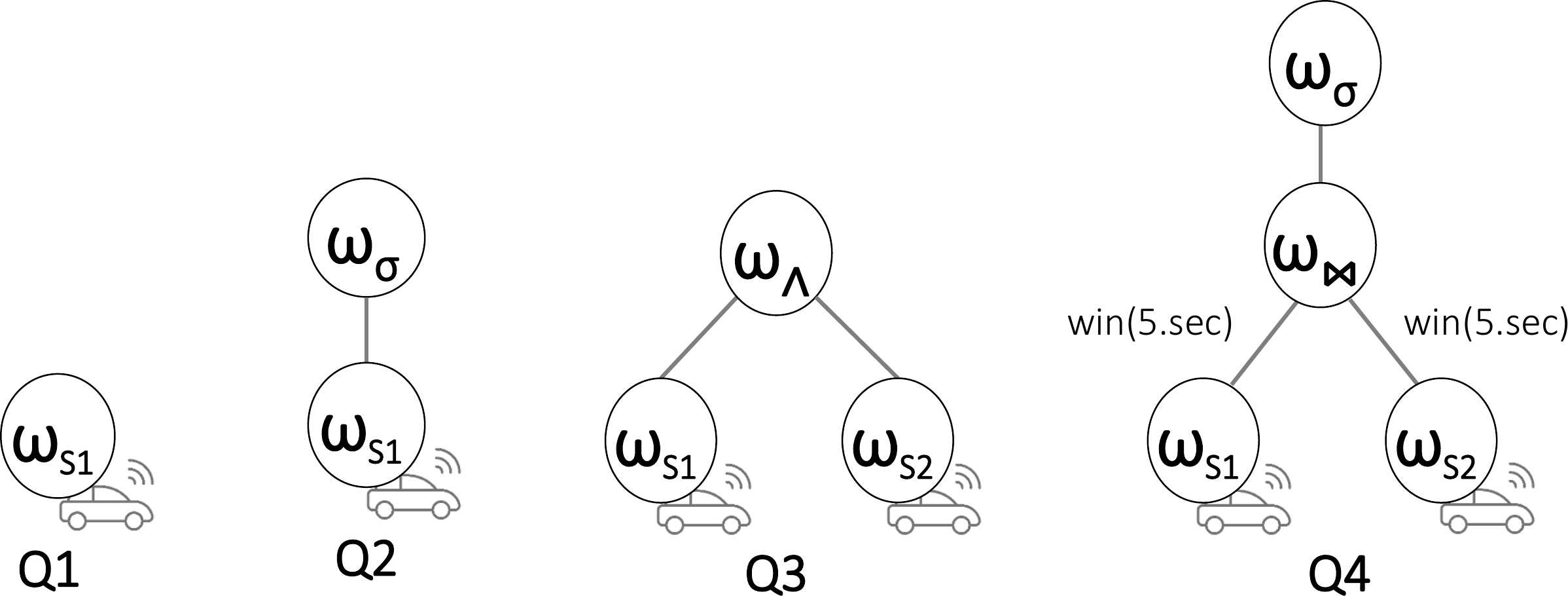}
     \includegraphics[width=0.8\linewidth]{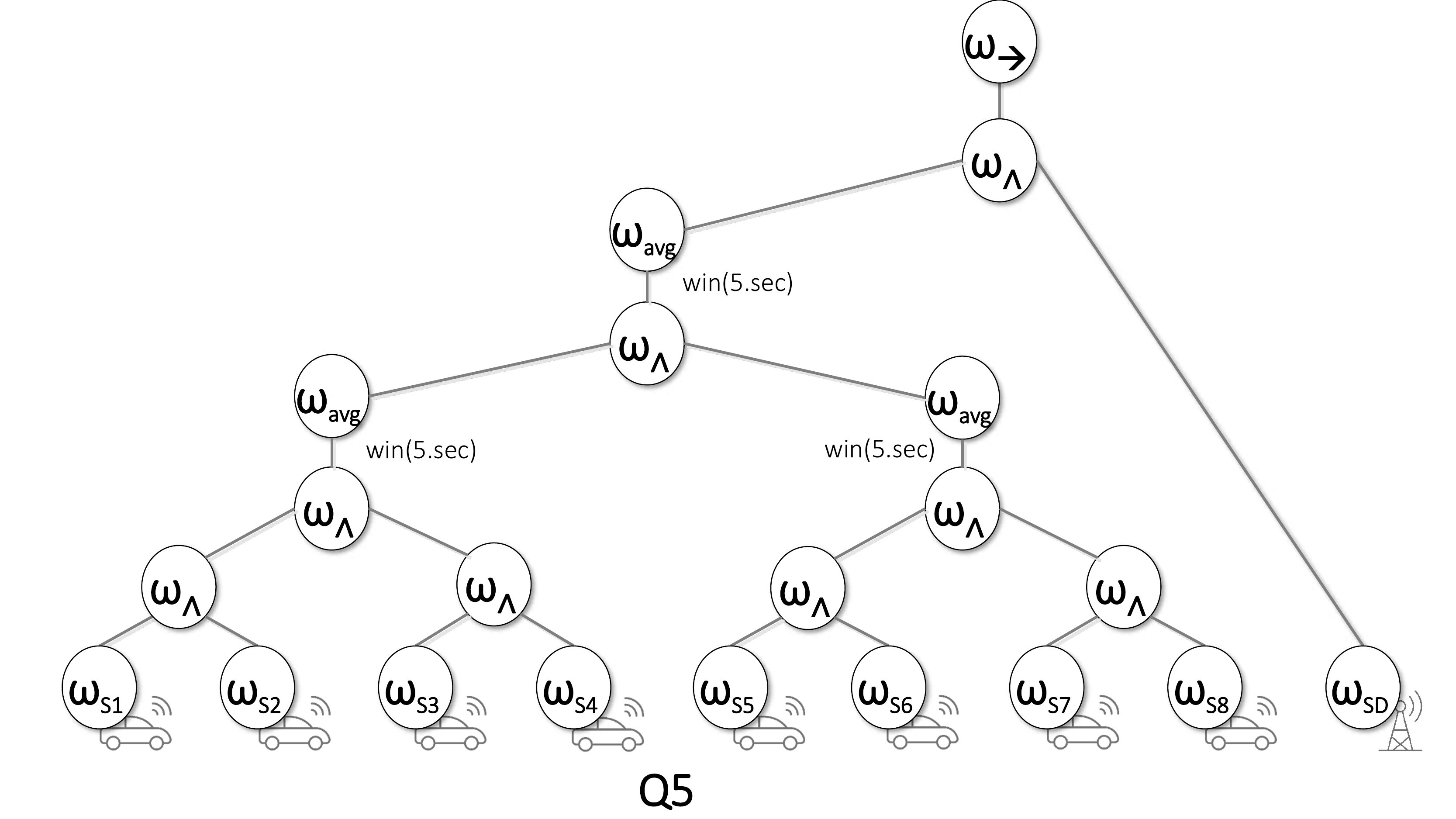}
    \caption{Operator graph for queries Q1 to Q5.}
    \label{fig:og-queries}
\end{figure}
 We build \system's  Docker image upon the Alpine Linux distribution,\footnote{Docker image upon Alpine Linux distribution. \url{https://github.com/gliderlabs/docker-alpine} [Accessed on 18.04.2021]}
  which is much smaller in size (base image size of only 5 MB) and lightweight than other Linux based images. For instance, a standard Ubuntu docker image is 129 MB in size.~\ml{R1: Much smaller and lightweight" than what? done}
   Furthermore, the lightweight Docker-based execution environment is contained such that it does not exceed 2 GiB of allocated memory which is a reasonable assumption for small devices available nowadays as in the edge IoT scenario. The Docker containers communicate over an overlay network using TCP (Transmission Control Protocol) as an underlying transport protocol. We use Akka v. 2.6.0~\cite{Akka}, the Esper CEP engine v. 5.5.0~\cite{Esper}  and Docker v. 19.03.8-ce~\cite{Docker}.

\paragraph{\textbf{Platform and Setup}} We deploy Docker services on 8 VMs with 8 GiB of memory and 8 processors per physical machine, as denoted in \Cref{fig:setup}.
We consider different physical machines comprising of network infrastructures of Geni~\cite{GeniArticle2014}, CloudLab~\cite{CloudLab}, and our onsite MAKI~\cite{MAKIwebsite} compute machines. Together, these resources  provide a realistic deployment environment similar to the IoT-fog-cloud infrastructure resource model introduced in \Cref{subsec:networkmodel} and hierarchically illustrated in \Cref{fig:setup}. With resources dispersed in North America (Ohio and UCLA) and Europe (Darmstadt), we have introduced geographical diversity and realistic network latencies, and packet loss environment for our experiments.
The Docker network is setup based on the services that connect using an overlay network.


\paragraph{\textbf{Queries}} We use multiple standard CEP queries defined below\footnote{The queries are specified in AdaptiveCEP DSL in Scala programming language.} (cf.  \Cref{tab:evaluation} and illustrated in \Cref{fig:og-queries}: Q1-Q4).
Besides the standard CEP queries, we use a traffic congestion detection query presented in  Section~\ref{sec:motivation}: \Cref{code:samplequery}.
We elaborate on the query, such as the generation of complex data streams \texttt{vehiclesAtSectionV1} and \texttt{vehiclesAtSectionV2} for the average values related to speed and density.
We illustrate the operator graph in \Cref{fig:og-queries}: Q5, comprising 8 publishers, each representing a Stream operator ($\omega_{S1}$ to $\omega_{S8}$). In the operator graph, the speed information related to the vehicle from the Stream operators is analyzed to get the average speed of the two road sections. Another Stream operator ($\omega_{SD}$) contributes the density information related to the two road sections, which is combined to detect a sequence for the congestion detection using a Sequence-Filter operator ($\omega_{\rightarrow}$). \\
\\
\i \texttt{Stream} Operator
    \begin{codenv}
    Stream => stream[StreamData](speedPublishers(1), demand QoS_DEMANDS)
    \end{codenv}
\ii \texttt{Filter} Operator
    \begin{codenv}
    Filter => stream[StreamData](speedPublishers(1), demand QoS_DEMANDS).where { v1 =>
    v1.avgVehiclesSpeed < NormalSpeedThreshold}
    \end{codenv}
\iii \texttt{Conjunction} Operator
    \begin{codenv}
    Conjunction => stream[StreamData](speedPublishers(0)).and(stream[StreamData](speedPublishers(1)), demand QoS_DEMANDS)
    \end{codenv}
\iv \texttt{Join} Operator
    \begin{codenv}
    Join => stream[StreamData](speedPublishers(0)).join(stream[StreamData](speedPublishers(1)), slidingWindow(5.seconds), slidingWindow(5.seconds)).where{ case (v1, v2) =>
    v2.time > v1.time }, demand QoS_DEMANDS)
    \end{codenv}



\paragraph{\textbf{Dataset}} We used a realistic dataset of the vehicular network scenario from Madrid~\cite{madrid/vhc/torre2017} comprising the input data stream of the form $<time, position, lane, speed>$. This is used to generate complex data streams of $\texttt{vehiclesAtSectionV1}$ and $\texttt{vehiclesAtSectionV2}$ and evaluate further the congestion detection query. Similarly, for other queries as well the same dataset is used. \mli{R1: vehiclesAtSectionV1 repeated twice
similarly for other queries as well the dataset is used - broken sentence. fixed}
We run each execution for $20$ minutes and initiate the measurements after $2$ minutes warm-up. Each measurement is taken at a regular interval of 5 seconds.
For some evaluations, we incrementally increase the query workload for up to $50$ queries.
The evaluation metrics are influenced by multiple parameters such as the number of queries and the window size.
To consider different environmental conditions, we perform a variability analysis on these parameters according to the ranges in Table~\ref{tab:evaluation}.
~\mli{R1: When it comes to the parameters, is it all possible combinations or only some combinations? we have summarized all the parameters of the evaluation in the same table, however, in the respective subsections of evaluations we have explicitly mentioned which parameters are used. Missing explanations are also included now in the major revision, for e.g., ... }

\subsubsection{Operator Placement Mechanisms} \label{subsubsec:opmechanisms}
To understand how the performance of \ac{OP} mechanisms, including those taken from the literature, differs in terms of \ac{QoS} fulfilment, we implemented several \ac{OP} mechanisms. In the following, we give a brief description of the design characteristics of the implemented mechanisms.

\begin{enumerate}
    \item \emph{Relaxation~\cite{Pietzuch2006}.} It is based on a so-called cost space that considers latency and bandwidth together as two dimensions. In the first step, the virtual operator placement is performed using the cost space, and in the second step, physical operator mapping is performed on the topology using KNN (K-nearest neighbours algorithm). The basic idea behind the first step, i.e., virtual operator placement, is a physics analogy revolving around springs. The distance by which a spring is extended resembles a link's latency, and the spring constant (specifying its stiffness) is the bandwidth of the link. The product of spring extension and spring constant is the force needed to extend the spring (Hooke's Law); the product of latency and bandwidth is the bandwidth-delay product (BDP). Note that Relaxation uses the squared latency to ensure a unique solution if the bandwidth observed is equal.  Operators are connected by springs that pull and push them into place inside the virtual coordinate space until the system has "relaxed" completely, \ie until the sum of forces inside the operator graph is zero. The operators are then mapped to the nodes closest to their respective virtual locations that are not overloaded. Through this heuristic, the overall BDP, \ie the total amount of data in transit through the network at a given moment, is minimized, better known as network usage.

    \item \emph{MOPA Algorithm~\cite{Rizou2010}.}
    MOP is a variant of the Relaxation algorithm to minimize the bandwidth-delay product; hence instead of squared delay, this algorithm considers delay as an optimization criterion\ml{R2: delay … criteria -> criterium done}. Besides the optimization goal, this algorithm finds the local optimal solution using a gradient method, terminating when the current network usage (given by the above optimization criteria) becomes smaller than a threshold.

    \item \emph{Global Optimal.} Compared to the above two algorithms that find a sub-optimal solution, we implemented a global optimal mechanism that chooses the best possible operator placement with  minimum network usage (bandwidth-delay product) based on an exhaustive search of the possible placements. This \ac{OP} mechanism requires global knowledge of the entire network.

    \item \emph{MDCEP~\cite{Starks2015}.} The placement decision in MDCEP is made locally, and no cost information is shared among the nodes resulting in lower communication overhead and achieving a stable operator placement near the data sources.
    The authors consider a scenario of highly mobile nodes for operator placement. Hence, a decentralized mechanism with optimization criteria of minimizing message overhead and latency is considered.

    \item \emph{Producer-Consumer.} For comparison to the above approaches, we consider placement on the randomly chosen producer or consumer \emph{only}. This is because the MDCEP mechanism considers stable operator placement that can be achieved by placing operators on producers or consumers where message loss can be minimal. Hence, this approach is also considered for comparison.

    \item \emph{Random.} This mechanism chooses a physical host for each operator randomly and serves as a naive comparison.
\end{enumerate}

\subsection{Performance of \ac{OP} mechanisms}~\label{subsec:performanceplacement}
To understand the design space of \ac{OP} mechanisms with distinct and conflicting optimization criteria, we evaluate them using the \system programming model presented in \Cref{subsec:prog-model}. We consider the QoS demands, queries, and \ac{OP} mechanisms as stated in \Cref{tab:evaluation} for comparison. The performance metrics, including the \ac{QoS} demands, are defined as follows:
\i~Mean end-to-end latency or simply latency: It is the time taken from the query subscription was first received at the consumer end until the complex event was received back to the consumer (cf. \Cref{def:e2elatency}). \mli{R1: In 6.2, Mean end-to-end latency or simply latency, is this the one that says time in the Y label? yes}
\ii~Mean message control overhead or message overhead: The number of messages (in MB) exchanged to perform the operator placement. This includes establishing the broker network, exchanging network or node information for placement, and performing the placement (cf. \Cref{def:cmOverhead}).
\iii~Mean network usage: The amount of data in transit through the network given by the bandwidth-delay product as introduced in the Relaxation mechanism above.
\iv~Mean number of hops: The number of hops or physical hosts used for an operator placement.

\begin{figure}[H]
  \includegraphics[width=0.6\linewidth,left]{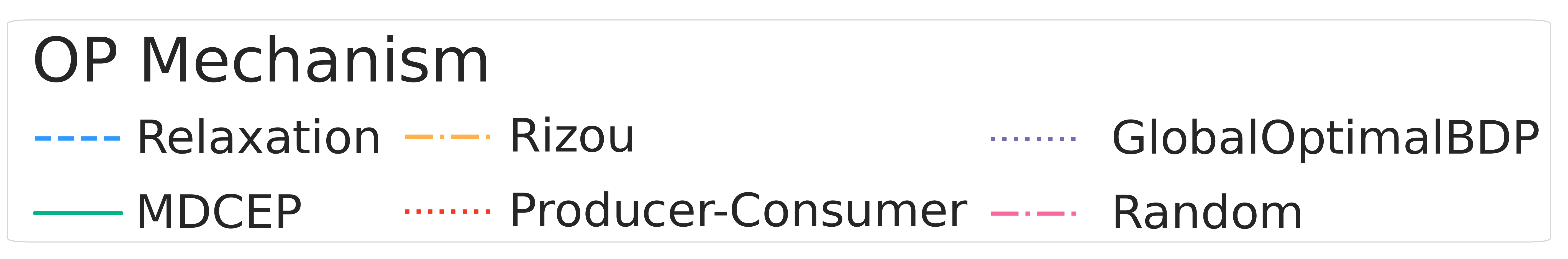}
  \newline
    \centering
    \includegraphics[width=0.49\linewidth]{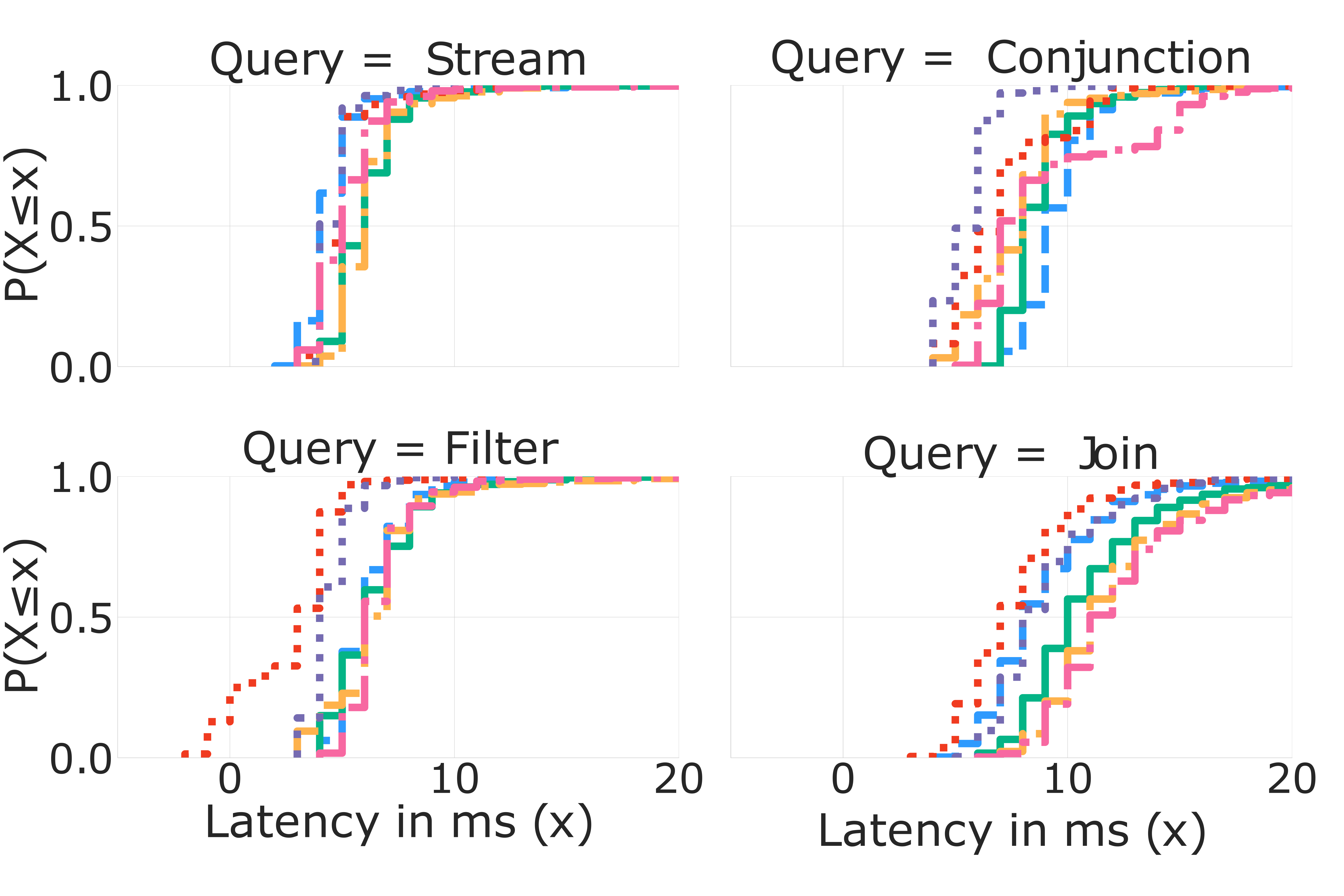}
    \includegraphics[width=0.49\linewidth]{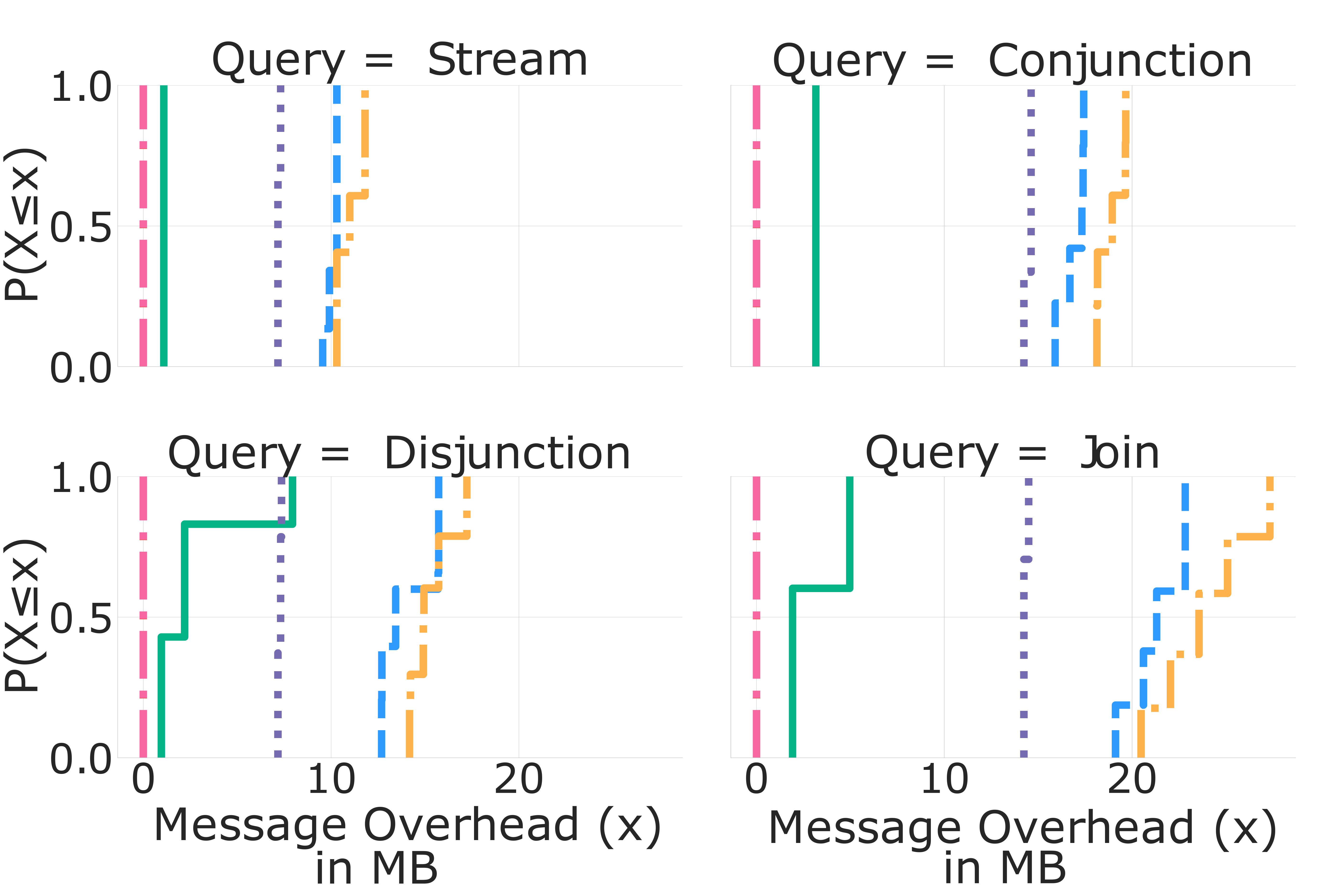}
    \includegraphics[width=0.49\linewidth]{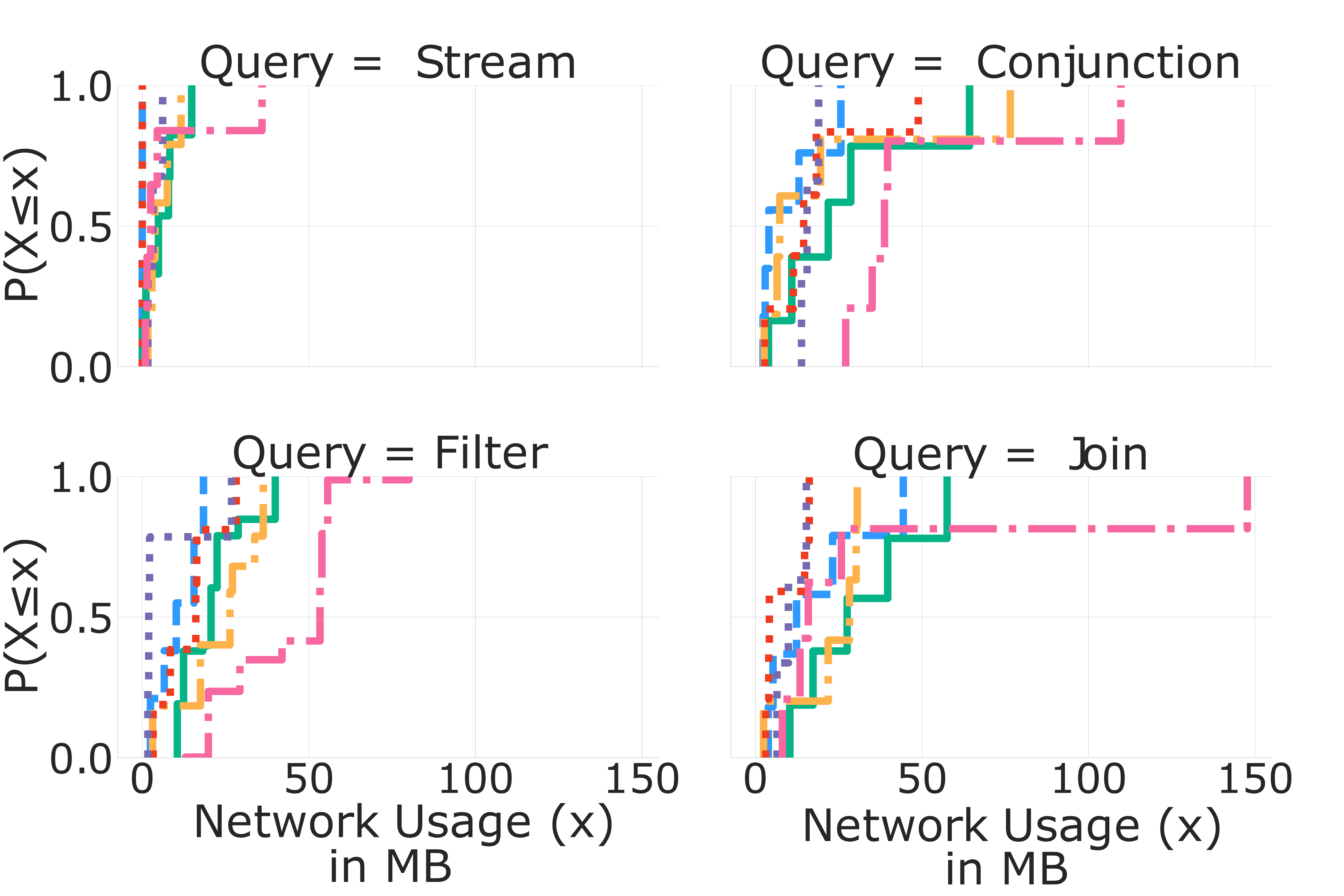}
    \hfill
    \includegraphics[width=0.49\linewidth]{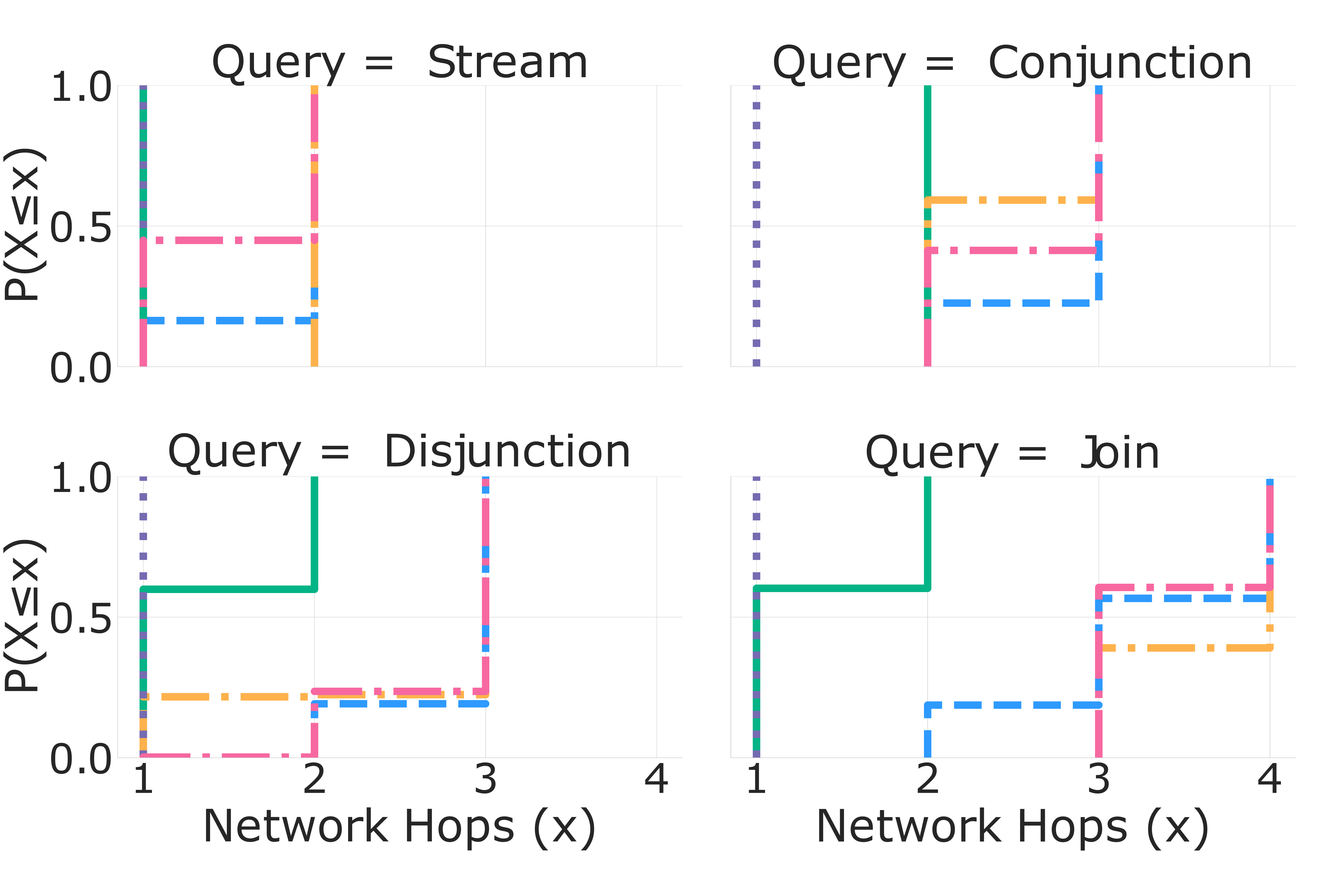}
    \caption{Performance evaluation of \ac{OP} mechanisms (CDF) using programming model of \system in terms of latency, message overhead, network usage, and the number of hops for the standard CEP queries listed in \Cref{tab:evaluation}. \emph{Here, for all the metrics the more to the left is the distribution, the better it is.}}
    \label{fig:placement/performance_all}
\end{figure}

\Cref{fig:placement/performance_all} presents the cumulative distribution function (CDF) for the different QoS metrics using the given \ac{OP} mechanisms (cf. \Cref{subsubsec:opmechanisms}) and standard CEP queries Q1 - Q4 (cf. \Cref{fig:og-queries}).
It can be seen that each \ac{OP} mechanism behaves differently for different queries. For instance, in terms of latency, Relaxation and Global Optimal mechanisms perform best for Stream and Conjunction operators, Producer-Consumer supersedes them when executing Filter and Join queries.
This is because the main objective of Relaxation and Global Optimal \ac{OP} mechanisms is to minimize overall latency.
The Producer-Consumer mechanism can also achieve similar performance because of its proximity to the event sources and the end-users.
In terms of message overhead, MDCEP and Random mechanisms perform the best because of the low management overhead in both \ac{OP} mechanisms.
In terms of network usage or the BDP product, we again see a difference in the performance of Relaxation and Global Optimal mechanisms in different queries.
While for simple operators like Stream, the Producer-Consumer mechanism supersedes the former by a small magnitude for more complex queries like Conjunction, the Global Optimal and Relaxation mechanisms are better.
Since we are focused on more complex queries, those applied in \ac{IoT} application scenarios, we further look into their performance for a traffic congestion detection query introduced in the setup (cf. \Cref{subsec:ev_setup}) in the next paragraph.

\begin{figure}[H]
 \includegraphics[width=0.7\linewidth,left]{figures/graphs/placement/legend}
  \newline
    \centering
    \includegraphics[width=0.5\linewidth]{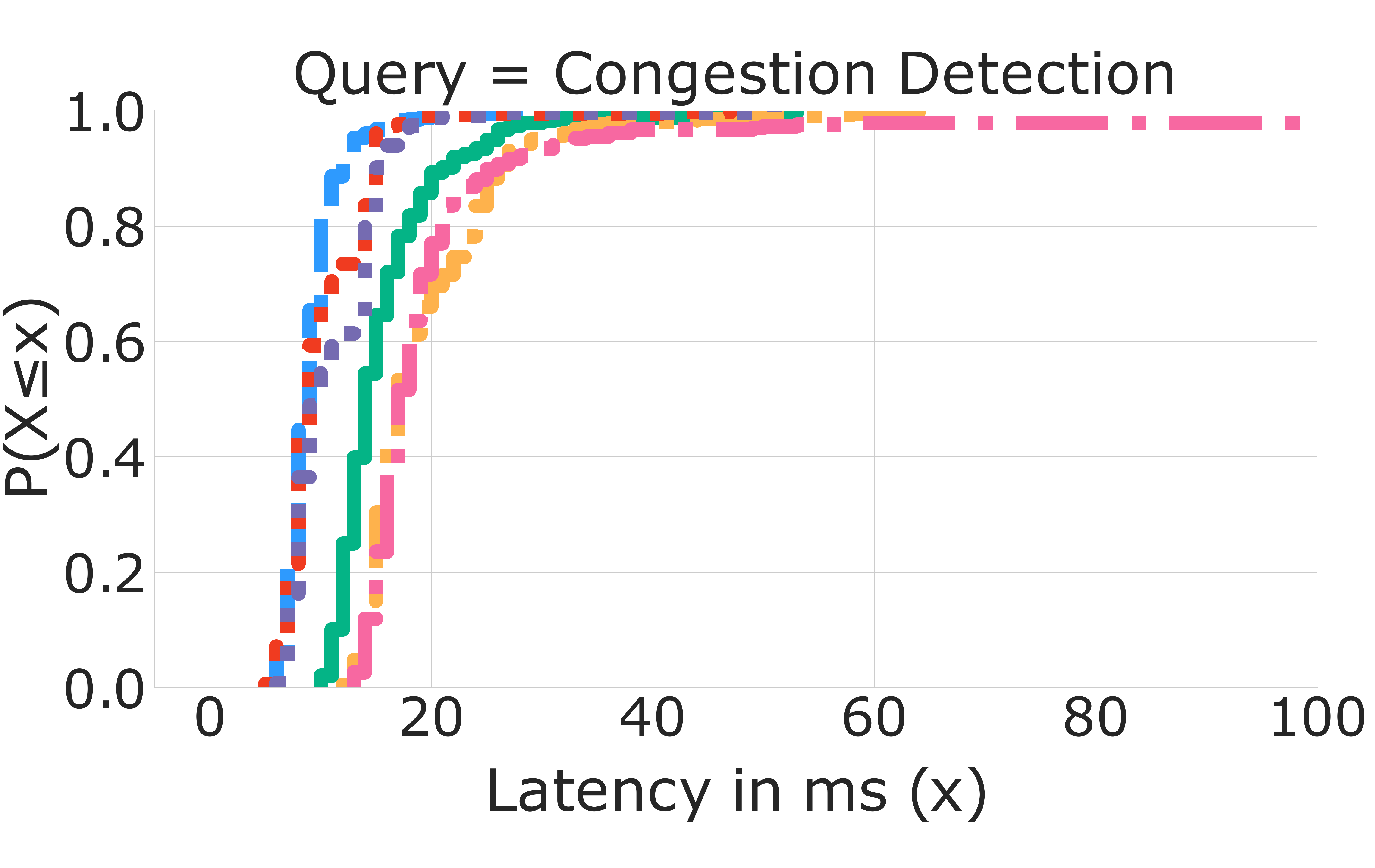}
    \includegraphics[width=0.49\linewidth]{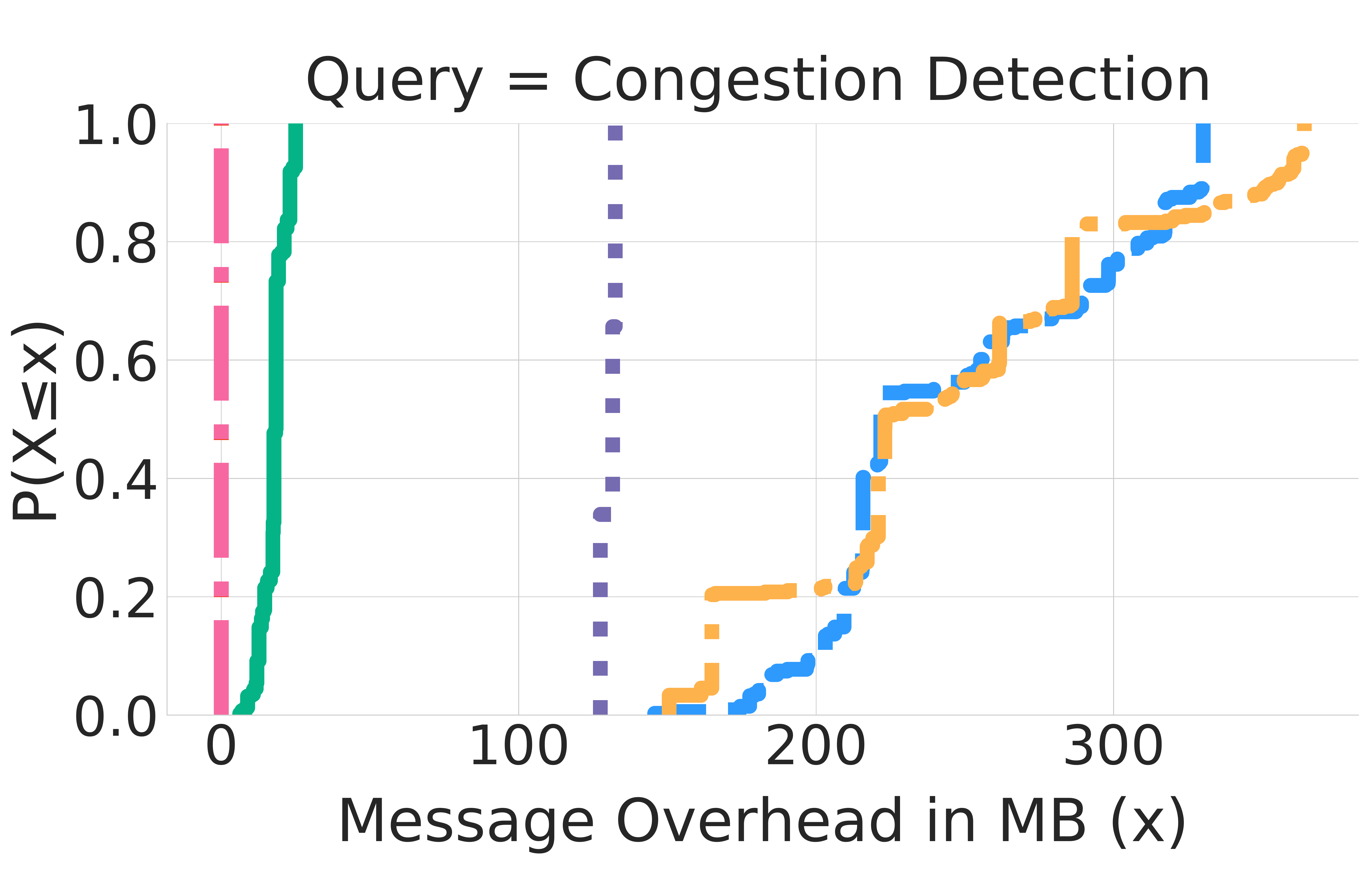}
    \includegraphics[width=0.49\linewidth]{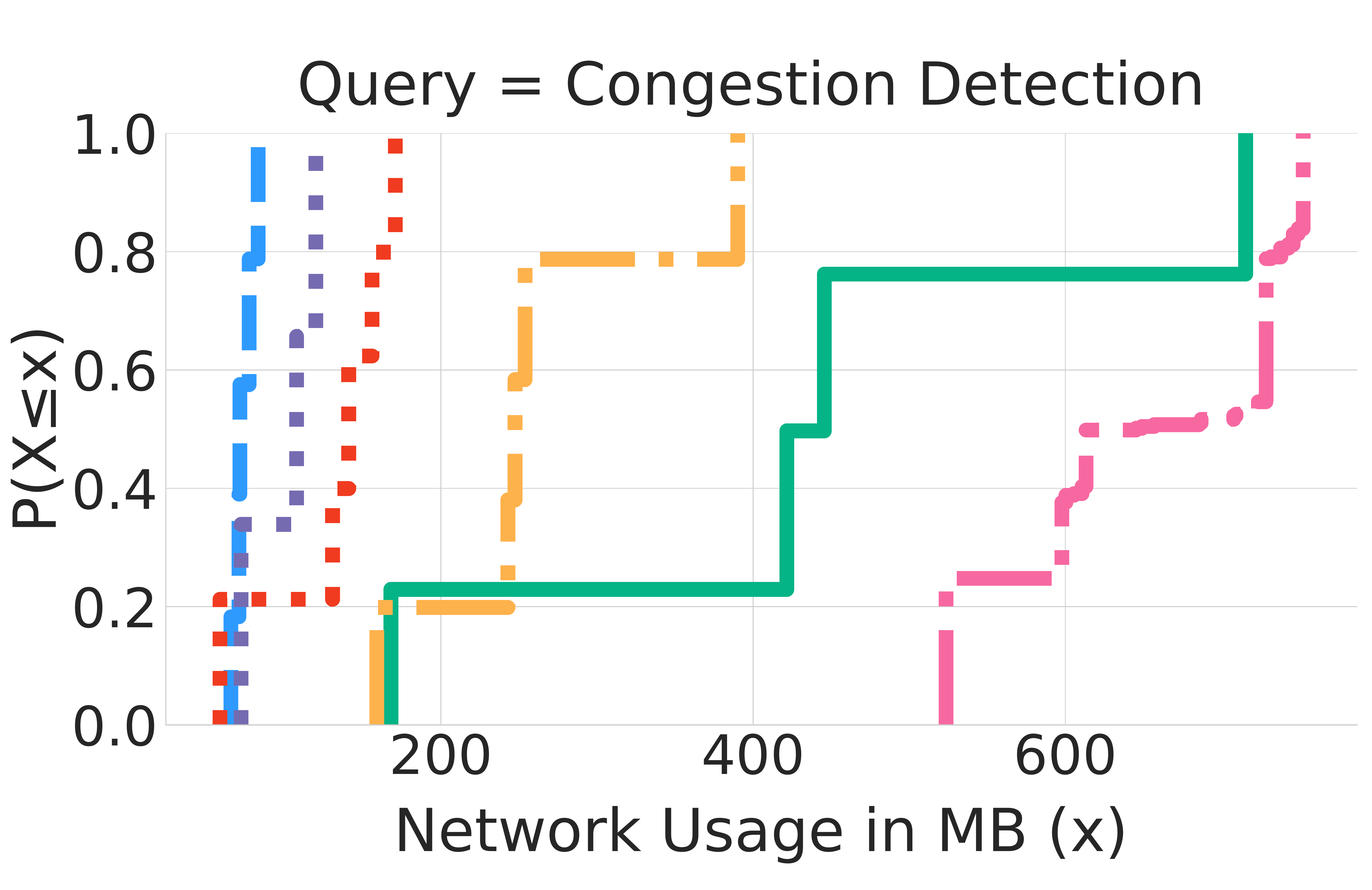}
    \includegraphics[width=0.49\linewidth]{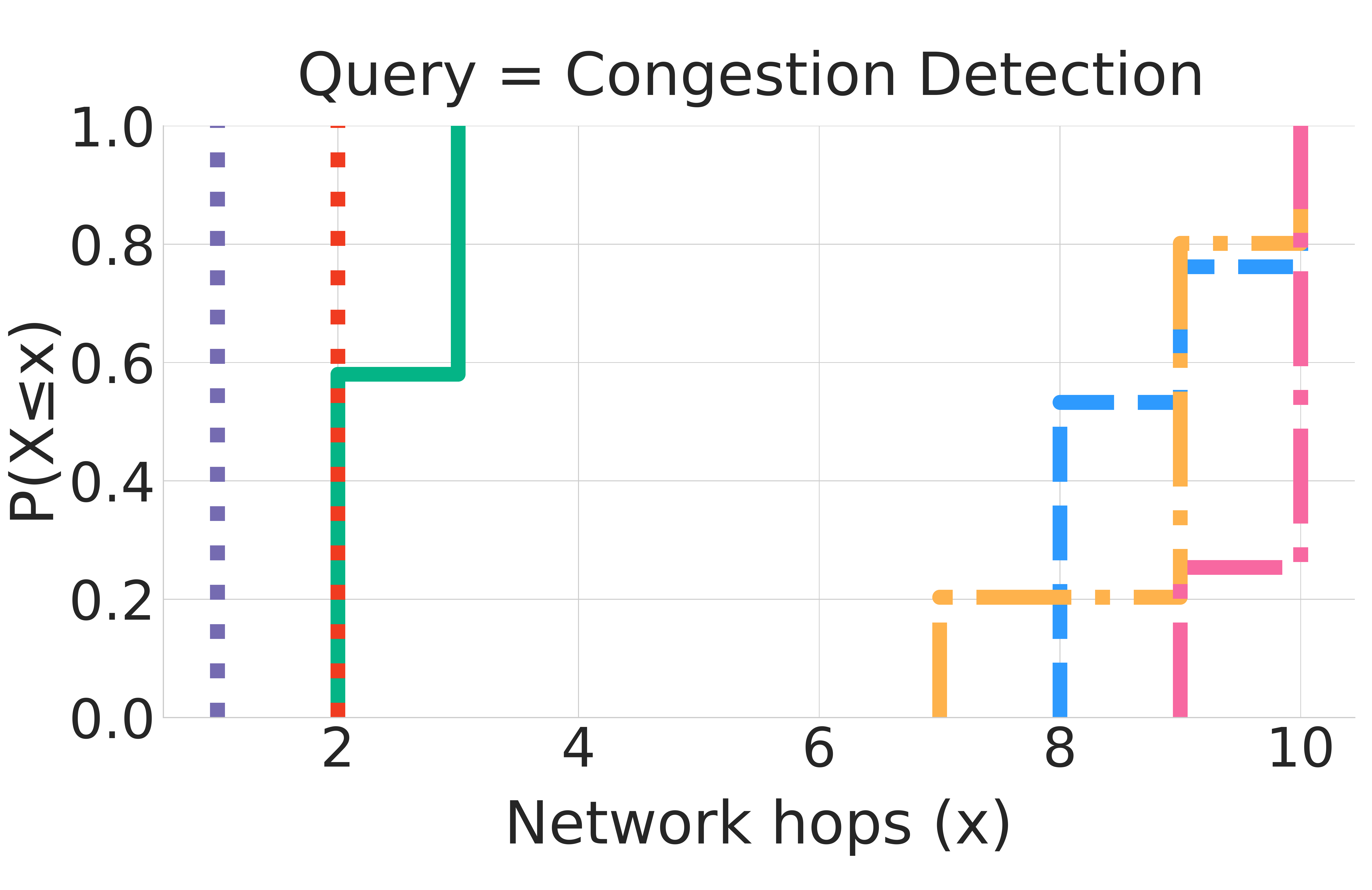}
    \caption{Performance evaluation of \ac{OP} mechanisms (CDF) using programming model of \system in terms of latency, message overhead, network usage, and the number of hops for the congestion detection query. \emph{Here, for all the metrics the more to the left is the distribution, the better it is.}
 }
    \label{fig:placement/performance}
\end{figure}

\Cref{fig:placement/performance}  presents the cumulative distribution function (CDF) for the different metrics using the given \ac{OP} mechanisms while executing a traffic congestion query.
Similar to the other queries analyzed above, Relaxation performs well in terms of latency.
However, it possesses much high message overhead due to the maintenance of the latency cost space.
In contrast, MDCEP possesses much low message overhead while it suffers from very high latency for a high workload of queries. The variant of Relaxation,
~\ml{R1: In The variant of (typo) done}
the MOPA and Optimal mechanisms also suffer in performance in terms of message overhead. Contrarily, the Producer-Consumer and Random mechanism suffer in terms of network usage.
This further solidifies our belief that no mechanism can satisfy both the optimization criterion network usage and message overhead at the same time because these two are inherently conflicting. \Cref{tab:op_algorithms} in Appendix \ref{sec:tableOP} summarizes the mean, minimum, maximum, and quantiles (90, 95, 99\%) of the metrics latency and message overhead important for the considered scenario. The table presents the results for Q1, Q4, and Q5 execution using the different \ac{OP} mechanisms. It can be derived from \Cref{fig:placement/performance} and \Cref{tab:op_algorithms} that Relaxation and MDCEP mechanisms stand representatives for the metrics latency and message overhead, respectively. Furthermore, as assumed in our hypothesis, there is no \emph{one size fits all} mechanism~\cite{op/TPDS/Nardelli2019}.

We have proposed mechanism transitions for such scenarios with dynamic environmental conditions and changing \ac{QoS} demands. In the rest of the evaluation, we will focus on the two representative \ac{OP} mechanisms, Relaxation and MDCEP and investigate the performance of mechanism transitions.

\subsection{Performance of \ac{OP} Mechanism Transitions}~\label{subsec:TCEPimpact}
\begin{figure}
    \centering
    \includegraphics[width=\linewidth]{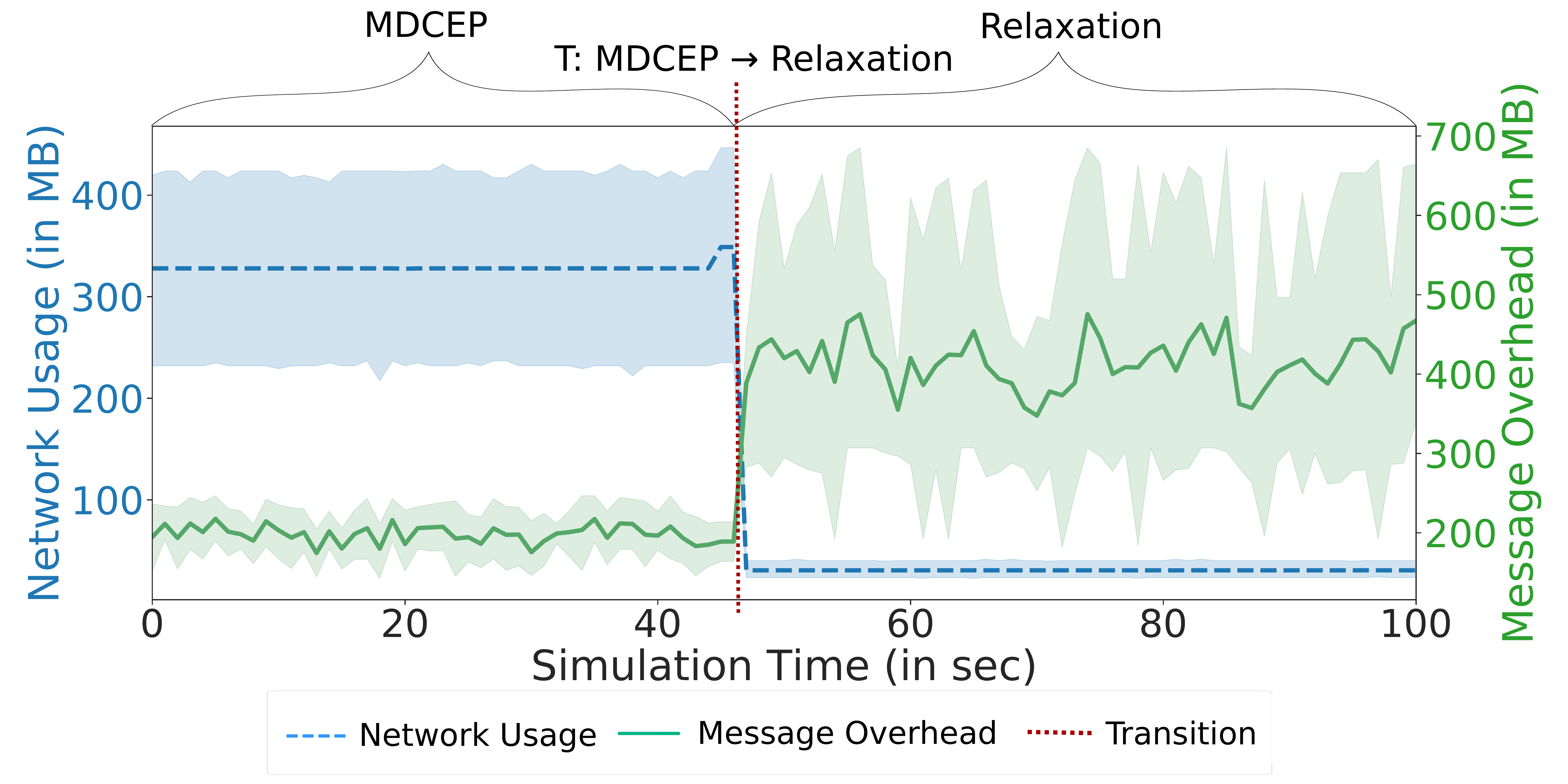}
    \caption{Network usage (y1-axis) and message overhead (y2-axis) measurement over a transition from MDCEP to Relaxation \ac{OP} mechanism. \system system seamlessly transits to a fresh \ac{OP} mechanism without incurring any overhead in terms of the specified \ac{QoS} demands.
    ~\mli{R1: * If you look at Figure 11, you see the transition takes a couple of seconds (which is not too much, but not negligible either). Transitions in these realms of time were already achieved in 2012 (StreamCloud) for states larger than 20 MB. We would like to clarify here the meaning of a transition is much more than only operator migrations, although, it being a major role in the lifecycle of a transition. While many existing streaming systems offer solutions for operator migrations already, none of them explicitly looked into the costs incurred due to migration and choosing optimal time points for migrations. Furthermore, achieving seamlessness in transitions, planning for operator migrations using optimal sequence and discrete time points, and exchanging \ac{OP} mechanism to fulfil varying \ac{QoS} demands were in our knowledge never done before us.}}
    \label{fig:TCEPevaluation}
\end{figure}

To understand whether the mechanism transition can fulfil changing \ac{QoS} demands for dynamic environmental conditions, we evaluate the performance of \system.
We consider the following metrics:
mean network usage (objective function for Relaxation), and
mean control message overhead (objective function for MDCEP) as defined before.
Furthermore, we consider a traffic congestion detection query
~\mli{R1: Congestion detection or Congestion detection? Listing one or Figure 1b}
for the rest of the evaluations because
\i~it is representative of our scenario (\Cref{sec:motivation}),
\ii~it captures the major standard CEP operators,
\iii~and since for this query, we have observed significant variation in the performance of the \ac{OP} mechanisms as shown in Figures~\ref{fig:placement/performance_all} and~\ref{fig:placement/performance}.
Figure~\ref{fig:TCEPevaluation} shows mean network usage on the first y-axis and control message overhead on the second for 5 runs in \system. At around 45 seconds (shown with an arrow), we observe a change in \ac{QoS} demand from message overhead to network usage.
\system handles this by executing a dynamic transition between MDCEP to Relaxation. \system triggers a transition automatically by selecting an appropriate placement mechanism that fulfils the \ac{QoS} demand. It is noticeable that \system does not induce any interruption or costs in terms of optimized metrics and while performing a transition to a completely new \ac{OP} mechanism. We further investigate the transition cost for the different algorithms for transition and selection of placement mechanism in the next sections.

\subsection{Performance of transition algorithms}~\label{subsec:TCEPperformance}

This section aims to understand how far the transitions are disruptive and the imposed cost in performing the transitions.
In the evaluation, we consider:
\i~the output event rate,
\ii~the required time for the transition, and
\iii~the transition overhead.
To evaluate the transition execution algorithms reasonably, we extend Algorithm~\ref{algo:seqTransition} to migrate the operators concurrently. Similarly, we extend Algorithm~\ref{algo:expoTransition} to migrate the operators sequentially. The four approaches are enlisted in Table~\ref{tab:evaluation}. Furthermore, we increase the query load to up to 10 queries to impose changes in the environmental conditions that trigger transitions in the \system system.

\subsubsection*{\textbf{Cost of transition algorithms with learning-based selection}}
We analyze the cost of the different transition algorithms proposed in \Cref{sec:transitionEngine}. The transition algorithm works together with the learning algorithm responsible for selecting the \ac{OP} mechanism for a transition.

Besides the different transition algorithms, we implemented a requirement-based algorithm that selects a placement mechanism by matching the \ac{QoS} demand with the optimization criteria for comparison with our learning algorithm. If there exists more than one mechanism matching the \ac{QoS} demand, there is a random selection. In contrast, the genetic learning-based selection algorithm takes into account the performance of the \ac{OP} mechanism, as explained in the design section.

\begin{figure}
    \centering
    \begin{subfigure}[H]{\linewidth}
        \centering
        \includegraphics[width=0.9\linewidth]{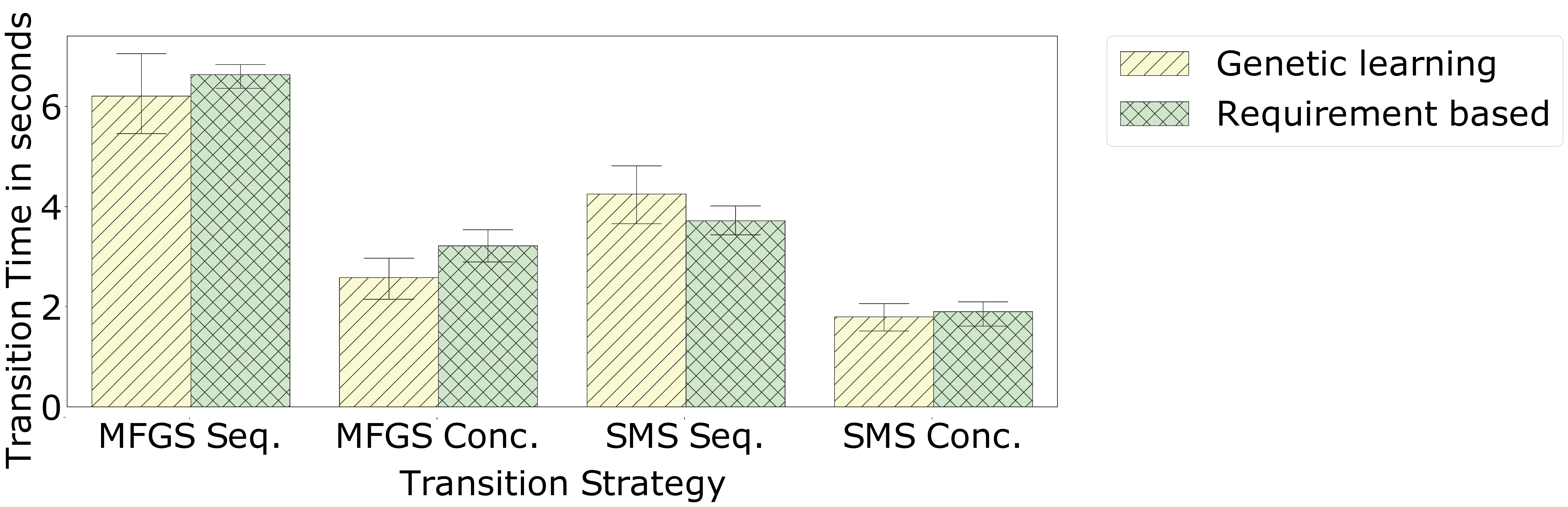}
        \caption{Transition time.}
    \end{subfigure}
    \begin{subfigure}[H]{\linewidth}
        \centering
        \includegraphics[width=0.9\linewidth]{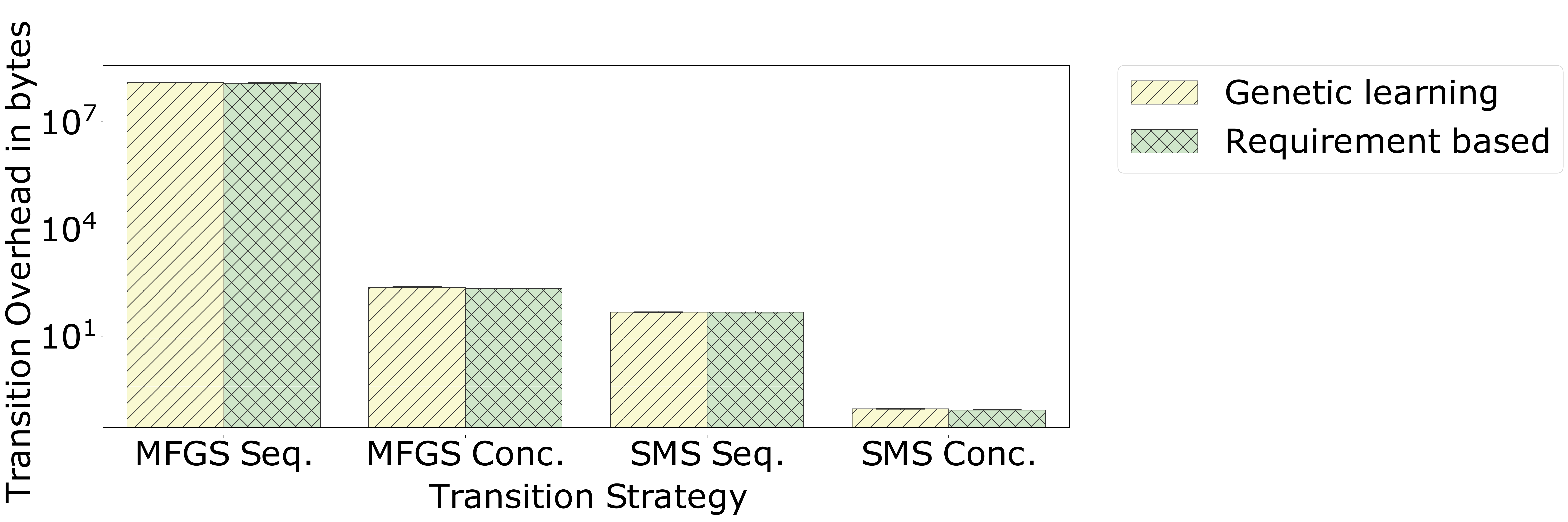}
        \caption{Transition overhead.}
    \end{subfigure}
    \caption{The plot shows the transition cost in terms of time and overhead for the proposed transition algorithms. \ac{SMS} transition algorithms require minimal state transfer during operator migrations, and hence, can perform the transition in a mean time of 1.82 seconds compared to 6.29 seconds required by MFGS Sequential algorithm. Moreover, the SMS Concurrent algorithm has only a negligible overhead of 0.72 bits, thanks to the cost-optimal algorithm (cf. \Cref{subsubsec:SMS}).
   }
    \label{fig:trstr_cost}
\end{figure}

\Cref{fig:trstr_cost} shows the transition time (a) and overhead (b) incurred by the transition algorithms using different selection algorithms. Here the costs of transition include the cost in time and overhead as represented earlier in \Cref{eq:cost_Ttime} and \Cref{eq:cost_Toverhead}, respectively, in \Cref{sec:problem}. It is noticeable that MFGS algorithms possess higher transition times than SMS algorithms. This is due to the state involved that is to be transferred by the MFGS algorithms, while the SMS algorithms optimize for the minimal amount of state transfer (cf. \Cref{eq:tr_cost_objective}). There is a substantial improvement in the transition time between Sequential and Concurrent algorithms. This is because  operators are migrated concurrently, which leads to lesser transition time. Finally, using the SMS Concurrent algorithm, we achieve an effective mean transition time of $1.82$~seconds for the load of 10 complex congestion detection queries involving multiple stateful operators. We see an effective reduction of around 4 seconds compared to MFGS Sequential transition algorithm that takes $6.29$~seconds to finish a transition with state transfer. The only cost parameter involved in SMS transition algorithms is in terms of selection of the OP mechanism and transition coordination costs in terms of communication between the distributed nodes. This is because there are no costs involved in terms of state migrations. 

In the second plot, we observe the total transition overhead in terms of selecting an \ac{OP} mechanism, transition coordination, and operator migrations due to transition (\Cref{eq:cost_Toverhead} in \Cref{sec:problem}). In consistent with the transition time, we observe a lower overhead of SMS algorithms due to the low amount of state involved in migration. Note the scale of the y-axis is logarithmic\ml{R2:  log -> logarithmic done} to show the amount of overhead involved for SMS algorithms that is substantially lower. In particular, we have only a mean overhead of $0.72$ bits for SMS Concurrent and $379.79$ bits for SMS Sequential algorithm, where the former is more than $2000\times$ better and the latter is $5\times$ better than the MFGS Concurrent algorithm.

A second observation from these plots is that the genetic learning-based selection algorithm equally performs like a requirement-based algorithm because of no training and minimal learning costs involved. In \Cref{subsec:learningcosts}, we elaborate on the learning costs of the algorithm. In conclusion, our results show that the
\ac{SMS}-Sequential and Concurrent algorithms perform better in both transition time and overhead, with the time within a range of $0.85-2.83$~seconds (for 10 queries) in comparison to $35$\,seconds (if the transition is performed naively using the stop and start migration algorithm) for the congestion detection query. We analyze costs per operator in the next section.

\subsubsection*{\textbf{Cost of transition algorithms for different operators}}~\label{subsubsec:single}
In this section, we analyze the cost incurred by the transition algorithms in detail.
The transition time comprises operator migration time and the time an operator has to wait for migration until the predecessor starts its operation at the target broker (cf. Section~\ref{sec:transitionEngine}). For example, Conjunction operator waits for migration until {Average} and {Stream} operators start their operation at the target brokers.
Leaf operators (Stream or producers) have no wait time as they have no predecessors.
The operator transition overhead involves the cost for
\i first and foremost the state involved in migration for stateful operators like Window-Aggregates, Joins and Sequences, and
\ii second the coordination overhead for the operator graph migration in terms of communication, such as acknowledgements (cf. \Cref{subsubsec:MFGS}: \Cref{algo:seqTransition} and \Cref{algo:expoTransition}). Stateful operators have costs in both dimensions, communication as well as migration costs depending on the transition algorithm -- MFGS or SMS, while stateless operators like Filter and Stream do not have any state migration costs, but do have a small communication cost again depending on the transition algorithm -- Sequential or Concurrent.

\begin{figure}[H]
  \centering
  \begin{subfigure}[H]{0.48\linewidth}
        \centering
        \includegraphics[width=\linewidth]{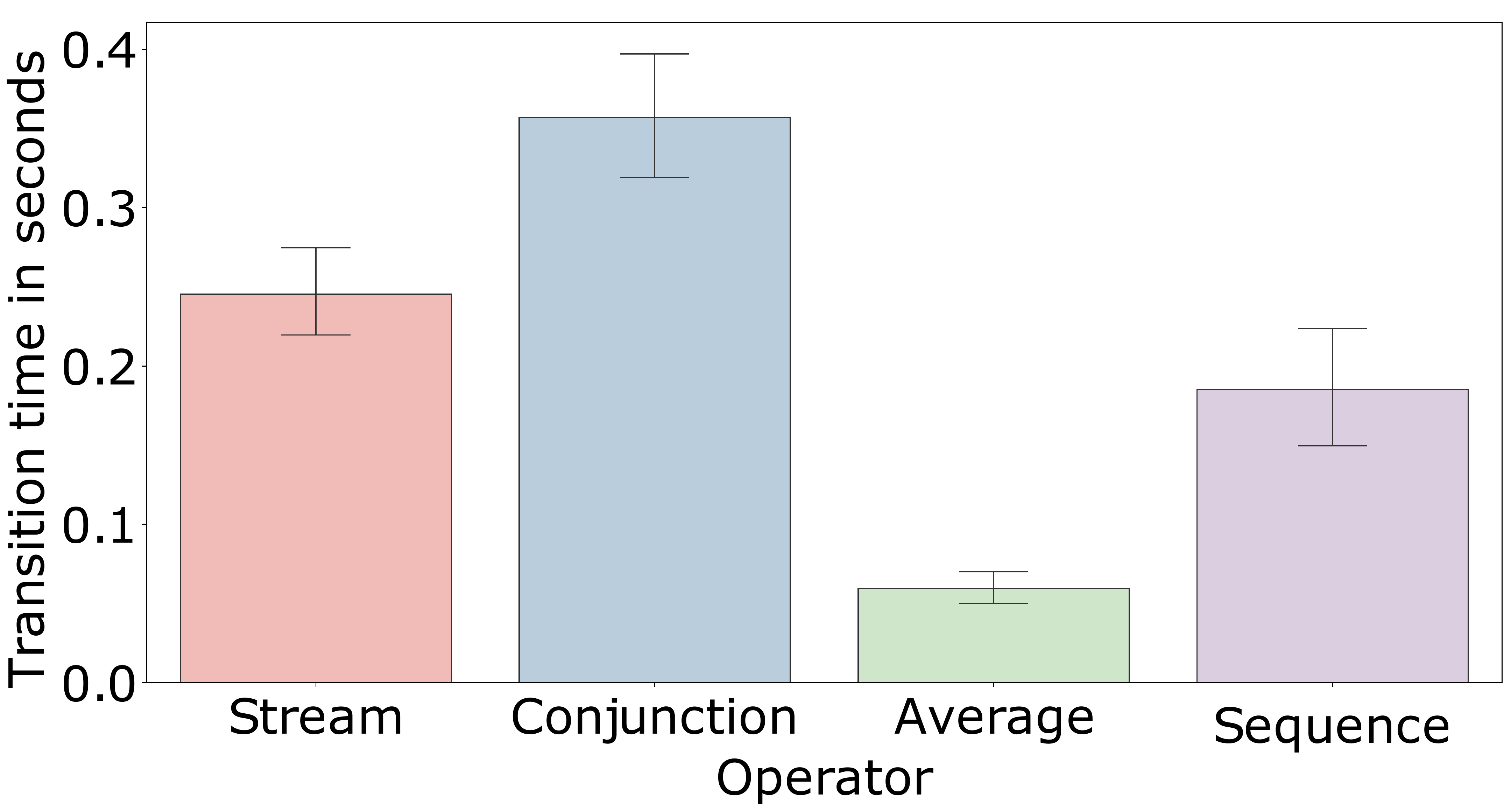}
        \caption{Transition time for the different operators in congestion detection query.}
   \end{subfigure}
   \begin{subfigure}[H]{0.48\linewidth}
        \centering
        \includegraphics[width=\linewidth]{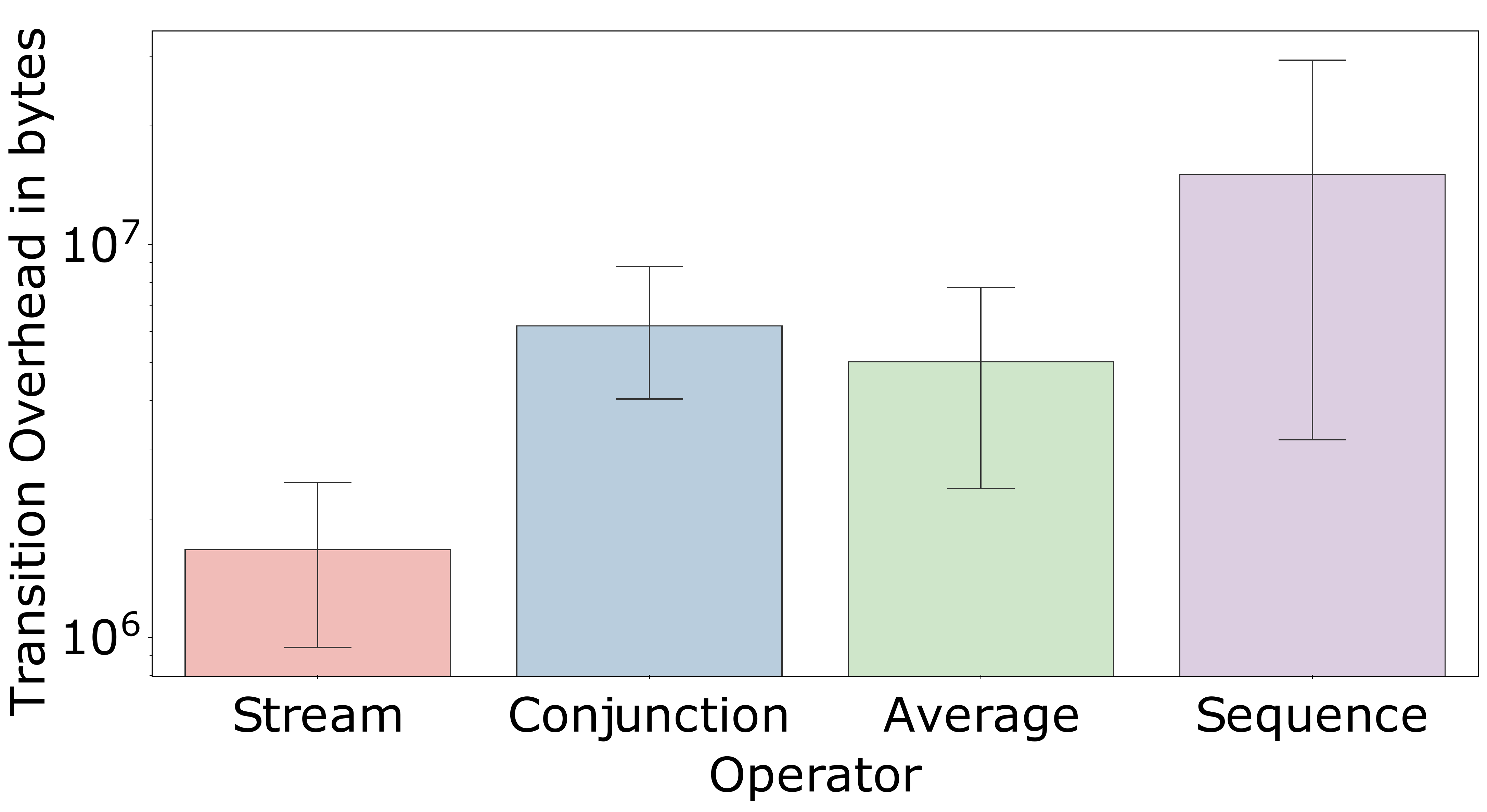}
        \caption{Transition overhead for the different operators in congestion detection query.}
    \end{subfigure}
    \caption{Operator transitions are performed in the order of few milliseconds and with very low overhead using our transition algorithms.}
    \label{fig:cost_operator}
\end{figure}
Figure~\ref{fig:cost_operator} shows the mean transition time (a) and overhead (b) using the transition algorithms for all the operators using 10 incrementally deployed congestion detection queries Q5 (cf. \Cref{fig:og-queries}).
The total migration time correlates to the number of operators, and the transition state denoted as transition overhead in the second plot.
It can be seen that the stateless operators like Stream, although high in number (90 operators), can be transited in $245.3$ ms. While other operators like Conjunction and Sequence need slightly higher mean transition times of $356.89$ and $185.32$ ms, respectively, with a mean and maximum transition overhead of $6.2-130.98$ MBs, and $15.06-129.9$ MBs, respectively. \Cref{tab:Ttime_overhead_op} in Appendix~\ref{sec:tableOP} summarizes the mean, minimum, and maximum values of the distribution.

\begin{figure}
    \begin{minipage}{0.62\linewidth}
        \begin{subfigure}[H]{\linewidth}
            \centering
            \includegraphics[width=\linewidth]{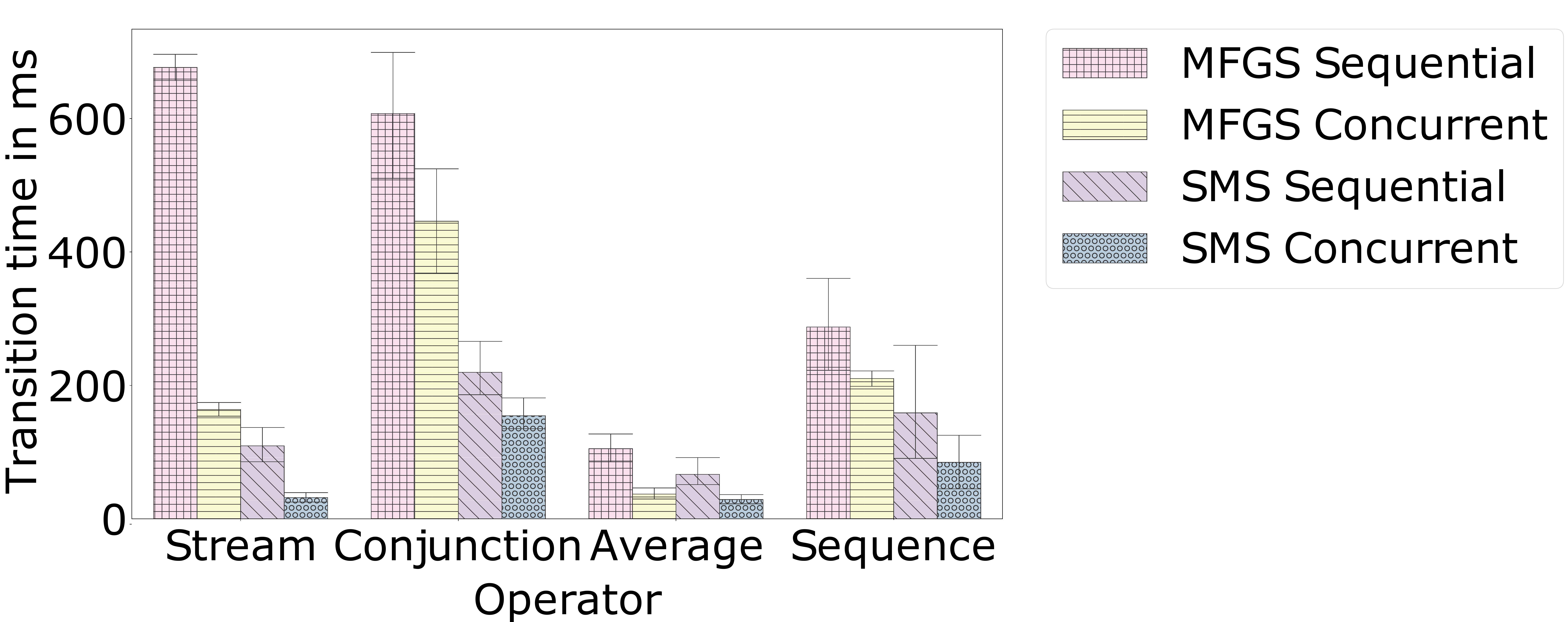}
          \caption{Transition time observed for different operators and transition algorithms.}
        \end{subfigure}
         \begin{subfigure}[H]{\linewidth}
            \centering
            \includegraphics[width=\linewidth]{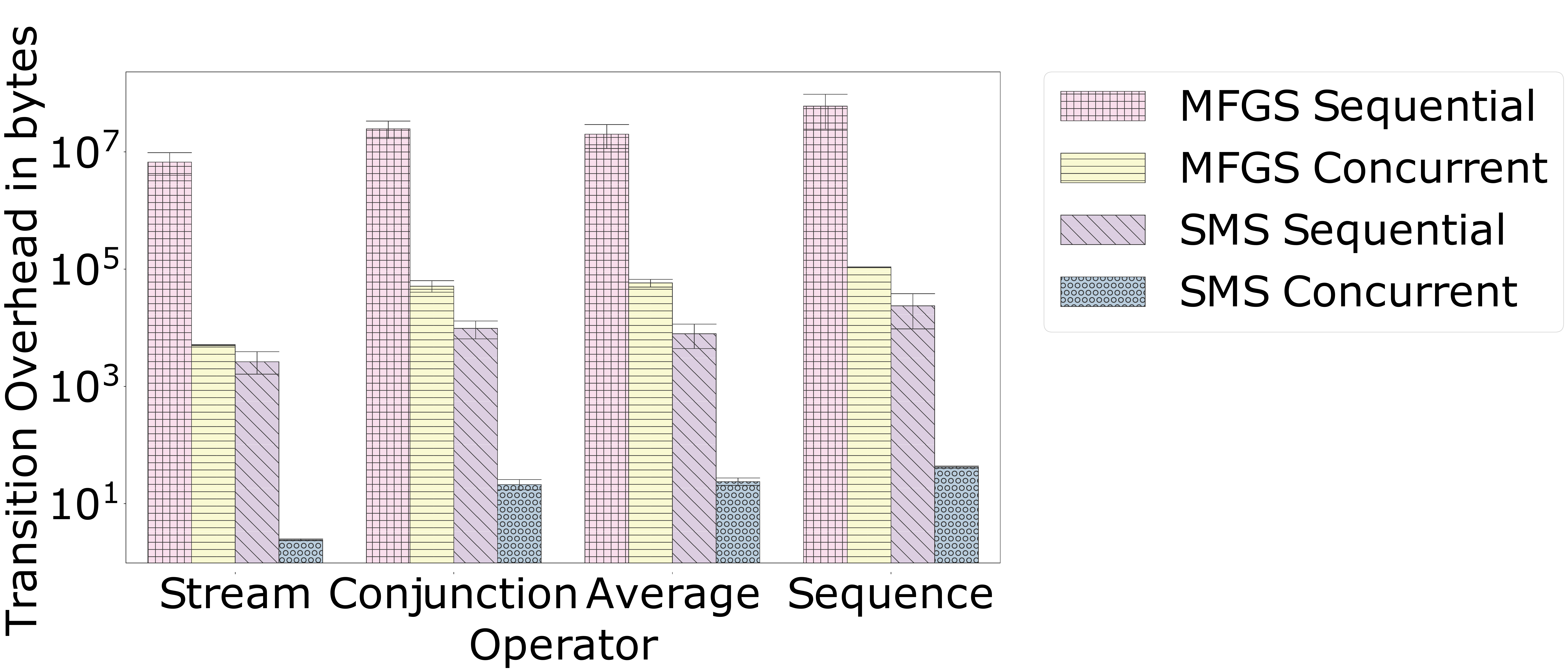}
            \caption{Transition overhead observed for different operators and transition algorithms.}
         \end{subfigure}
        \caption{Transition time and overhead measurement for different operators for 10 incrementally deployed Q5 queries. SMS algorithms possess minimum migration time and overhead due to the minimal amount of transition overhead.}
        \label{fig:trstr_cost_operator}
    \end{minipage}
    \hspace*{10mm}
    \begin{minipage}{0.35\linewidth}
        \scriptsize
        \begin{tabular}{|lll|}
            \toprule
            Operator    & \# pQ5. &  \# tot. \\
            \midrule
            Stream      & 9  &  90   \\
            Conjunction & 8  &  80     \\
            Average     & 3  & 30    \\
            Sequence      & 1 & 10    \\
            \bottomrule
        \end{tabular}
        \captionof{table}{Number of operators per congestion detection query (Q5 in \Cref{fig:og-queries}) and total in a single simulation run for 10 queries. \mli{R1: Why is the number of operators in Figure 15 so high? Is the query being executed in parallel?}}
        \label{tab:operators}
    \end{minipage}

\end{figure}

\Cref{fig:trstr_cost_operator} classifies the transition costs further based on the transition algorithms. MFGS Sequential algorithm performs the worst clearly because of the high amount of transition overhead (mean value for Sequence operator $60.12$ MB).
In contrast, the SMS algorithms require a very short time to transit, a mean time of $41.1$ ms for 30 Average operators and $85.1$ ms for 90 Stream operators.

\Cref{tab:operators} shows the number of respective operators per congestion detection query and the total number of operators in a single run. The {Conjunction} operator takes the highest amount of time to migrate, although the state involved is less due to a high number of operators involved. The same applies to the {Stream} operator. The number of Sequence operators to be migrated is less; however, it takes longer to transit due to the high amount of state (\textasciitilde{60}~MB) transfer. \Cref{tab:Ttime_overhead_strategy} in Appendix~\ref{sec:tableOP} summarizes the mean transition time and overhead required by the different algorithms shown earlier in \Cref{fig:trstr_cost_operator}.
\mli{R1: The formulation "state of filter" is strange, because filters are stateless}

In \Cref{fig:trstr_cost_operator}, we analyze the cost per operator for the transition algorithms. In consistent with our findings in the previous section, with MFGS algorithms, operators take longer to migrate than SMS algorithms. The SMS Concurrent algorithm performs\ml{R2: algorithm perform -> algorithm performs done} the best.



\subsubsection*{\textbf{Seamless execution of transitions}}

\begin{figure}[H]
	\centering
	\includegraphics[width=0.9\linewidth]{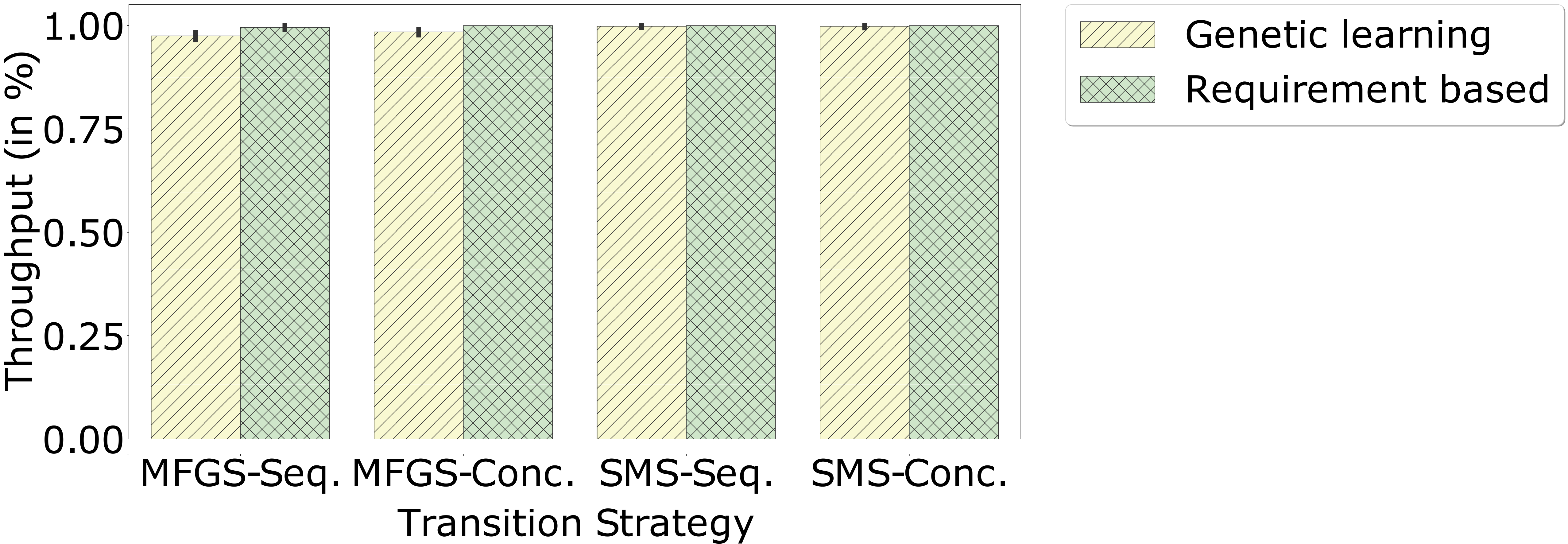}
	\caption{Throughput measurements using the different transition algorithms and selection algorithms for \ac{OP} mechanisms. \ac{SMS} transition algorithms consistently deliver output events enabling seamless execution of a transition. \mli{R1: Figure 15 is introduced after Figure 13 in page 44 but it looks like Figure 14 is not referenced.  done}}
	\label{fig:cdf_opeventrate}
\end{figure}

To verify the seamless execution of transitions, we measured the throughput rate produced while \system's transition algorithms were executed (cf. Figure~\ref{fig:cdf_opeventrate}).
A minor output disruption for \ac{MFGS}-Sequential and Concurrent algorithms was observed in Figure~\ref{fig:cdf_opeventrate} (around $0.02\%$). However, \ac{SMS}-Sequential and Concurrent algorithms do not exhibit any disruption and continuously deliver output events with an output event rate of 100\% for both the selection algorithms.

\subsection{Learning Costs of Placement Selection} \label{subsec:learningcosts}

This section aims to understand the learning costs of the adaptive placement selection algorithm introduced in \Cref{sec:placement_performance}. We consider the following metrics to determine the costs: \i the time taken to learn the performance characteristics, in other words, to update the learning model, and \ii communication cost for the placement selection. The genetic learning-based learning algorithm has no training costs since the algorithm is based on online learning. Hence, it induces only a negligible overhead in time within a range of $2.5 - 3.15$~ms (95\% confidence interval). Often the update of the learning model induces no overhead at all. Furthermore, the algorithm does not induce any communication overhead due to the local handling of operator placement selection.

\begin{figure}[H]
    \centering
     \begin{subfigure}[H]{0.49\linewidth}
        \centering
        \includegraphics[width=\linewidth]{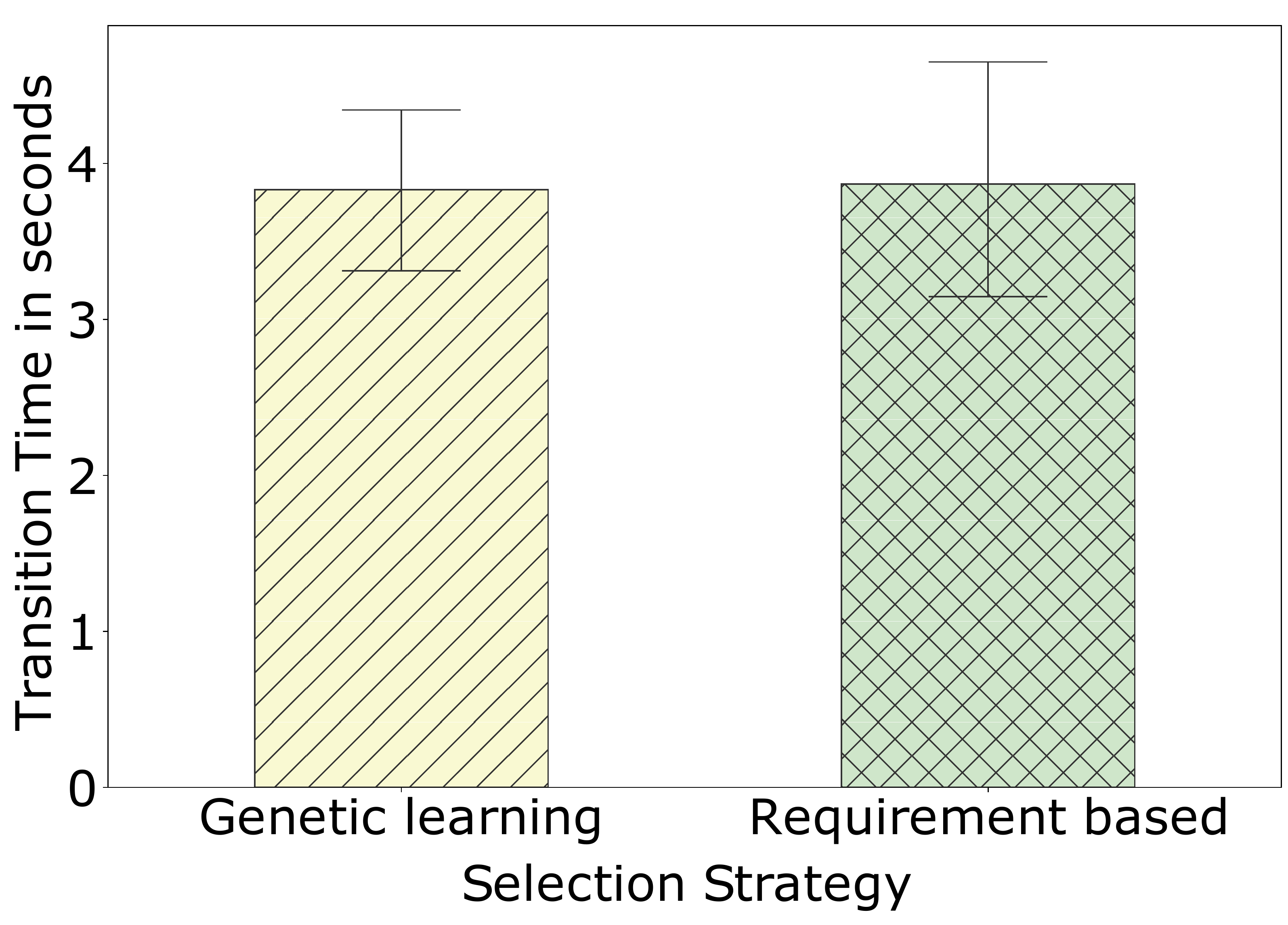}
        \caption{Transition time.}
    \end{subfigure}
     \begin{subfigure}[H]{0.49\linewidth}
        \centering
        \includegraphics[width=\linewidth]{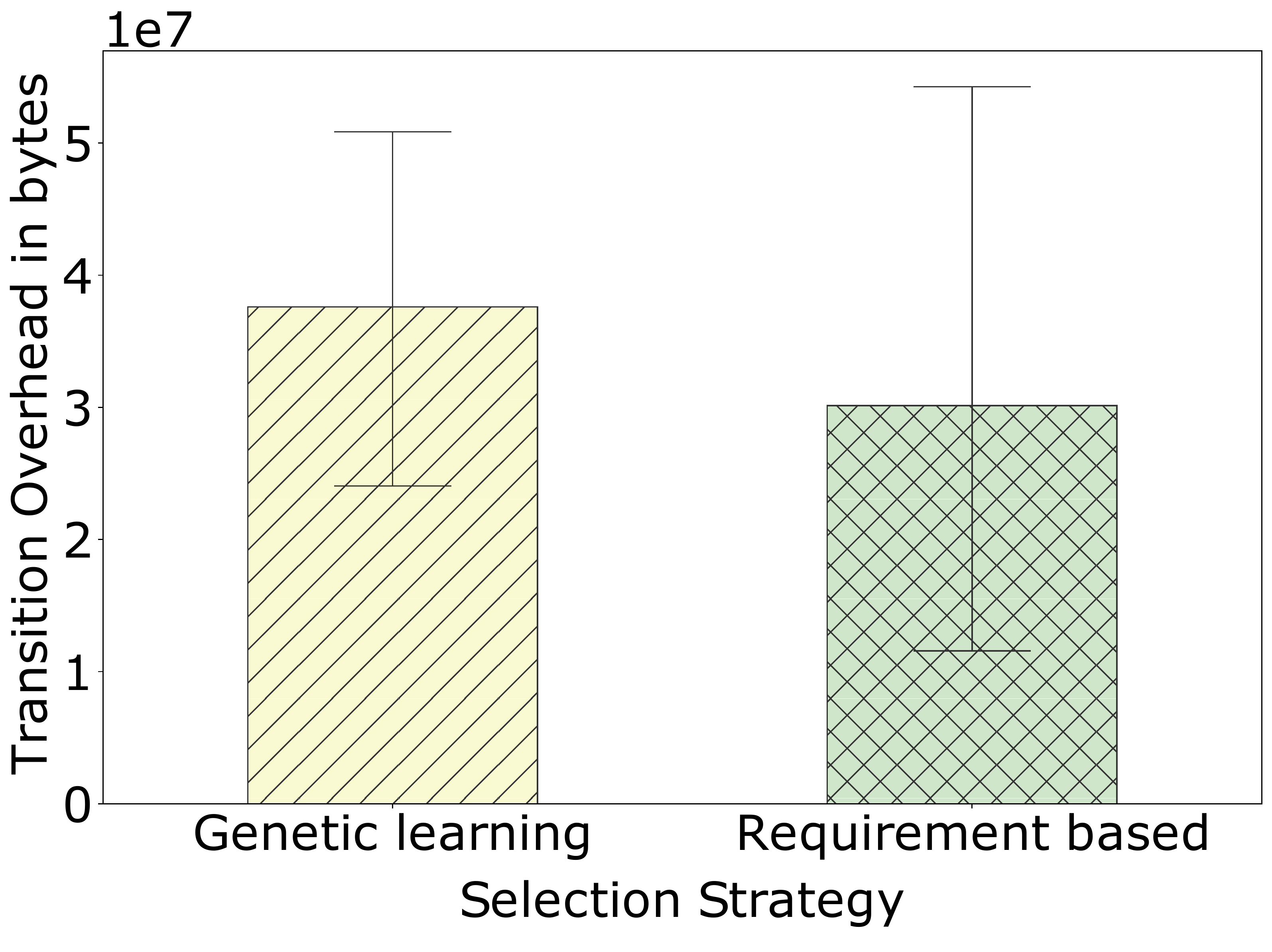}
        \caption{Transition overhead.}
    \end{subfigure}
    \caption{Transition cost comparison with a requirement-based selection algorithm. The genetic learning-based algorithm does not impose high cost in terms of learning and is at par to the requirement-based algorithm.}
    \label{fig:learningCosts}
\end{figure}
Finally, to understand the influence of genetic learning-based selection algorithm on the performance mechanisms transitions, in \Cref{fig:learningCosts}, we analyze the transition costs in time (a) and overhead (b). We compare the learning algorithm with a requirement-based algorithm where the selection of a mechanism is based on \ac{QoS} demands.

From the figure, we observe that due to the negligible overhead of the genetic learning-based algorithm, the cost induced by it is comparable to the requirement-based algorithm. In fact, the transition time observed using the genetic learning-based is slightly less than the requirement-based algorithm. In terms of overhead, we see a slight increase due to the exploration of a suitable placement algorithm that is not performed in the requirement-based algorithm.

\section{Related Work} \label{sec:relatedwork}

It is highly important to fulfil \ac{QoS} demands in a \ac{DCEP} system for a wide range of application domains~\cite{Rahmani2018}.   
By enabling transitions, \system allows changing OP mechanisms, and in this way, fulfil \ac{QoS} demands under dynamic environmental conditions.
In this section, we analyze and compare related work in four key areas: programming models, operator placement and migration, adaptive event processing systems, and existing methods for mechanism transitions.  

\subsection{CEP Programming models}
Many CEP languages have been developed in the past years, such as CQL~\cite{cql/vlbd/arasu2006}, 
Cayuga~\cite{cayuga/springer/demers2006}, SASE~\cite{sase/sigmod/eugene2006}, TESLA~\cite{tesla/debs/cugola2010} for specifying complex events and detecting them by triggering notifications. Modern CEP programming models like Apache Flink \cite{flinkCarbone}, Heron~\cite{Kulkarni2015}, and Beam~\cite{googleDataflow} provides extensive APIs to specify complex events for both batch and stream processing. Recently proposed benchmarking frameworks such as~\cite{Karimov2018} and DCEP-Sim~\cite{Starks2017} unify CEP systems~\cite{C2:Luthra2019c} and simulate CEP environment and operator placement, respectively. However, none of the above programming models has enabled the specification of distinct operator placement mechanisms in a heterogeneous environment of physical machines we do in this work. DCEP-Sim~\cite{Starks2017} has enabled the development of operator placement but only in a simulation environment. While we study the effect of distinct operator placement mechanisms, perform adaptations between them, and analyze the cost of adaptations in real-world network infrastructure and dynamic environment.

\subsection{Operator Placement and Migration}
 \ac{OP} mechanisms\ml{R2: please avoid multiple definition of OP done} are widely studied to fulfil \ac{QoS} demands while incurring minimum cost in performance~\cite{Lakshmanan2008,Starks2017}. A wide range of \ac{OP} mechanisms has been proposed considering different \ac{QoS} demands, such as to achieve low latency~\cite{Ahmad2004}, to minimize bandwidth~\cite{Pietzuch2006,Rizou2010,Schilling2011}, to lower message overhead~\cite{Starks2015}, as well as to preserve trust and privacy~\cite{PhDWermund2018}. 

The fulfilment of \ac{QoS} demands, however, is only feasible under limited changes in  environmental conditions. For instance, most existing work~\cite{op/TPDS/Nardelli2019, placement/sigmetrics/cardellini2017, Ahmad2004,Cardellini2016,Rizou2010, C2:Luthra2019a} builds on stationary networks.
Approaches considering dynamic changes, \eg in the cause of mobility, introduces
\i redundancy by means of duplication~\cite{Starks2015} or checkpointing~\cite{Koldehofe2013}, 
\ii placement update at regular intervals~\cite{Pietzuch2006}, or
\iii operator migrations when changing the placement~\cite{Keeffe2015,Koch2010,Ottenwalder2014,PhDWermund2018}.

Overall, it is essential to note that current approaches for \ac{DCEP}, so far, build on a \emph{single} placement mechanism. 
In contrast, \system enables to benefit from adaptive use of \emph{multiple} existing \ac{OP} mechanisms by supporting transitions while increasing the range at which a \ac{DCEP} system can adapt to meet a specific \ac{QoS} demand.

Another critical mechanism that contributes to the cost in mechanism transitions is the operator migration mechanism. Operator migration has been extensively studied for data stream processing and complex event processing systems. Existing work can be characterized into the following three state migration strategies: \\
\i Stop and restart. A naive way to proceed with state migration is to stop the execution of the source broker, safely transfer the state, and start the execution on the target source broker. Such strategies were used in early stream processing systems like CAPE~\cite{Zhu2004} for dynamic query plan migration or runtime query optimization. 
Moreover, this method is most commonly used across fault-tolerance mechanisms, such as global state checkpoints. 
It is widely used by modern stream processing systems like Spark~\cite{Zaharia2010} and Flink~\cite{Carbone2015ApacheFS}. \\
\ii Partial pause and resume. In the face of dynamic environmental conditions,  streaming systems only have to migrate state for stateful operators, and hence stopping the entire streaming system is not necessary. 
This approach was introduced by a streaming system called Flux~\cite{Shah2003/ICDE/adaptiveOp}, which was later adopted and improved by multiple streaming systems, including StreamCloud~\cite{Gulisano2010}, Chi~\cite{Mai2018/pvldb/Chi}, Seep~\cite{CastroFernandez2013/sigmod/statemgmt}, and FUGU~\cite{Heinze2014} that only pauses the stateful operator in the operator graph. \\
\iii Seamless migration. After our initial work on seamless operator migration for transitions~\cite{Luthra2018}, several other authors addressed similar concerns for different problems for state recovery~\cite{PhDWermund2018},  state migrations in streaming systems~\cite{Megaphone2019/pvldb/statemigration, Monte2020/sigmod/statemgmt, C2:Luthra2020}. 
In contrast to the above mechanisms, we aim towards a cost-efficient transition capable system that integrates and benefits from multiple \ac{OP} mechanisms through operator migrations.

\subsection{Adaptive Event Processing Systems}
In this section, we review approaches that have so far considered the adaptive exchange of mechanisms in the context of event processing systems. 
For example, Weisenburger et al.~\cite{Weisenburger2017} proposed \textsc{AdaptiveCEP}, 
a programming model and CEP system that supports specifying QoS demands at run time. 
This work is complementary to \system since \textsc{AdaptiveCEP} does not focus on the adaptive selection and execution of transition strategies. However, in \system, the query language is used to specify changes in the QoS demands to instantiate a transition.

Heinze et al. proposed an elastic 
data stream processing system
(DSPS)~\cite{Heinze2014}, where the number of active hosts can be scaled up and down, and operator migration is coordinated accordingly. Later, authors utilized an online learning approach~\cite{Heinze2014b} for auto-scaling.
Based on this work, the same authors proposed an adaptive replication scheme for DSPS~\cite{Heinze2015} that performs adaptation at runtime between active replication and upstream backup schemes for fault tolerance.
Furthermore, the authors looked into the trade-off between monetary costs against the offered \ac{QoS}~\cite{Heinze2015b}.
Similar to the work of Heinze et al.~\cite{Heinze2015}, Martin et al.~\cite{Martin2018} also looked into the trade-off of active vs passive replication techniques for a fault-tolerant and elastic stream processing system.
Proactivity in elasticity was proposed by Matteis et al.~\cite{Matteis2017} using the \emph{Model Predictive Control} method, which accounts for system behaviour over a future time horizon to predict the best reconfiguration to be executed.

\mli{R1: I think it can be beneficial to check some survey papers about elasticity. There have been a couple in the last year (one in the encyclopedia on Big Data Technologies from Springer, for instance).done}
Several surveys on elasticity~\cite{Lorido-Botran2014, Martin2018, Roger2019, Marcos2018} highlight the importance of adaptivity of a streaming system towards the changing workload in terms of adding and removing resources on runtime.
For instance, Lorido et al.~\cite{Lorido-Botran2014} argue that the auto-scaling process in elastic streaming systems resembles the MAPE loop for autonomous systems similar to our work. 
Assunção et al.~\cite{Marcos2018} discuss the advantages of the online approach over static approaches in adaptivity. 
Finally, Röger et al.~\cite{Roger2019} highlight the importance of distributed elasticity solutions using multiple operator approaches. 

Furthermore, adaptivity in the OP mechanism has been investigated before. 
Aniello et al.~\cite{Aniello2013} proposed an adaptive online scheduling algorithm for Apache Storm using two placement mechanisms. 
Sutherland et al.~\cite{Sutherland2005} developed an adaptive scheduling selection framework for continuous queries in DSPS. 
\mli{R2: In the related work, the following reference can be discussed. R2: An Adaptive Online Scheme for Scheduling and Resource Enforcement in Storm done}
Liu et al.~\cite{Liu2019} advance the work on state migration to look into the problem of colocating stateful and stateless operators.

Although the aforementioned approaches benefit from integrating multiple mechanisms, the adaptation between the mechanisms is heavily dependent on the internals of the specific mechanisms in use. Therefore, integrating new alternative mechanisms is a complex task. By offering the abstraction of a transition, \system is highly extensible and can easily integrate new mechanisms.
Furthermore, no previous work up today has studied the idea of adapting between distinct \ac{OP} mechanisms.

\subsection{Mechanism Transitions}
The idea of mechanism transitions origins from the collaborative research centre MAKI, in which researchers investigate mechanism transitions for the Future Internet~\cite{MAKI2015}. Within MAKI, mechanism transitions are investigated in the context of a wide range of communication mechanisms~\cite{Richerzhagen2016,Richerzhagen2015,Richerzhagen2014TransitionsIL,Frommgen2015a,Richerzhagen2018b}.
For example, in publish-subscribe systems, mechanism
transitions between filtering schemes~\cite{Richerzhagen2016} and event dissemination mechanisms~\cite{Richerzhagen2015} are studied.
Another line of work by Froemmgen et al.~\cite{Frommgen2015b,Frommgen2015a}
proposed transition strategies that always execute the best suitable search overlay. 
Richerzhagen et al.~\cite{Richerzhagen2018b} recently proposed a transition-enabled monitoring service that executes transition on different monitoring mechanisms. 
Our work builds on and extends the concept of transitions proposed in prior work~\cite{Richerzhagen2016,Frommgen2015b}.
By focusing on transitions for \ac{OP} mechanism, our contribution is the design and understanding of transition strategies that can support highly dynamic and stateful mechanism transitions comprising many dependent distributed entities. The proposed strategies deal with the specific challenges for coordinated state migration as part of the \ac{SMS} and \ac{MFGS} transition strategies.

\section{Conclusion and Future Work}~\label{sec:conclusion}

In this work, we proposed \system, a transition-capable CEP system.
\system is capable of dealing with changing \ac{QoS} demands caused by dynamic network environment conditions.
\system allows integration of state-of-the-art \ac{OP} mechanisms using the programming model and dynamically executes the best matching \ac{OP} mechanism to meet the \ac{QoS} demands of IoT applications.
To this end, we have explored how to perform transitions and analyzed their cost and performance.
We proposed two transition execution algorithms for efficient migrations of operator state during a transition that is adaptively selected using the online learning algorithm.
The learning algorithm possesses very low learning costs due to the online nature of performance analysis of operator placement mechanism.
Our evaluation in the context of an IoT scenario and based on the state-of-the-art \ac{OP} mechanisms
shows that \system fulfils changing \ac{QoS} demands by seamlessly performing transitions, i.e., without any output disruption.

The cost analysis shows that the transition execution time and overhead can be decreased to the range of $0.85-2$ seconds for the presented use case using our proposed transition strategies. Moreover, the learning cost of the lightweight selection algorithm proposed in this work is negligible.
As future work, we consider trade-offs between different learning algorithms for the adaptive selection of an optimal learning algorithm.

\section*{Acknowledgements}
This work has been co-funded by the German Research Foundation (DFG) as part of the project C2 within the Collaborative Research Center (CRC) 1053 -- MAKI. We would like to express our gratitude to the anonymous reviewers for their detailed and very helpful feedback to improve this manuscript.

\balance
\bibliography{Bibliography-clean}

\begin{appendices} 
\section{Selection Method for \ac{OP} mechanism}\label{proof:selcmethod}

\begin{definition} Selection pressure $(\mathcal{S})$.
It is used to characterize the strong or high respectively weaker or small emphasis of selection on the best \ac{OP} mechanisms. The selection pressure  $\mathcal{S}$ for the fitness disrtribution $\overline{s}(f)$ is defined as follows.

\begin{equation} \label{eq:selectionpressure}
 \mathcal{S} = \frac{\mathcal{\overline{M^*}} - \mathcal{\overline{M}}}{\overline{\sigma}}
\end{equation}

In \Cref{eq:selectionpressure}, the selection pressure depends on the fitness distribution of the population. Therefore, for different fitness distributions will generally lead to different selection pressure even for the same selection method. In order to define it specifically, we assume that the fitness distribution follows a Gaussian distribution $\mathcal{G}(0,1)$. In our evaluation, we have empirically validated this fact that the fitness distribution of all \ac{OP} mechanisms follows a Guassian distribution, which leads to the following definition.
\label{def:selection_pressure}
\end{definition}

\begin{definition} Standardized Selection Pressure $(\mathcal{S_R})$.
The standardized selection pressure $\mathcal{S_R}$ is the expected average fitness value of the \ac{OP} mechanism distribution after applying the linear ranking based selection method to the normalized Guassion distribution $\mathcal{G}(0,1)(f) = \frac{1}{\sqrt{2\pi}} e{^{-\frac{f^2}{2}}}$

\begin{equation}
    \mathcal{S_R} = \int_{-\infty}^{\infty} f(\overline{R}^*) (\mathcal{G}(0,1))(f)df
\end{equation}
\label{def:stand_selection_pressure}
The effective and average fitness value of a Gaussian distribution with mean $\mu$ and variance $\sigma^2$ can be easily derived as $\mathcal{M}^* = \sigma\mathcal{S_R} + \mu$.
\end{definition}

\begin{theorem}
The selection pressure using a linear ranking method can be derived as follows.

\begin{equation} \label{eq:selectionpressuremethod}
\mathcal{S_R (\eta^-)} =  (1 -  \eta^-) \frac{1}{\sqrt{\pi}}
\end{equation}

\begin{proof}
Using the definition of standardized selection pressure in \Cref{def:stand_selection_pressure} and the Gaussian function for the initial fitness distribution, one can obtain

\[ \mathcal{S_R (\eta^-)}=\int_{-\infty}^{\infty} x \frac{1}{\sqrt{2\pi}} \exp \left( - \frac{x^2}{2} \right) \left(\eta^- + 2 (1 - \eta^-) \int_{-\infty}^{x} \frac{1}{\sqrt{2\pi}} \exp \left( - \frac{y^2}{2} \right) dy \right) dx\\ \]

 \[                  =\frac{\eta^-}{\sqrt{2\pi}}  \int_{-\infty}^{\infty} x \exp \left( - \frac{x^2}{2} \right) dx + \frac{1 - \eta^-}{\pi}  \int_{-\infty}^{\infty}  x \exp \left( - \frac{x^2}{2} \right) \int_{-\infty}^{x} \exp \left( - \frac{y^2}{2} \right) dy dx \]
Using \[ \int_{-\infty}^{\infty} x \exp \left( - \frac{x^2}{2} \right) = 0 \] \\
and

 \[ \int_{-\infty}^{\infty} x \exp \left( - \frac{x^2}{2} \right) \left( \int_{-\infty}^{x} \exp \left( - \frac{y^2}{2} \right) dy \right)^2 dx = \sqrt{2}\pi
 \]

\Cref{eq:selectionpressureRmethod} (and \Cref{eq:selectionpressuremethod}) follows.
\end{proof}

\end{theorem}

\section{Additional Insights into the Performance Evaluation} \label{sec:tableOP}

\paragraph{OP mechanism} In this section, we report additional insights into the performance of OP mechanisms analysed in \Cref{subsec:performanceplacement}: \Cref{fig:placement/performance_all}.
\Cref{tab:op_algorithms} summarizes the mean, minimum, maximum, and quantiles (90, 95, 99\%) of the metrics latency and message overhead for the different OP mechanisms. The table presents the results for Q1, Q4, and Q5 (cf. \Cref{tab:evaluation}) execution using the different \ac{OP} mechanisms.
\begin{landscape}
It can be derived from \Cref{fig:placement/performance} and \Cref{tab:op_algorithms} that Relaxation and MDCEP mechanisms stand representatives for the metrics latency and message overhead, respectively.

\begin{table}[H]
\footnotesize
\centering
\begin{tabular}{p{2cm}p{1.5cm}|l|l|l|l|l|l|l|l|}
\cline{3-10}
\multicolumn{2}{l}{}                                                                                                     & \multicolumn{4}{|l|}{Latency (ms)}                                                      & \multicolumn{4}{l|}{Message Overhead (MB)}                                                   \\ \hline
\multicolumn{1}{|l|}{OP mechanism}                                                                  & Query              & mean  & min & max  & \begin{tabular}[c]{@{}l@{}}quantiles \\ (90, 95, 99)\end{tabular} & mean   & min    & max    & \begin{tabular}[c]{@{}l@{}}quantiles \\ (90, 95, 99)\end{tabular} \\ \hline
\multicolumn{1}{|l|}{\multirow{3}{*}{Relaxation}}                                                  & Stream            & \textbf{4.52}    & \textbf{2}           & \textbf{24}          & \textbf{6, 6, 10}                                                        & 11.03          & 10.3          & 10.3           & 10.3, 10.3, 10.3                                                  \\ \cline{2-10}
\multicolumn{1}{|l|}{}                                                                              & Join               & \textbf{8.98}           & \textbf{4}           & \textbf{34}          & \textbf{12, 14, 19.86}                                                   & 21.38          & 19.12         & 22.83          & 22.83, 22.83, 22.83                                               \\ \cline{2-10}
\multicolumn{1}{|l|}{}                                                                              & Congestion Detection & \textbf{9.23}           & \textbf{6}           & \textbf{34}          & \textbf{12, 13, 19.96}                                                     & 249.52         & 145.74        & 330.16         & 329.77, 330.16, 330.16                                            \\ \hline
\multicolumn{1}{|l|}{\multirow{3}{*}{MOPA}}              & Stream            & 6.19  & 3.0 & 24   & 7, 9, 12               & 11.03  & 10.3   & 11.8   & 11.8, 11.8, 11.8       \\ \cline{2-10}
\multicolumn{1}{|l|}{}                                   & Join               & 12.33 & 7   & 49   & 16, 19, 33.04          & 23.84  & 20.48  & 27.34  & 27.34, 27.34, 27.34    \\ \cline{2-10}
\multicolumn{1}{|l|}{}                                   & Congestion Detection & 19.42 & 12  & 64   & 26, 29.15, 50.32       & 246.23 & 150.6  & 364.15 & 354.23, 263.27, 364,15 \\ \hline
\multicolumn{1}{|l|}{\multirow{3}{*}{\begin{tabular}[c]{@{}l@{}}Global \\ Optimal\end{tabular}}}    & Stream            & 4.62  & 3   & 10   & 5, 6, 8.29             & 7.21   & 7.16   & 7.31   & 7.31, 7.31, 7.31       \\ \cline{2-10}
\multicolumn{1}{|l|}{}                                   & Join               & 9.05  & 5   & 23   & 12.3, 14, 16.86        & 14.31  & 14.24  & 14.49  & 14.49, 14.49, 14.49    \\ \cline{2-10}
\multicolumn{1}{|l|}{}                                   & Congestion Detection & 11.17 & 6   & 51   & 15, 17, 20.68          & 130.5  & 127.45 & 132.48 & 132.48, 132.48,132.48  \\ \hline
\multicolumn{1}{|l|}{\multirow{3}{*}{MDCEP}}                                                         & Stream            & 6.07  & 4  & 38 & 8, 8, 12                                                   & \textbf{1.08}  & \textbf{1.08} & \textbf{1.08}  & \textbf{1.08, 1.08, 1.08}                                         \\ \cline{2-10}
\multicolumn{1}{|l|}{}                                                                              & Join               & 11.07 & 6  & 45 & 15, 17, 24                                                 & \textbf{3.13}  & \textbf{1.92} & \textbf{4.96}  & \textbf{4.96, 4.96, 4.96}                                         \\ \cline{2-10}
\multicolumn{1}{|l|}{}                                                                              & Congestion Detection & 15.68 & 10 & 53 & 21, 25.25, 41.10                                        & \textbf{17.97} & \textbf{6.22} & \textbf{25.04} & \textbf{23.19, 25.04, 25.04}                                      \\ \hline
\multicolumn{1}{|l|}{\multirow{3}{*}{\begin{tabular}[c]{@{}l@{}}Producer \\ Consumer\end{tabular}}} & Stream            & 4.82  & 3   & 14   & 6, 7, 11               & -      & -      & -      & -                      \\ \cline{2-10}
\multicolumn{1}{|l|}{}                                   & Join               & 7.7   & 3   & 24   & 11, 12, 17             & -      & -      & -      & -                      \\ \cline{2-10}
\multicolumn{1}{|l|}{}                                   & Congestion Detection & 10.22 & 5   & 47   & 15, 15, 19             & -      & -      & -      & -                      \\ \hline
\multicolumn{1}{|l|}{\multirow{3}{*}{Random}}            & Stream            & 5.22  & 3   & 23   & 7, 8, 12               & -      & -      & -      & -                      \\ \cline{2-10}
\multicolumn{1}{|l|}{}                                   & Join               & 12.41 & 6   & 36   & 17, 20, 26.03          & -      & -      & -      & -                      \\ \cline{2-10}
\multicolumn{1}{|l|}{}                                   & Congestion Detection & 34.54 & 13  & 1036 & 25.6, 33, 1022.3       & -      & -      & -      & -                      \\ \hline
\end{tabular}
\caption{Performance results of \ac{OP} mechanisms Relaxation and MDCEP are among the best in comparison to its alternative mechanisms.}
\label{tab:op_algorithms}
\end{table}
\end{landscape}
\paragraph{Transition cost per operator} \Cref{tab:Ttime_overhead_op} elaborates on the  statistics of the transition cost for different operators presented in \Cref{subsec:TCEPperformance}. It summarizes the mean, minimum and maximum values for the transition cost in time and overhead for the different operators in Q5: congestion detection query (cf. \Cref{fig:cost_operator}).
Intuitively, the stateful operators like Sequence require a higher amount of state to be migrated compared to stateless operators like Stream.
\begin{table}[H]
\footnotesize
\begin{tabular}{llllllll}
\toprule
\multirow{2}{*}{Operator} & \multicolumn{3}{l}{Transition time (in ms)} & \multicolumn{3}{l}{Transition overhead (in MB)} &  \\
 & mean & min & max & mean & min & max &  \\
 \toprule
Average & 59.43 & 15 & 411 & 5.02 & 13.679 bytes & 64.72  &  \\
Conjunction & 356.89 & 119 & 1404 & 6.2  & 5.53 bytes &  \textbf{130.98} &  \\
Sequence & \textbf{185.32} & 15 & 556 & \textbf{15.06} & 35.507 bytes &  \textbf{129.9} &  \\
Stream & 245.38 & 8 & 913 & 1.67 & 1.315 bytes & 64.72 & \\
\bottomrule
\end{tabular}
\caption{Mean, min and max values of transition time and overhead per operator for 10 incrementally deployed Q5 queries.}
\label{tab:Ttime_overhead_op}
\end{table}

\paragraph{Transition cost for different transition algorithms}
\Cref{tab:Ttime_overhead_strategy} elaborates on the statistics of transition cost for different transitions algorithms as presented in \Cref{fig:trstr_cost_operator}. It summarizes the mean transition time and overhead required by the different algorithms. Clearly, the SMS strategies supersede both in terms of cost in time and overhead.

\begin{table}[H]
\footnotesize
\centering
\begin{tabular}{llll}
\toprule
 & Operator & \begin{tabular}[c]{@{}l@{}}Mean transition time\\ (in ms)\end{tabular} & \begin{tabular}[c]{@{}l@{}}Mean transition \\ overhead (in MB)\end{tabular} \\
 \toprule
\multirow{4}{*}{MFGS Sequential} & Average & 105.13 & 20 \\
 & Conjunction & 607.49 & 24.74 \\
 & Sequence & 287.50 & 60.12 \\
 & Stream & 676.50 & 6.67 \\
 \midrule
\multirow{4}{*}{MFGS Concurrent} & Average & 37.06 & 0.05 \\
 & Conjunction & 445.98 & 0.05 \\
 & Sequence & 210.30 & 0.107 \\
 & Stream & 163.78 & 0.005 \\
 \midrule
\multirow{4}{*}{SMS Sequential} & Average & 66.66 & 0.007 \\
 & Conjunction & 219.56 & 0.009 \\
 & Sequence & 243.7 & 0.023 \\
 & Stream & 352.6 & 0.002 \\
 \midrule
\multirow{4}{*}{SMS Concurrent} & Average & \textbf{41.1} & \textbf{23.74 Bytes} \\
 & Conjunction & \textbf{158.2} & \textbf{21.13 Bytes} \\
 & Sequence & \textbf{158.3} & \textbf{41.86 Bytes} \\
 & Stream & \textbf{85.1} & \textbf{2.41 Bytes} \\
 \bottomrule
\end{tabular}
\caption{Mean values for transition time and overhead for MFGS and SMS strategies. SMS strategies clearly supersede both time and overhead required to transfer the operator.}
\label{tab:Ttime_overhead_strategy}
\end{table}

\end{appendices}

\end{document}